\documentclass[twocolumn,longbibliography,amsmath,amssymb,aps,pre]{revtex4-2}
\usepackage{dcolumn}
\usepackage{amsmath}
\usepackage{amssymb}
\usepackage{bm}
\usepackage{graphicx,subfigure}
\graphicspath{{./figures/}}
\usepackage{epsf,epsfig}
\usepackage{hyperref}
\usepackage{color}
\usepackage{enumerate}
\usepackage{lineno}

\begin{document}
\title{Enhanced extracellular matrix remodeling due to embedded spheroid fluidization}
\author{Tao Zhang$^1$, Shabeeb Ameen$^2$, Sounok Ghosh$^2$, Kyungeun Kim$^2$, Minh Thanh$^1$, Alison E. Patteson$^1$, Mingming Wu$^3$, and J. M. Schwarz$^{1,4}$}
\affiliation{$^1$School of Chemistry and Chemical Engineering, Shanghai Jiao Tong University, Shanghai 200240, China\\
$^2$Department of Physics and BioInspired Institute, Syracuse, Syracuse University, Syracuse, NY 13244, USA\\
$^3$Department of Biomedical Engineering, Cornell University, Ithaca, NY 14850, USA\\
$^4$Indian Creek Farm, Ithaca, NY, 14850, USA}
\date{\today}

\begin{abstract}

Tumor spheroids are {\it in vitro} three-dimensional, cellular collectives consisting of cancerous cells. Embedding these spheroids in an {\it in vitro} fibrous environment, such as a collagen network, to mimic the extracellular matrix (ECM) provides an essential platform to quantitatively investigate the biophysical mechanisms leading to tumor invasion of the ECM. To understand the mechanical interplay between tumor spheroids and the ECM, we computationally construct and study a three-dimensional vertex model for a tumor spheroid that is mechanically coupled to a cross-linked network of fibers. In such a vertex model, cells are represented as deformable polyhedrons that share faces. Some fraction of the boundary faces of the tumor spheroid contain linker springs connecting the center of the boundary face to the nearest node in the fiber network.  As these linker springs actively contract, the fiber network remodels. By toggling between fluid-like and solid-like spheroids via changing the dimensionless cell shape index, we find that the spheroid rheology affects the remodeling of the fiber network. More precisely, fluid-like spheroids displace the fiber network more on average near the vicinity of the spheroid than solid-like spheroids. We also find more densification of the fiber network near the spheroid for the fluid-like spheroids. These spheroid rheology-dependent effects are the result of cellular motility due to active cellular rearrangements that emerge over time in the fluid-like spheroids to generate spheroid shape fluctuations. These shape fluctuations lead to emergent feedback between the spheroid and the fiber network to further remodel the fiber network with, for example, lower radial alignment of the higher-tensioned fibers given the breaking of spheroidal radial symmetry, which can then further remodel the spheroid. Our results uncover intricate morphological-mechanical interplay between an embedded spheroid and its surrounding fiber network with both spheroid contractile strength {\it and} spheroid shape fluctuations playing important roles in the pre-invasion stages of tumor invasion. 
\end{abstract}

\maketitle

\section{Introduction}
\label{sec_introduction}

Cancer, in its diversity of forms, is a complex phenomenon. It is therefore fitting to deconstruct it into pieces, understand those pieces, and then put the pieces back together again, keeping in mind that strong interactions between the pieces can yield unexpected behaviors. One way to deconstruct cancer is to focus on {\it in vitro} models. A key \textit{in vitro} model system for studying the combined effects of cell-cell and cell-extracellular matrix interactions in a cancer-like setting is an embedded spheroid~\cite{Haeger2014,Tevis2017}.  More specifically, a spheroid consisting of cancerous cells is surrounded by a collagen network. One of the major goals for studying such a system is to be able to predict whether or not the tumor cells will invade the surrounding collagen. It has been shown that stiffer collagen fibers and collagen density influence tumor invasion ~\cite{Suh_2019}. Moreover, tumor spheroids under perfusion demonstrate that interstitial flow can also affect tumor invasion as can mechanical compression~\cite{Pandey2023,Huang2020}. In terms of good predictors of tumor invasion potential, some have argued that cell shapes in the spheroid, as opposed to two-dimensional cell migration assays, may be a good predictor~\cite{Baskaran2020}. 

Despite the recent increase in experimental work on embedded spheroids, many theoretical and numerical models for cells interacting with the extracellular matrix (ECM) have focused on single cells, potentially missing important features associated with collective cell behavior \cite{Hall2016,Steinwachs2016, Kim_2018, Han_2018,Doyle2021}. And yet, models for bulk tissue and collagen networks abound. One such bulk tissue model is a cell-based, vertex model \cite{Honda_2001, Farhadifar_2007,Bi_2015,Okuda2013,Zhang2022}, which predicts a density-independent rigidity transition in disordered confluent tissues and micro-demixing in tissue mixtures ~\cite{Bi_2015,Sahu_2019}. These predictions have been verified experimentally~\cite{Park_2015,Malinverno_2017, Sahu_2019}. As for modeling the ECM, collagen networks are well-approximated by a spring network model containing energetic costs to stretching and bending, otherwise known as fiber networks ~\cite{Licup_2015,Sharma_2016,Jansen_2018}. Strain-stiffening in under-constrained fiber networks has been predicted at finite strain and has been experimentally confirmed~\cite{Sharma_2016, Jansen_2018}. 

Given the successes of the vertex model and the fiber network model, a natural next step to approximate an embedded spheroid is to couple the two models. In fact, there exists prior work in that direction in two dimensions using a vertex model with interfacial tension at the boundary of a spheroid that couples to a stretchable, three-fold coordinated spring network~\cite{Parker2020}.  Detailed analysis of this two-dimensional model found two regimes with different global spheroid shapes resulting from competition between interfacial tension and tension in the network. In the interfacial tension-dominated regime, the spheroid remains compact with compression-induced fluidization. Interestingly, compression-induced fluidization has also been predicted by others studying the rheology of a similar two-dimensional vertex model in response to oscillatory shear~\cite{Tong2022}.  In the spring network tension-dominated regime, a cavitation-like instability leads to the emergence of gaps at the spheroid-spring network interface to minimize the energy of the coupled system via cell rearrangements at the boundary. Both the compression-induced fluidization and the cavitation-instability are experimentally-testable predictions. The latter prediction may require inhibiting the pathway for cells to make ECM. Moreover, the experimental finding that cells fluidize when surrounded by cancer-associated fibroblasts is consistent with the compression-induced fluidization prediction~\cite{Barbazan2023}. As for how does the spheroid affect the mechanics of the spring network, one finds that the location of the spring network's rigidity transition is sensitive to the spheroid size, mechanical properties and surface tension~\cite{Parker2024}. 

While the two-dimensional embedded spheroid model indeed provides insights into the richness of the system, one wonders about the mechanical interplay between the spheroid and the ECM when the spring network is a fiber network with bending stiffness and when there are explicit interactions between the cells and the ECM. To be specific, what happens when there is an extra degree of freedom coupling the cell to the collagen fibers with a molecular clutch-like mechanism as has been determined in single cells coupling to the ECM~\cite{Chan2008}. Moreover, one wonders how a three-dimensional coupled model would behave as it is not necessarily obvious that one cross-section of a spheroid behaves similarly as another cross-section.  Our new computational model addresses these issues.

\begin{figure*}[t]
    \centering
    \includegraphics[height=0.4\textwidth]{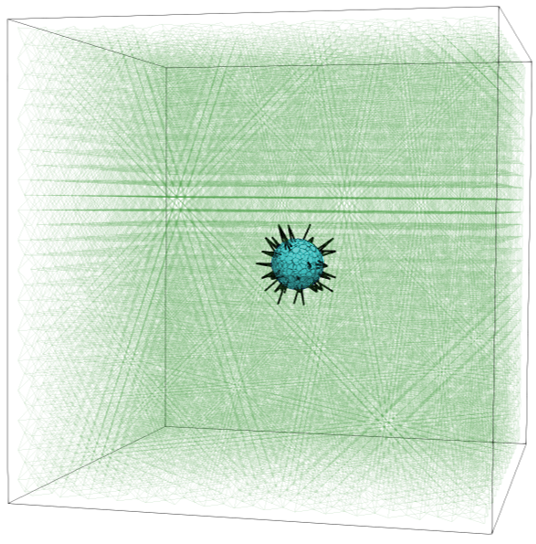}
    \hspace{0.5cm}
    \includegraphics[height=0.4\textwidth]{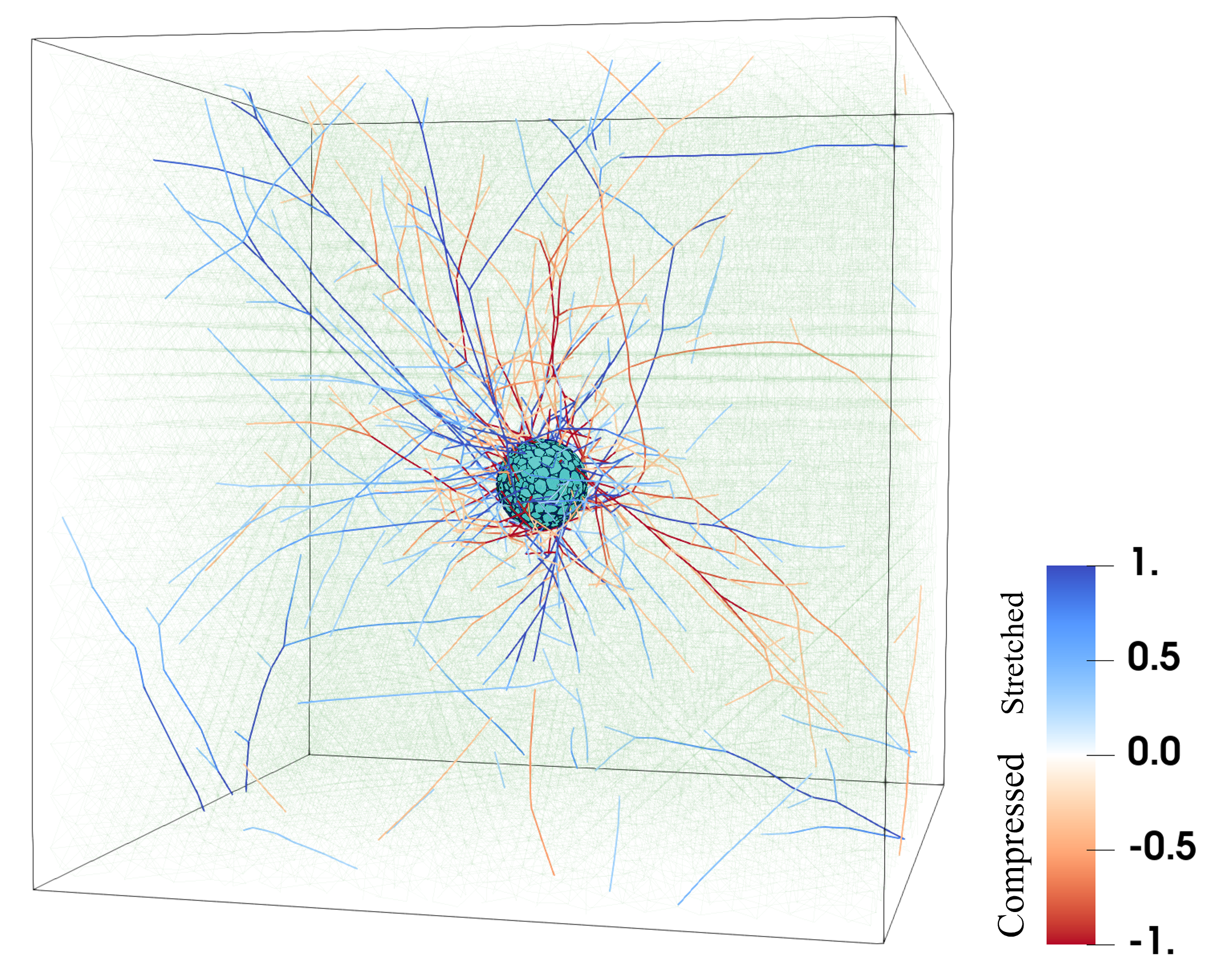}
    \caption{\textit{Three-dimensional computational model for an embedded spheroid}. Left: Initial simulation configuration in which the cells making up the spheroid are shown in cyan polyhedrons, collagen fibers are shown with thin, light green lines and initially occupy a FCC lattice, and the active linker springs connecting the spheroid to the fiber network are denoted with thick, black lines. Right: The embedded spheroid system at the final time of the simulation. The color scale of the collagen fibers denotes the normalized tension $\tau/\tau_0$ in the fibers with blue indicating extension and red indicating compression and $\tau_0$ denoting the square root of the variance in the tension.} 
    \label{fig:Simulation_Snapshot_1}
\end{figure*}

Here we couple, for the first time to our knowledge in three dimensions, a cell-based vertex model to a surrounding fiber network model with surface tension at the interface via active linker springs in three dimensions. These active linker springs contract the fiber network by decreasing their rest length as a function of time once they attach. We ask the question: {\it Given that tissues can toggle between solid-like behavior and fluid-like behavior, can such changes in spheroid rheology affect the remodeling of the collagen network?}  To begin to answer this question, we explore how the fiber network remodels for different parts of the parameter space of the model. The extent of the remodeling can subsequently impact spheroid rheology, to greater or lesser extents, and potentially cell breakout, particularly given potential feedback between the two. Finally, in three-dimensions, there is work coupling an elastic spheroid to a fiber network~\cite{Mark2020}. This model has been used to determine the forces in the fiber network that contains bending, stretching, and buckling in the presence of a spheroid with a specific rheology.

\section{Computational Model} 
 As our model is composed of a vertex model embedded within a fiber network (see Figure 1 for a snapshot of the initial configuration of the computational model and a final configuration), let us first discuss the energy functional for each constituent piece. We will use this energy functional to compute forces on each vertex, in both the spheroid and the fiber network, to determine how they move in response to each type of material interacting with the other.

\subsection{A vertex model for the spheroid} \label{sec_vertex_model_review}

The biomechanical properties of the spheroid are described by the energy functional:
\begin{equation}
\label{eq:energy}
E_{VM}= K_V\sum_{j}(V_{j} - V_{0})^2+K_A\sum_{j}(A_{j} -A_{0})^2 +\Gamma\sum_{\alpha}\delta_{\alpha,B} A_{\alpha},
\end{equation}
where $A_{j}$ and $V_{j}$ represent the area and volume of the $j$th cell, respectively. The terms $K_V$ and $K_A$ are volume and area stiffness coefficients. The volume term, with target volume $V_0$, corresponds to the cell's bulk elasticity, while the area term, with target area $A_{0}$, relates to the acto-myosin cortex's isotropic contractility. A larger $A_{0}$ implies reduced isotropic contractility, suggesting that all cell faces are equivalent in this aspect. This contrasts with two-dimensional models, where a larger effective target perimeter $P_0$ (assuming $A_0=1$) indicates a balance between cell-cell adhesion and contractility. Cell-cell adhesion and contractility are coupled, as demonstrated by changes in contractility upon the disruption of E-cadherin in keratinocytes~\cite{Hoffman2015, Sahu2020a}. However, due to shared edges in the vertex model, direct tuning of cell-cell adhesion is not feasible, necessitating the assumption of isotropic contractility. Reduced isotropic contractility might lead to anisotropic contractility, influenced by factors like stress fibers~\cite{Warmt2021}.

When modeling spheroids, it is important to consider their boundaries. 
To address this, we construct
a confluent cellular collective, or a clump of confluent cells, by
making a cut-out of the bulk periodic
system that contains cells with empty cells beyond the boundary
between cells and empty space (See our prior work for more details~\cite{Zhang2022}). The cut-out is initially a sphere.  For those cells at the boundary of the spheroid, the interfacial vertices contain an additional
interfacial surface tension $\Gamma$. 
$\alpha$ indexes the cell faces. The Kronecker delta $\delta_{\alpha,B}$ equals 1 for boundary faces and 0 otherwise. 

Lengths in the model are nondimensionalized with $l=V_0^{1/3}$. Time in the model is nondimensionalized with $t_0$. The dimensionless shape index $s_{0} = A_{0}/(V_{0})^{2/3}$ is a key parameter. For instance, a regular tetrahedron has $s_0\approx 7.2$.

\subsection{Fiber network model for the collagen matrix} \label{sec_spring_network_model_review}

To create a disordered network of crosslinked fibers, we first occupy a face-centered cubic lattice with bonds. Each bond is assigned an extensional spring constant denoted as $K_{\text{S}}$. Additionally, every pair of neighboring bonds aligned by 180 degrees is endowed with a bending modulus $K_{\text{B}}$. 
To introduce a finite fiber length $L$ within this framework, we randomly and independently dilute each bond in the lattice with probability $1-p$. In other words, if $p=1$, then the FCC lattice is fully occupied with bonds. At the junctures where two fibers intersect, a freely-rotating crosslink is established. The crosslink prevents the fibers from sliding relative to each other, thereby maintaining the structural integrity of the network.
However, since the FCC lattice contains twelve nearest neighbors, or six fibers, to crosslink only two fibers, one fiber network node is broken up into 3 separate phantom nodes such that not more than two fibers are crosslinked. Although the different pairs of cross-links may overlap geometrically, they do not constrain each other. 

Given these ingredients, the energy functional of the fiber network is   
\begin{eqnarray}
E_{FB}=\frac{K_{\text{S}}}{2}\sum_{<ij>} f_{ij}\;(l_{ij}-l_0)^2+\\ \nonumber\\
\frac{K_{\text{B}}}{2}\sum_{m=1}^{3}\sum_{<ijk>=\pi_{0}}f_{ij}f_{jk}\;(\theta_{ijk}^m-\pi)^2,
\end{eqnarray}
where $f_{ij}=1$ if a bond is occupied and $f_{ij}=0$ if not, $l_{ij}$ represents the length of each bond, $\sum_{\langle ij \rangle}$ represents sum over all nearest neighbor bonds, $\theta_{ijk}^m$ represents the angle between nearest neighbor bonds that are crosslinked by the same phantom node with phantom node index $m$. Moreover, $\sum_{\langle ijk \rangle=\pi_0}$ represents sum over pairs of nearest neighbor bonds sharing a node and only for those aligned along of the principle axes of the initial FCC lattice. The first term in $E_{FN}$ corresponds to the energy cost of extension or compression of the bonds, while the second term to the energy penalty for the bending of fibers segments made of the three possible pairs ($m=1-3$) of adjacent collinear bonds forming whose initial angle is 180 degrees. Note that there is no energy cost to fiber twisting.

\subsection{Embedded spheroid as a vertex model coupled to a fiber network}
The spheroid consists of $N_c$ total cells, while the fiber network comprises $N_f$ nodes. To integrate these two systems, we establish a coupling mechanism. There are $N_{LS}$ linker springs with each connecting a fiber network node to the nearest boundary spheroid cell's surface polygon center. The equilibrium length of each linker spring $i$, $l'_{0i}$, is time-dependent, decreasing progressively at a constant strain rate until reaching a minimum equilibrium length, $l'_{0min}$.  This time dependence captures the molecular clutch framework to focal adhesions in which integrins attach to the ECM fibers~\cite{Chan2008}. Since the focal adhesion is coupled to the acto-myosin cortex, contractile activity emerges.  Changing equilibrium spring lengths has been used previously to encode acto-myosin contractility~\cite{Lopez2014,Mayett2017}. As one boundary cell cannot have more than one linker spring, there is an upper bound on $N_{LS}$. For the parameters we are working with generally about 50 percent of the boundary, or surface cells, contain active linker springs. Given these assumptions, the total energy of the active linker springs is quantified by:
\begin{equation}
    E_{LS}=\frac{K_{LS}}{2}\sum_{i=1}^{N_{ls}} (l'_i-l'_{0i}(t))^2,
\end{equation}
where $K_{LS}$ represents the stiffness of each linker spring. Initially, $l'_{0i}$ is set at $1.5\,V_0^{1/3}$, which is approximately the spacing of the initial FCC lattice, and then linearly decreases to a minimum of $0.2\,V_0^{1/3}$ over time interval $5000\,t_0$. The active linker spring is, therefore, dynamic over this time after which time it is not dynamic, unless the active linker spring is removed or created in several specific cases. 

If a boundary cell undergoes a reconnection event to become an interior cell, its associated active linker spring is removed. If an active linker spring is removed, another one is created so that $N_{LS}$ is a conserved quantity. The new active linker will be chosen from the boundary cells that do not yet contain one.  In addition, if the surface polygon of a boundary cell shrinks to an area less than 0.1 or deforms into a concave shape where the polygon center lies on a polygon edge, and the associated active linker spring is consequently removed. These processes for the creation and removal of active linker springs ensure continuous adaptation and restructuring of the linker springs in response to the evolving configuration of the embedded spheroid.

\subsection{Embedded Spheroid System}

Finally, we arrive at the total energy for the embedded spheroid system is given by:
\begin{equation}
    E_{ES}=E_{VM}+E_{FB}+E_{LS}. 
\end{equation}
To study this energy functional, we will use a dynamical approach, as opposed to energy minimization, which is detailed in the next section. 

\section{Methods}

\subsection{Simulations}

Cell dynamics are integral to the model. Indeed, cells are capable of moving past each other while maintaining the confluence of the tissue. In two dimensions, such movements are termed T1 events, and they contribute to understanding the rigidity transition~\cite{Bi2015}. In three dimensions, these movements are referred to as reconnection events~\cite{Okuda2013,Zhang2022}. Prior work has developed
an algorithm for such reconnection events focusing on edges becoming
triangles and vice versa that can occur for edges below a threshold length $l_{th}$, or more precisely, if the maximum length of the edges involved in the reconnection event becomes shorter than the threshold length~\cite{Okuda2013}. There are additional 
conditions that must be met to ensure that the reconnection event is
physically plausible, namely, that the event is geometrically, energetically,
and topologically reversible~\cite{Okuda2013}. For instance, to ensure topologically reversibility, the constraint that no two polyhedral cells can simultaneously share two or more polygonal faces was recently added to the list of conditions~\cite{Zhang2022}. Other types of reconnection events may be possible to explore using the graph vertex model where the topology of the network is stored in a knowledge graph~\cite{Sarkar2023}.

Beyond reconnection events, the model incorporates explicit Brownian dynamics for each cell vertex. The equation of motion governing the position ${\bf r}_I$ of a cell vertex $I$ is expressed as:
\begin{equation}
\dot{\mathbf{r}}_{I} = \mu \mathbf{F}_{I} + \mu \mathbf{F}_{I}^B, \label{eq:dyn}
\end{equation}
where $\mathbf{F}_{I}$ represents the conservative force and $\mathbf{F}_{I}^B$ the random force acting on the $I$th vertex. The conservative force $\mathbf{F}_I$ arises from area and volume constraints, encompassing cell-cell interactions, as well as the interfacial tension between the spheroid and the fiber network and the active linker spring interaction. Furthermore, each cell vertex $I$ undergoes random fluctuations as encoded in an effective diffusion coefficient $\mu k_B T_{eff}$, with $T_{eff}$ denoting an effective temperature. This effective temperature captures the force fluctuations due to the internal, active mechanisms of a cell and is much smaller than the conservative force contribution. Alternatively, one can interpret these fluctuations as a means to probe the complex energy landscape of the conservative forces. Unless otherwise specified, the mobility $\mu=1$. 
The positional updates for each cellular vertex in the model are executed using the Euler-Murayama integration method. 

As for the nodes of the fiber network, they are updated us-
ing Euler’s method, based on the forces acting on them due to the other fibers and the active linker springs. We employ overdamped dynamics. There is no ac-
tive force fluctuation contribution to the fiber network nodes.  

A spheroid is then created from a bulk, random Voronoi tessellation from which cells are cut out from. Similarly, edges of the fiber network are removed from the center so that the spheroid can be inserted.  The active linker springs are then determined (for a given fraction of boundary faces). After a specific number of time steps, the linker springs begin to contract to their smaller target spring length. Figure 1 shows an initial configuration of the simulation and a final configuration of the simulation. For comprehensive details on the parameters employed in our simulations, including their specific values, refer to Table I. To study this coupled system, the parameters are chosen based on prior simulations of fiber networks and on computational efficiency. To convert simulation units to biophysical units, one simulation length unit $V_0^{1/3}$ is approximately 10 microns, and one simulation time unit $t_0$ is approximately 0.144 seconds (the simulation time in total $t_f=25000t_0$ is approximately 1 hour), while one simulation force unit is approximately one nanoNewton.  We have chosen to plot quantities in their dimensionless form. While there are a number of aspects of this computational model to be explored, {\it here we will study the system as a function of $s_0$ and as a function of the interfacial tension.} In future work, we will vary the fiber density to be compared with experiments that vary collagen density~\cite{Suh2019}.

\begin{table*}
\begin{center}
\begin{tabular} { l | c | c }
{\bf Quantity} & {\bf Symbol} & {\bf Value }\\
 
 Simulation timestep & $\frac{dt}{t_0}$ & $0.005$ \\   
 Simulation time in total & $\frac{t_f}{t_0}$ & $25000.$ \\
 Cell area stiffness & $\frac{K_A}{K_V\,V_0^{2/3}}$ & $0.01$ \\   
 Cell target surface area & $s_0$ & 5.2-5.8 \\  
 Boundary interfacial tension & $\frac{\Gamma}{K_V\,V_0^{4/3}}$ & 0.25,1.0 \\
 Reconnect. Event threshold edge length & $\frac{l_{th}}{V_0^{1/3}}$ & $0.02$ \\
 Damping & $\frac{K_V\,V_0^{4/3}\,t_0}{\mu}$ & $1$\\   
 Active force fluctuation energy & $\frac{k_BT_{eff}}{K_V\,V_0^2}$ & $10^{-4}$ \\
 Individual fiber bending stiffness & $\frac{K_B}{K_S l'_{0}(0)^{2}}$ &  $10^{-4}$\\
 Individual fiber diameter & $\frac{d_f}{l'_0(0)}$& 0.02\\
 Fiber network pore size & $\frac{\xi}{l'_0(0)}$ & 1.0\\
 Edge occupation probability for fiber network & $p$ & $0.8$\\
 Active linker spring stiffness & $\frac{K_L}{K_S \,l_0(0)'^2}$ & 1.0\\
 Final active linker spring target length & $\frac{l'_{t0}}{l'_0(0)}$ & 0.2\\
Number of spheroid cells & $N$ & $~400$ \\
 Number of active linker springs & $N_L$ & 100\\
 Number of fiber network nodes & $N_f$ & $~1000$\\
 Number of realizations & $N_R$ & $20$\\
\end{tabular}
\caption{Table of the dimensionless parameters used in the simulations.}
\end{center}
\end{table*}

\subsection{Measurements}
\begin{figure*}[t]
    \centering
    \includegraphics[width=0.45\textwidth]{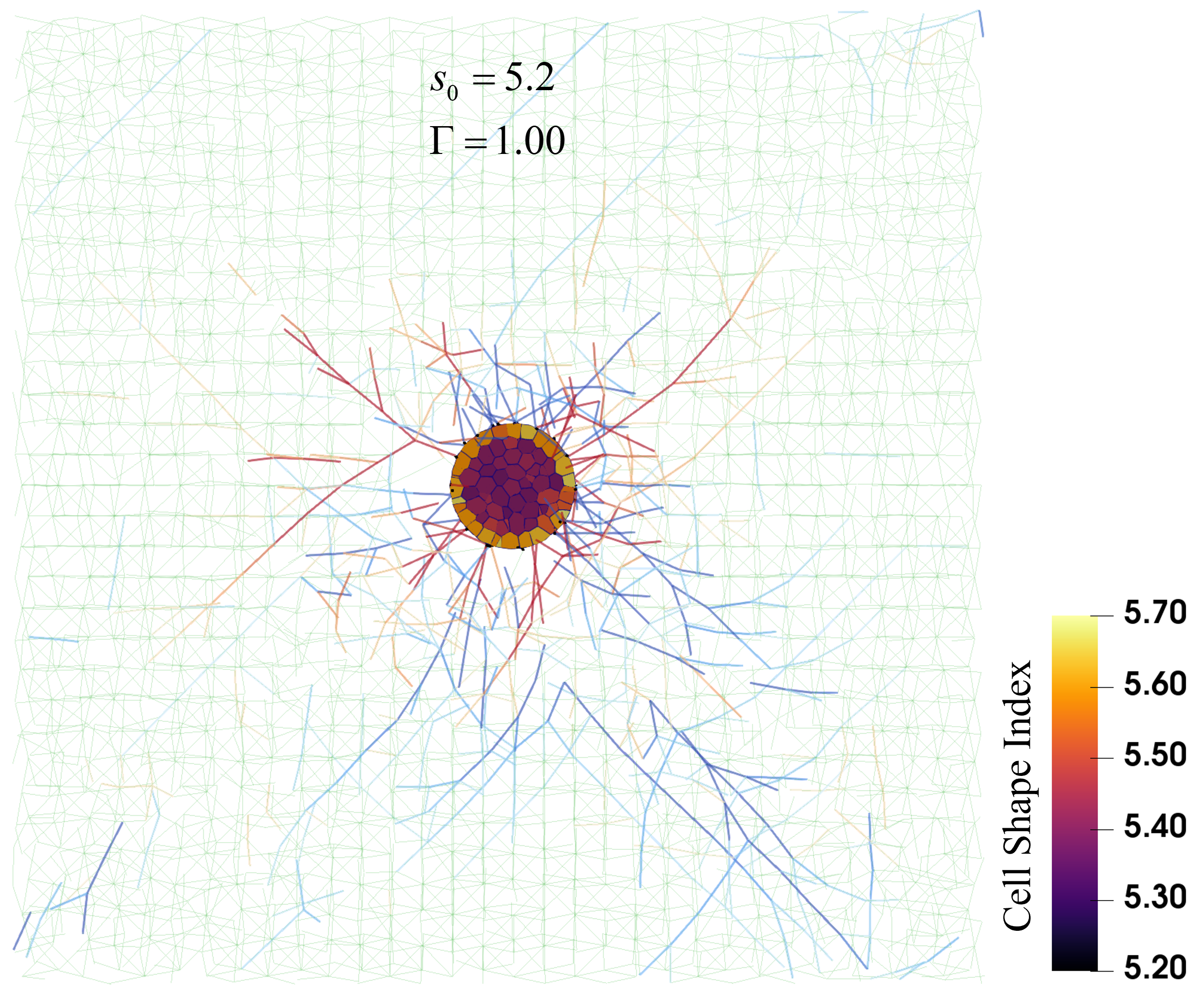}
    \hspace{0.5cm}
    \includegraphics[width=0.45\textwidth]{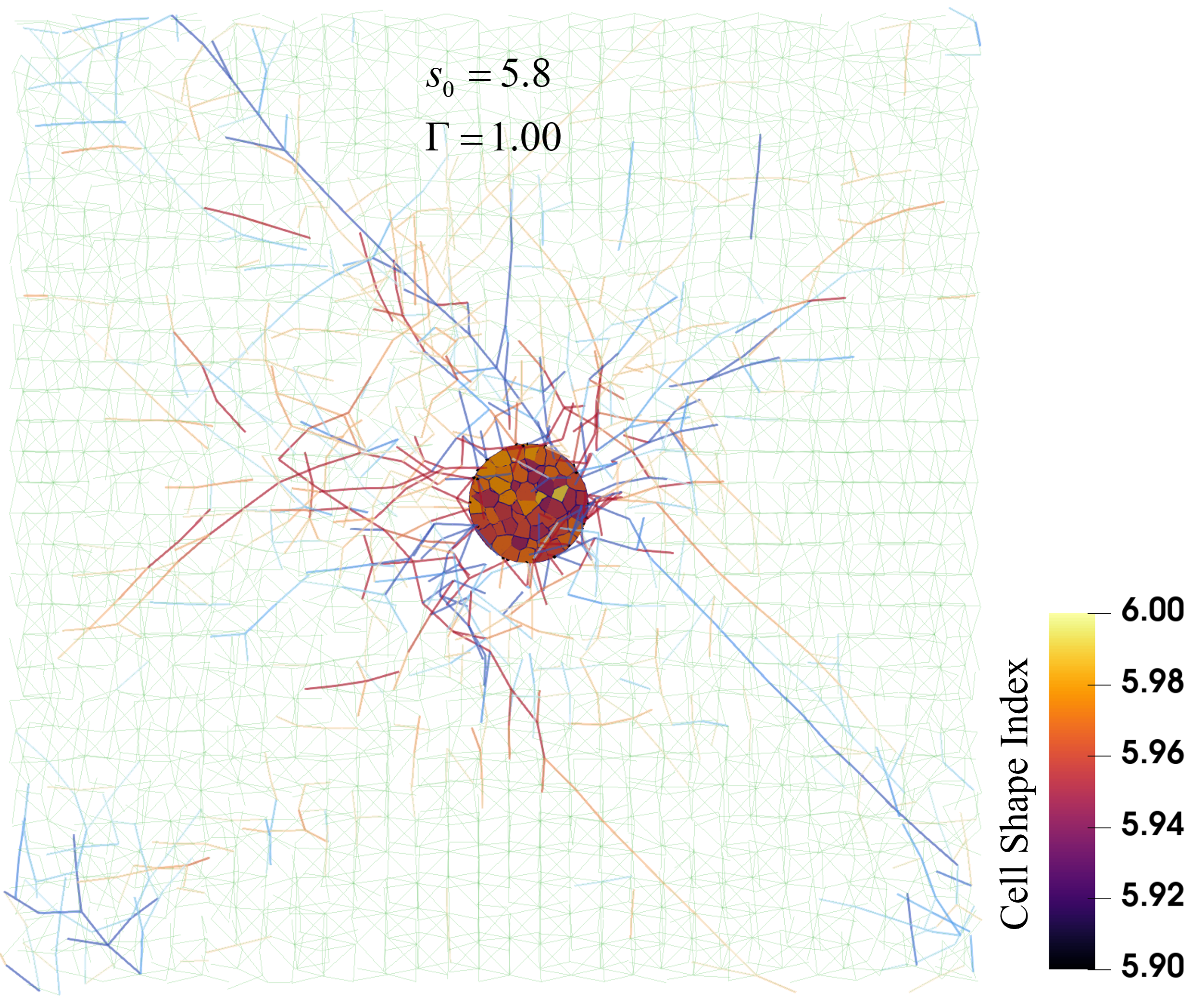}
    \caption{\textit{Cross-section of computational model and cell shape.} Left: Cross-section with cell shape index color map and fiber tension for $s_0=5.2$ and $\Gamma=1.0$. Right: Same image as left but with $s_0=5.8$.} 
    \label{fig:Simulation_Snapshot_2}
\end{figure*}

\begin{figure}[!htbp]
    \centering
    \includegraphics[width=0.45\textwidth]{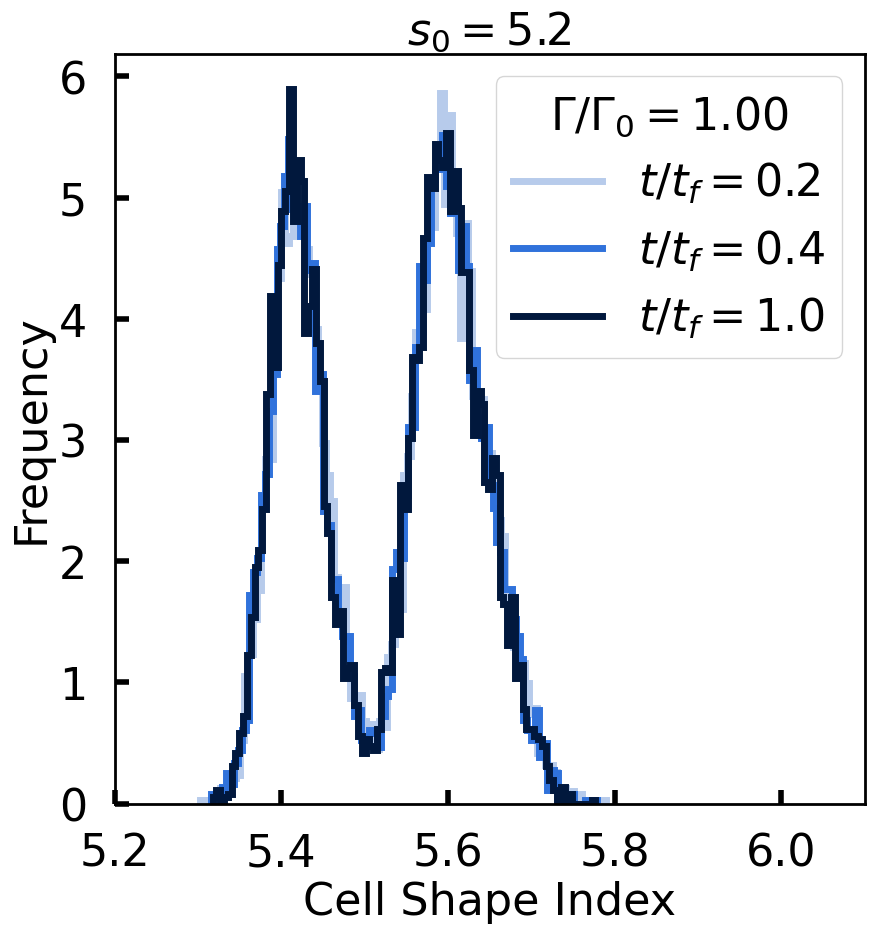}
    \includegraphics[width=0.45\textwidth]{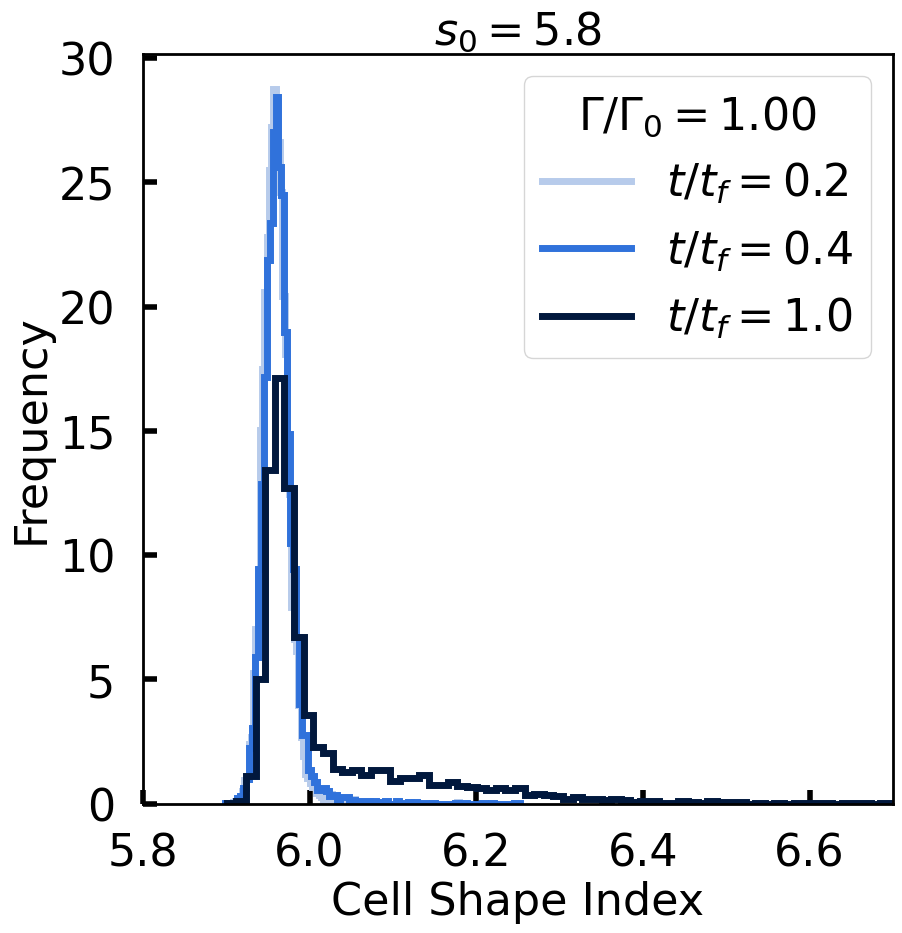}
    \caption{\textit{Individual cell shape index histogram is two-peaked for solid-like spheroids and single-peaked for fluid-like spheroids} Top: Cell shape index histogram for $s_0=5.2$ at different time points in the simulations. The two-peaked individual cell shape index distribution indicates two different populations of cells, higher cell shape index cells at the boundary and smaller cell shape index cells in the bulk. Bottom: Cell shape index histogram for $s_0=5.8$ at the same time points in the simulations as above.  } 
    \label{fig:cell_shape_index_distribution}
\end{figure}

After performing the simulations for a range of target cell shape indices and for two different interfacial tensions, the following quantities were then measured. To quantify how the fiber network is remodeled, we compute the magnitude of the fiber displacement as a function of the distance from the center of mass of the spheroid. We also measure the fiber density as a function of the distance from the center of mass of the spheroid to further quantify the fiber remodeling extent by the spheroid. Finally, in terms of spatial fiber network reorganization, we compute the fiber orientation tensor using spherical coordinates. More specifically, the fiber orientation tensor $\Omega_{\alpha\beta}$ is given by 
\begin{equation}
    \Omega_{\alpha \beta}=\sum \frac{L_{f}n_{\alpha\beta}}{L_{tot}},
\end{equation}
where $\alpha$ and $\beta$ denote the component in three dimensions, $L_f$ denotes fiber length, $L_s$ denotes the total system length, and $\hat{n}$ denotes the unit vector along the fiber axis.  Should the fiber network be isotropically organized, $\Omega_{11}=\Omega_{22}=\Omega_{33}=1/3$. 

We explore how certain components of the fiber orientation tensor in spherical coordinates, such $\Omega_{rr}$, behaves on average as a function of distance from the center of mass of the spheroid. Moreover, as the fiber network is spatially remodeled, tension and compression will build up in the fiber network, particularly as the active linker springs contract to pull on the fiber network. We record the amount of tension/compression in the fiber network as the coupled system evolves in time. To understand how the active linker springs mediate the interaction between the spheroid and the fiber network, we also record the tension/compression in each of the linker springs over time. 

Given that we have cellular-scale resolution of the spheroid, we will not only keep track of the overall spheroid shape index to determine how far from spherical it deviates, we will also study the shape index for the individual cells as a function of time. To quantify the amount of cell motion within the spheroid, we will compute displacements of the cell center of masses over time, which gives us a measure of the spheroid displacement. Finally, we also record the fraction of cells that lose two or more neighbors over time~\cite{Zhang2022}. More fluid-like spheroids undergo more cellular rearrangements than solid-like ones. 

\begin{figure}[!htbp]
    \centering
    \includegraphics[width=0.45\textwidth]{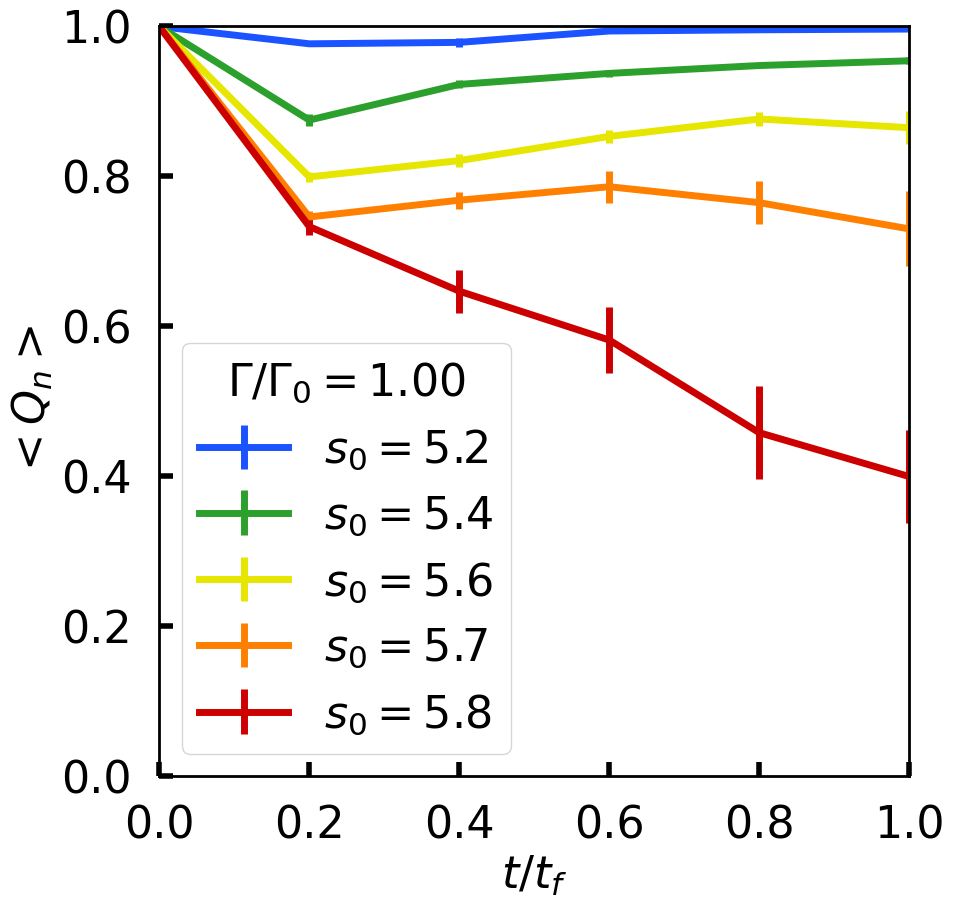}   
    \caption{\textit{Fluid-like spheroids indicate a larger fraction of cells  undergoing cellular rearrangements as a function of time}. Plot of $<Q_n>$ as a function of time for different $s_0$s.} 
    \label{fig:overlap_function}
\end{figure}

\section{Results}

\begin{figure}[!htbp]
    \centering
    \includegraphics[width=0.457\textwidth]{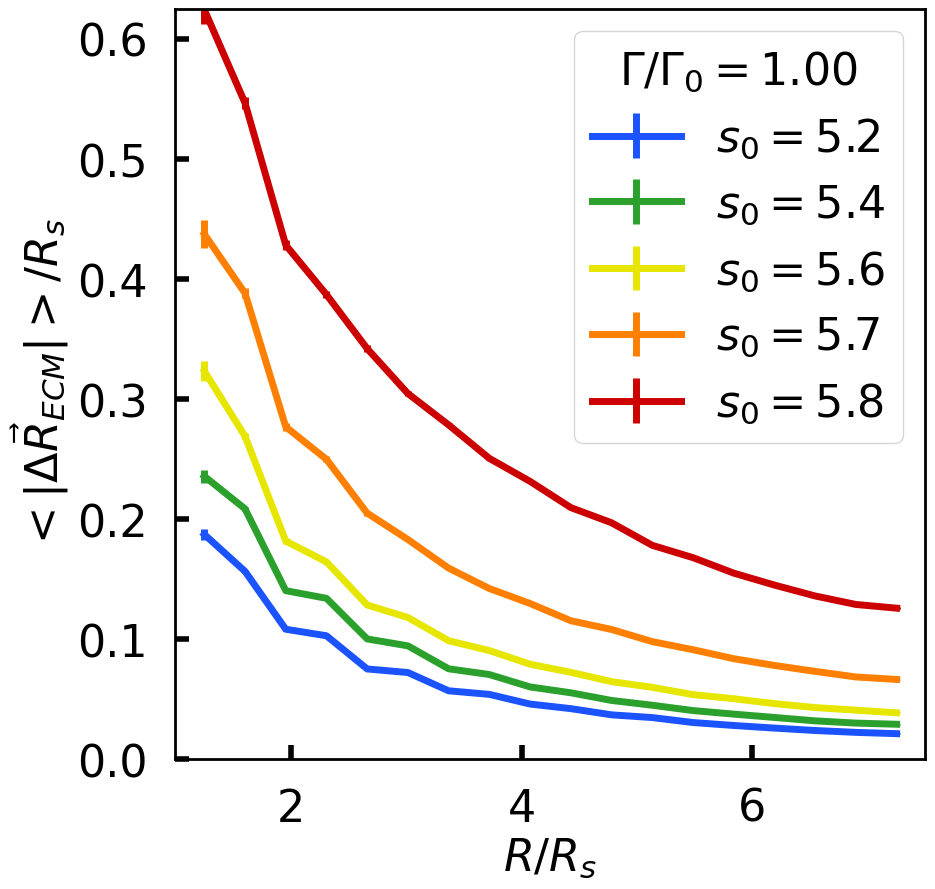}
    \includegraphics[width=0.45\textwidth]{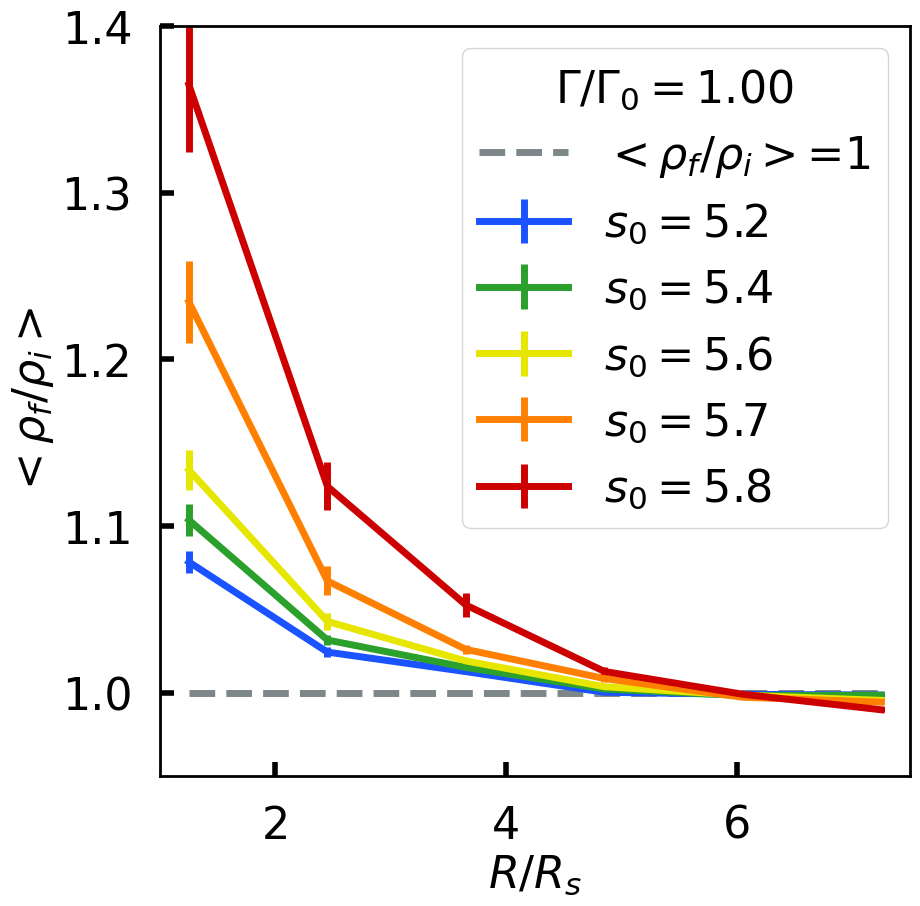}
    \caption{\textit{Fluid-like spheroids displace and densify the fiber network near the spheroid more than solid-like spheroids} Top: Magnitude of fiber displacement as a function of radial distance $R$ from the center of mass of the spheroid. As more fluid-like spheroids are quantified by a larger target cell shape index, $s_0$s, the fiber positions are displaced more the larger the target cell shape index. Bottom: The larger the target cell shape index $s_0$, the more the fibers are densified towards the center of the system. Note that $\rho_i$ denotes initial fiber density, $\rho_f$ denotes the final fiber density, and $R_s$ denotes the radius of the embedded spheroid system. These results are a dimensionless interfacial surface tension, $\Gamma/\Gamma_0=100$.  } 
    \label{fig:fiber_displacement_density_100}
\end{figure}

\begin{figure}[!htbp]
    \centering
    \includegraphics[width=0.45\textwidth]{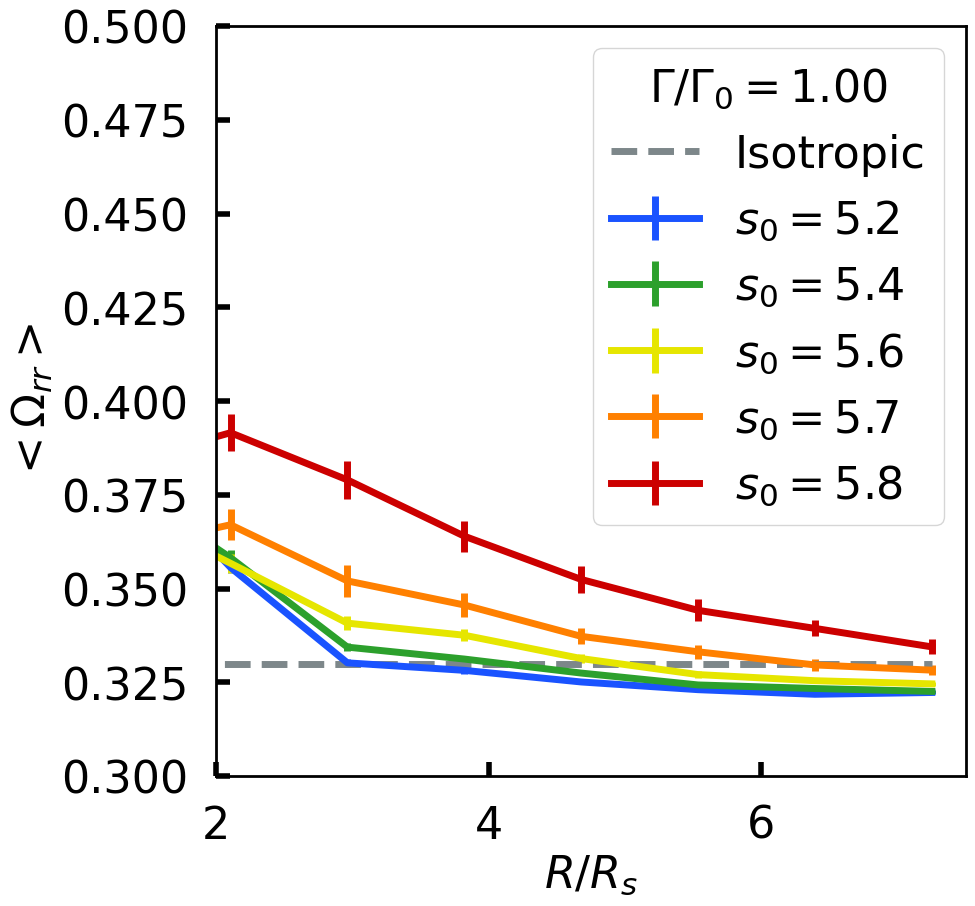}
    \includegraphics[width=0.45\textwidth]{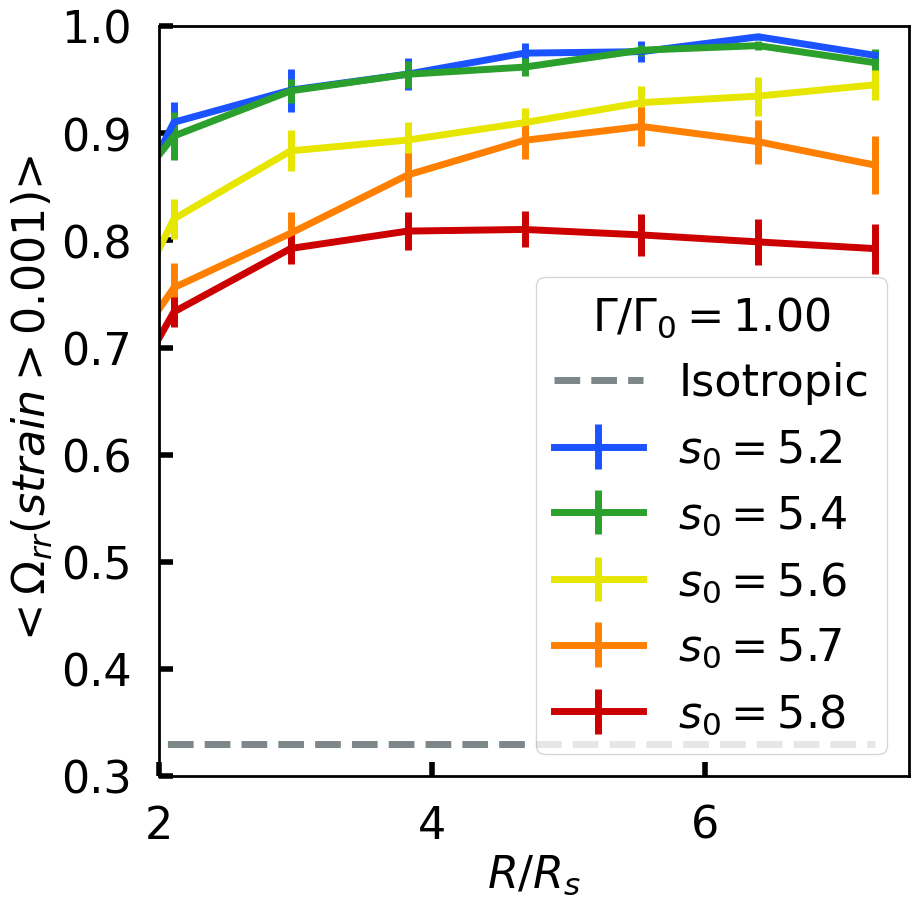}
    \caption{\textit{Fluid-like spheroids radially orient fibers overall more so that solid-like spheroids with a different trend for high-tensioned fibers.} Top: Plot of the radial component of the fiber orientation tensor as a function of radial distance $R$ from the center of mass of the spheroid for all fibers. Bottom: Plot of the radial component of the fiber orientation tensor for tensioned edges that exceed a strain of 0.1\%. For these high-tensioned edges, the trend is more fluid-like spheroids exhibit less radial alignment than the solid-like spheroids.  } 
    \label{fig:fiber_radial_orientation}
\end{figure}

{\it Cell shape within the spheroid}: Figure 2 shows cross-sectional snapshots of the shape cell and accompanying linker springs and fiber networks at $t/t_f=0.8$ for two different $s_0$s. For $s_0=5.2$, we find that the cells near the boundary of the spheroid have a larger shape index than the cells in the core. We do not readily observe such a trend for the higher $s_0=5.8$ case. Given this marked difference in cell shape index for $s_0=5.2$ and $s_0=5.8$ in Figure 2, we study the distribution of shape index for all cells. In Figure 3 we plot the histogram for the cell shape index for two different target cell shape indices, $s_0=5.2$ (top) and $s_0=5.8$ (bottom). There does not appear to be much change with time in the histogram for $s_0=5.2$.  The histogram is double-peaked with the peak at the larger cell shape index describing the boundary cells and the peak at the smaller cell shape index characterizing the bulk cells.  A similar histogram was observed for a spheroid/organoid studied earlier~\cite{Zhang2022}. Note that for the more solid-like spheroid, the bulk cells are not able to achieve their target cell shape index as they are not able to rearrange as readily within the spheroid via reconnection events. However, the boundary cells, as they interact with the fiber network do take on larger cell shape indices. For the fluid-like spheroid, on the other hand, the cell shape indices evolve more in time with the histogram becoming more broad, though still dominated by a single peak slightly higher than $s_0=5.8$. The broadening occurs for the boundary cells as they interact with the surrounding fiber network.  Therefore, one can readily distinguish between the two rheologically different spheroids by looking at the cell shape index distribution. 

Interestingly, the shift from a more solid-like spheroid to a more fluid-like spheroid occurs at an $s_0\approx 5.7$, which is higher than the bulk result~\cite{Zhang2022}. This finding is not surprising, given earlier calculations for the two-dimensional case of a spheroid with an interfacial boundary tension where the transition point shifts to due to the interfacial boundary tension~\cite{Parker2020}.  When we explore the difference in cell shape index distributions between the more solid-like and fluid-like spheroids for small interfacial tension, we find the two peaks for the solid-like spheroid start to merge into one as there is less of a distinction between bulk cells and boundary cells. See Figure S1.  Moreover, the single peak in the fluid-like case is more broad.

In Figure S2 we plot the histogram of individual cell volumes for $\Gamma/\Gamma_0=1.0$ demonstrating that the cell volumes are approximately 4 percent less than the target volume of one for the fluid-like spheroids, while the cell volumes in the solid case differ more so from their target volume in a two-peaked fashion. Indeed, as a cell is not a closed system, its volume is not expected to be perfectly conserved.  Earlier work with tumor spheroids indicated the addition of dextran to the environment of a tumor spheroid to exert mechanical stress on it~\cite{Delarue2014}. The volume of cells near the core of the tumor spheroid in response to this applied mechanical stress decreased within minutes~\cite{Delarue2014}. We observe a decrease in the volume of the cells for both the solid-like and fluid-like spheroids. 

{\it Cellular Rearrangements within the spheroid}: We also plot the fraction of cells who have {\it not} lost two or more of their neighbors on average as a function of time, denoted by $<Q_n>$ (Figure 4).  In other words, if no cells exchange neighbors, then $<Q_n>=1$, and the spheroid is a solid. However, if $<Q_n>=0$, then all cells are performing neighbor exchanges within the spheroid. For smaller $s_0$ values, we observe that most cells are not finding new neighbors. As the active linker springs contract until $t/t_f=0.2$, the fraction of cells rearranging decreases somewhat and then increases to a fraction that is close to unity. These spheroids are more solid-like.  However, for $s_0=5.7,5.8$, the fraction of cells undergoing neighbor exchanges continues to decrease even after the active linker springs stop contracting.  The spheroid is becoming increasingly fluid-like as it interacts with the fiber network.  This increasing fluidization could eventually set the stage for tumor invasion. 

Given such a difference in the cell shape index distributions and in the frequency of cellular rearrangements between the two cases, we now ask whether or not such a difference translates into differences in remodeling the fiber network. 

\begin{figure}[!htbp]
    \centering
    \includegraphics[width=0.45\textwidth]{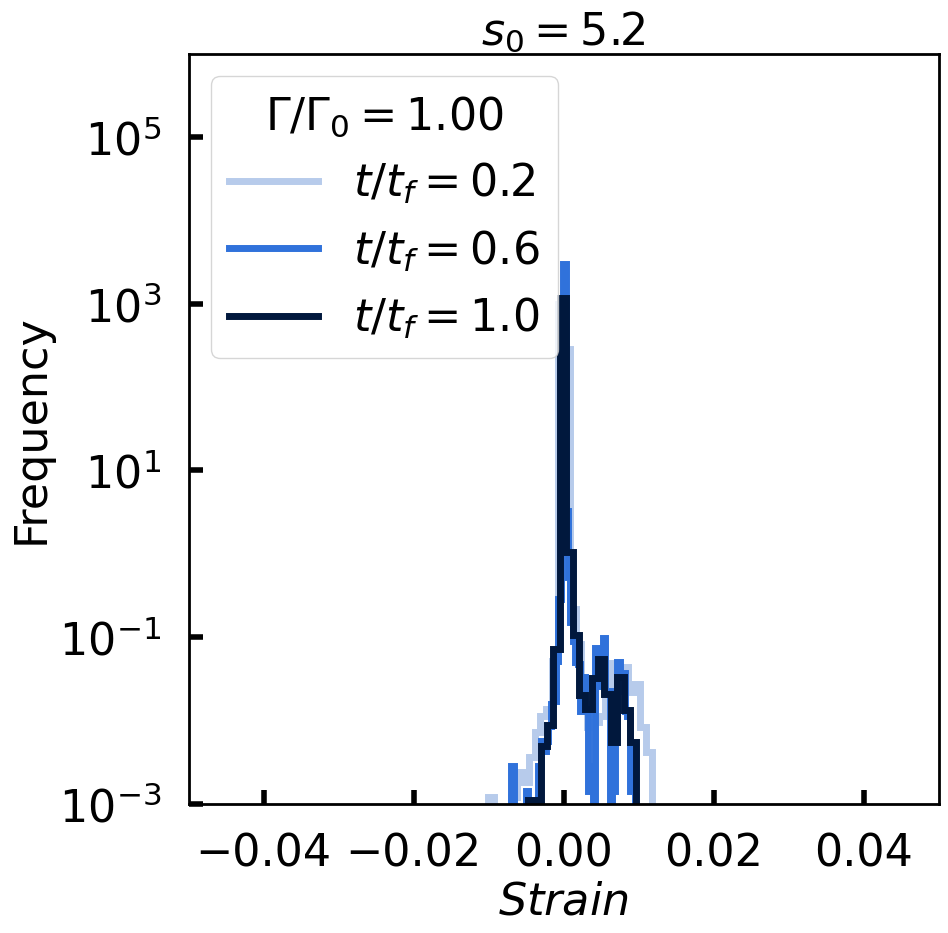}
    \includegraphics[width=0.45\textwidth]{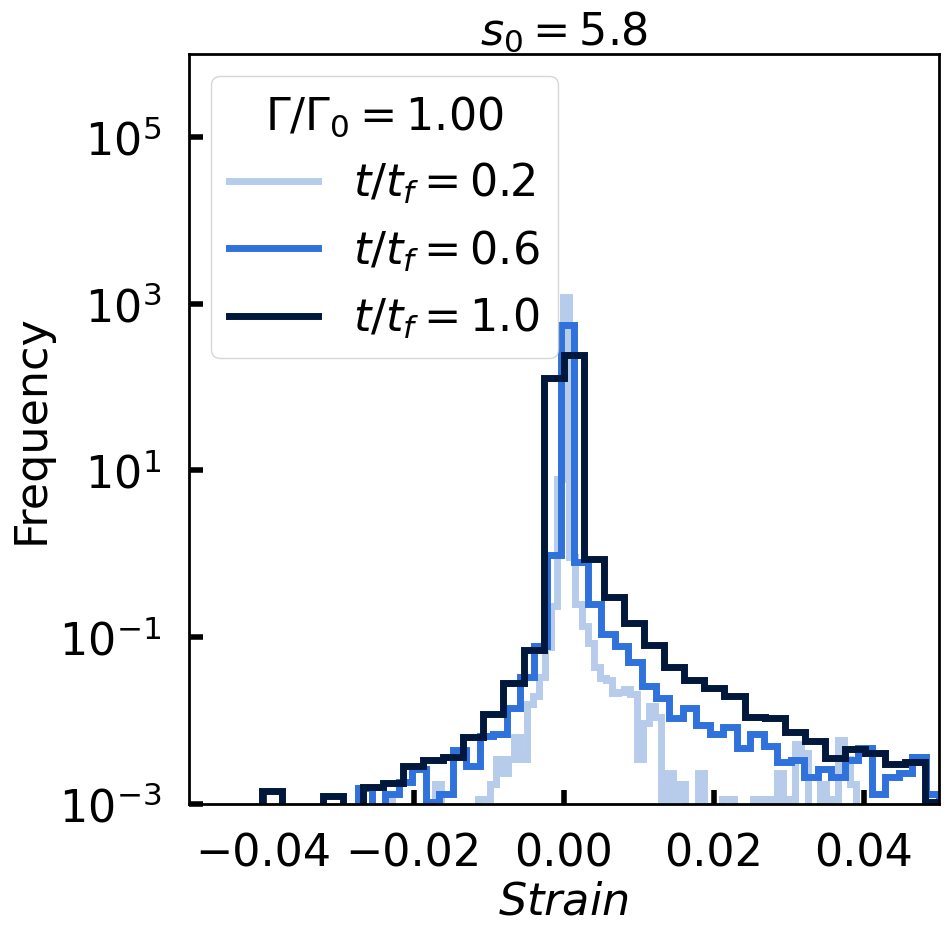}
    \caption{\textit{Fiber strain histogram develops an asymmetric boardening over time that is more pronounced in fluid-like spheroids}: Top: Log-linear histograms of the tension and compression, the latter of which corresponds to negative tension, in the fibers at different times $t$ and for $s_0=5.2$. Bottom: Similar plot as above but for $s_0=5.8$.} 
    \label{fig:fiber_tension_asymmetry}
\end{figure}

{\it Fiber network remodeling}:  We observe remodeling of the fiber network for different target shape indices. To better understand how the spheroid remodels the fiber network, we first report on the magnitude of fiber displacement as a function of radial distance from the center of mass of the spheroid in Figure 5(top).  The displacement is taken from $t/t_f=0$ to $t/t_f=1$. We find for all target cell shape indices, $s_0=5.2-5.8$, that the magnitude of the average fiber displacement within concentric shells emanating from the spheroid is larger closer to the spheroid in comparison to further away from it. This trend is consistent for the system sizes we study, which include smaller system sizes than the one shown in Figure 5(top). We also observe that the smaller $s_0$ spheroids do not displace the fiber network as much as the larger $s_0$ spheroids. There is a slightly more dramatic difference in the magnitude of displacement for $s_0=5.7$ and $s_0=5.8$, indicating that the spheroid rheology may differ between $s_0=5.6$ and $s_0=5.7$, which is consistent with the change in the cell shape index histogram as discussed above.

In addition to the magnitude of the displacement of the fiber network, as a result of being coupled to the spheroid, one can ask about the direction of the displacement. Given the spherical symmetry of this bi-material system combined with prior experimental observations, we instead focus on the density of the fiber network as a function of the radial distance from the center of mass of the spheroid.  Should the density of the fiber network that is closer to the spheroid increase, then the spheroid, with its active linker springs, has radially pulled/contracted the fiber network towards it and vice versa. We find that the fiber network is displaced radially toward the spheroid with an enhancement of fiber density closer to the spheroid in comparison to further away from it (see Figure 5(bottom)). Prior two-dimensional simulations of a cellular Potts model coupled to a fiber network also demonstrate such radial densification behavior~\cite{Tsingos2023} as do experiments~\cite{Mark2020}. Finally, as the more fluid-like spheroid can displace the fiber network more so than the solid-like spheroid, there is more radial densification of the fiber network for the fluid-like spheroids.

Now that we have evidence for the spatial remodeling of the fiber network by the spheroid, let us probe the orientation of the fibers as a result of it by computing the fiber orientation tensor $\Omega_{xy}$. Figure 6 plots $\Omega_{rr}$ as a function of radial distance from the center of mass of the spheroid. Should there exist a bias of fiber modeling along the radial direction, then $\Omega_{rr}$ should be larger than $1/3$.  We find that for fluid-like spheroids that can remodel the fiber network to a greater extent, closer to the spheroid, the fibers are oriented more radially, while further away they are more uniformly oriented (see Figure 6 top).  Figure 6 bottom, however, demonstrates that for high-tensioned edges, or edges that are strained beyond 0.1\%, there is a very high degree of radial ordering for the solid-like spheroids.  As the spheroids become more fluid-like, the high-tensioned fibers become less radially oriented. Interestingly, a high degree of radial alignment in high-tensioned fibers was found in earlier work focusing on a radially-contractile monopole in a fiber network in both two and three dimensions~\cite{Ronceray2016}. Here, the spheroid has many degrees of freedom, including the linker springs, which makes the problem more complex indeed. 

We interpolate between the radially-contracting monopole~\cite{Ronceray2016} and our work to briefly study, for simplicity, a triangular lattice of fibers with six active linker springs, which has 13 more degrees of freedom than a contractile monopole, via energy minimization we also observe the same trend of high-tension fibers as more radially aligned and compression fibers being oriented circumferentially near the spheroid. See Figure S3.  For this simpler two-dimensional model, at full occupation probability, the smaller the final target equilibrium spring length, the larger the average displacement in the fiber network, not surprisingly. For smaller occupation probabilities, this trend is less clear due to the flipping of nodes. Such flipping of nodes does not readily occur in three dimensions.  So for a spheroid with a few degrees of freedom, the contractile monopole is a reasonable approximation. It is therefore feasible that the solid-like spheroids better approximate the contractile monopole case as there are far fewer cellular rearrangements and so additional degrees of freedom represent an elastic object that radially contracts the fiber network. 

To understand how interfacial tension affects the fiber network remodeling trends in terms of average displacement and densification as a function of distance from the center-of-mass of the spheroid, in Figure S4, we plot both quantities for smaller interfacial tension. We find similar trends, though the average displacement is not quite as dramatic as for the larger interfacial tension case as the spheroid is not as effectively strong as a contractile object on the fiber network. To provide additional interpretation, the stronger a contractile force, the more densification of the fiber network around the spheroid.  According to a simplified analysis, the contractile force allows one to probe the stretching and bending of the fiber network and to what spatial extent it can do so. To assess the spatial extent of the contractile spheroid, when looking at the average displacement curves, we numerically compute the first derivative to look for a crossover between bending (scaling with $r/R_s$) and stretching (scaling with $1/(r/R_S)^2$). We find that for $s_0=5,2,5.4,5.6$, the slope of the curves approaches a constant approximately near $R/R_s=4$. However, for $s_0=5.7, 5.8$, the slope does not approach a constant indicating a different effective lengthscale over which the spheroid is acting. More system-size studies will need to be conducted to quantify this length scale, which appears to be larger for more fluid-like spheroids. 

As we have mentioned high-tension edges and a tension distribution, or histogram, let us construct the histogram of strains in the edges of the fiber network as it evolves with time. We will use strain, as opposed to tension, as strain is a dimensionless quantity. Note that negative strain denotes compression. In the top plot of Figure 7, we observe that after $t/t_f=0.2$ for $s_0=5.2$, the strain histogram does not evolve with time, noting that at $t/t_f=0$ all edges in the fiber network exhibit zero strain.  After $t/t_f=0.2$, the active linker springs have finished with their contraction and so the system remains somewhat static in terms of forces with few new active linker springs being created or destroyed as their is very little exchange between boundary cells and bulk cells. Also note that an asymmetry in the strain histogram develops for both the top and bottom plots of Figure 7, i.e., for both types of spheroids. The radial alignment of high-strain fibers facilitates higher strain as compared to the negative strain circumferentially-oriented fibers. Furthermore, these high-strain fibers exhibit strain stiffening~\cite{Sharma_2016}.  For the more fluid-like spheroids, the strain asymmetry becomes even larger with more fiber network remodeling. 
\begin{figure}[!htbp]
    \centering
    \includegraphics[width=0.45\textwidth]{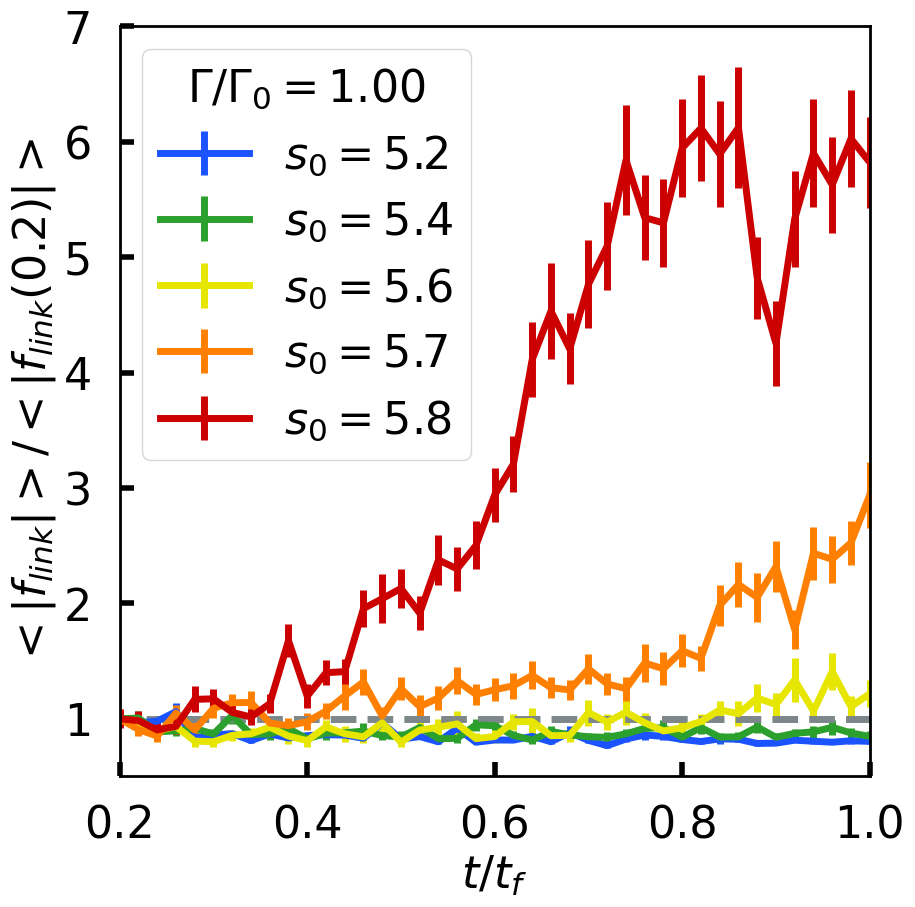}
    \caption{\textit{Average linker spring force increases at later times more significantly for fluid-like spheroids as compared to solid-like spheroids}: Plot of the average linker spring force as a function of time for different $s_0$s.  } 
    \label{fig:average_linker_spring_force}
\end{figure}

\begin{figure}[!htbp]
    \centering
    \includegraphics[width=0.45\textwidth]{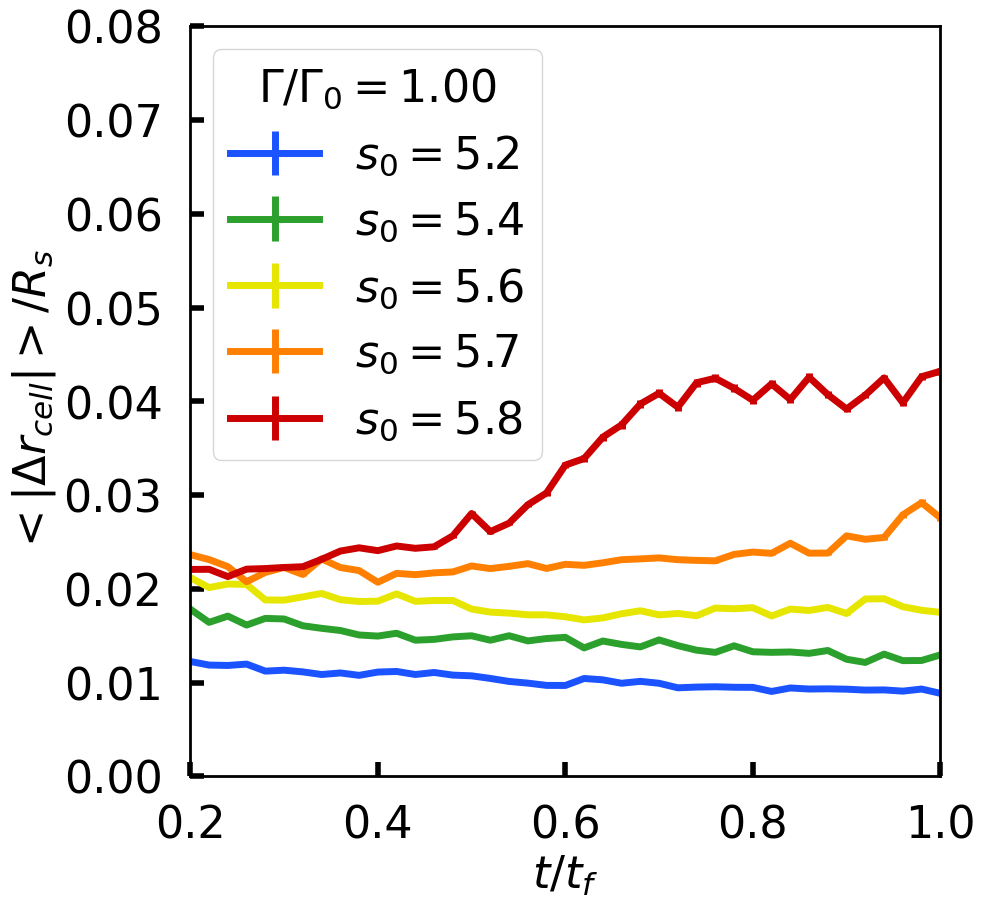}
    \caption{\textit{Cells displace more in the fluid-like spheroids than in the solid-like spheroids with the difference accentuated for $t/t_f>0.6$}. The magnitude of the cell displacement $|\Delta r_{cell}|$ (by tracking cell centers) over a fixed time window as a function of time for different values of the target shape index. Here, $\Gamma/\Gamma_0=1$.} 
    \label{fig:cell_displacement}
\end{figure}

\begin{figure}[!htbp]
    \centering
    \includegraphics[width=0.45\textwidth]{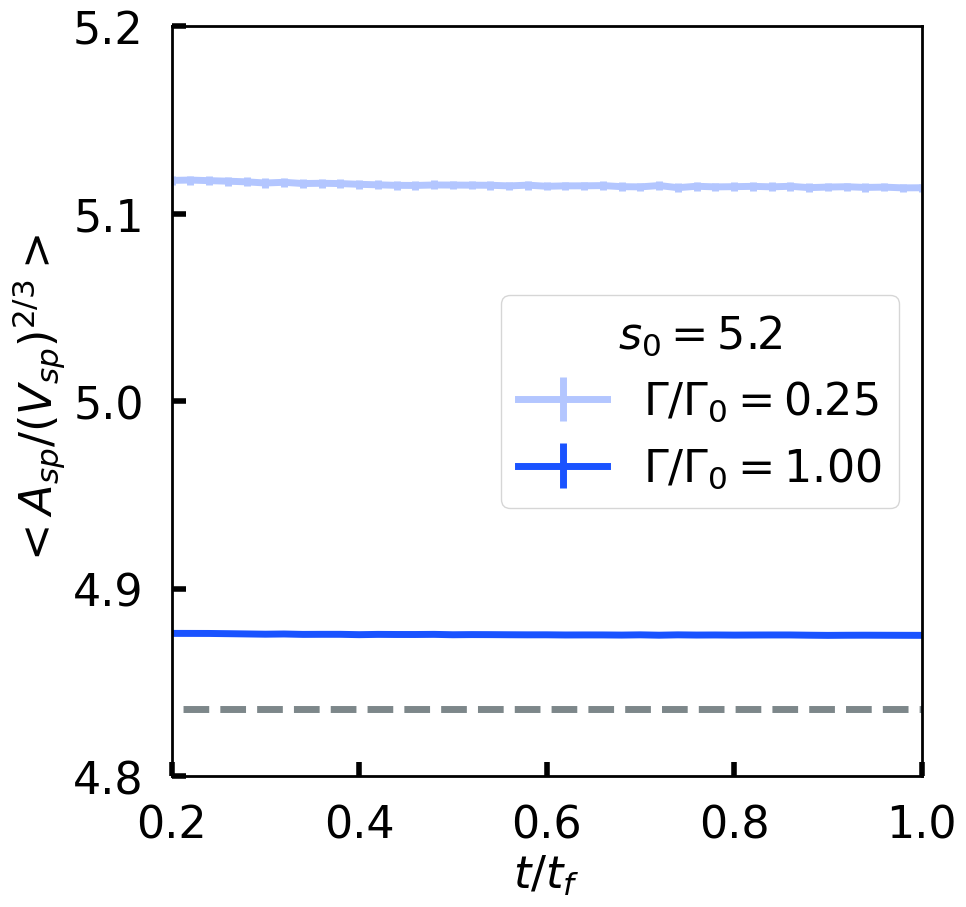}
    \includegraphics[width=0.45\textwidth]{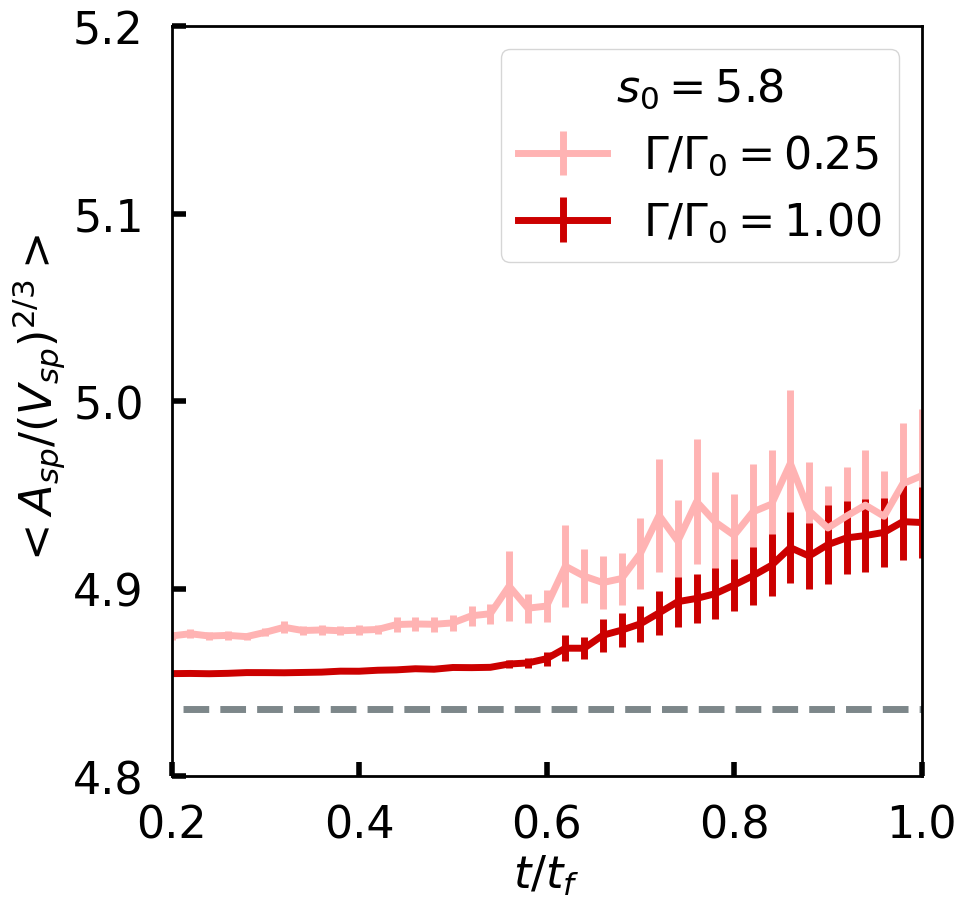}
    \caption{\textit{Spheroid shape index eventually increases over time for fluid-like spheroids.} Top: Spheroid shape index ($A_{sp}/V_{sp}^{2/3}$) as a function of time for $s_0=5.2$ and for two different values of the dimensional interfacial tension. Bottom: Same as the above plot, but for $s_0=5.8$.  Note that the dashed line in both plots denotes the spheroid shape index of a perfect sphere. } 
    \label{fig:spheroid_shape_index}
\end{figure}

\begin{figure*}[!htbp]
    \centering
    \includegraphics[width=0.45\textwidth]{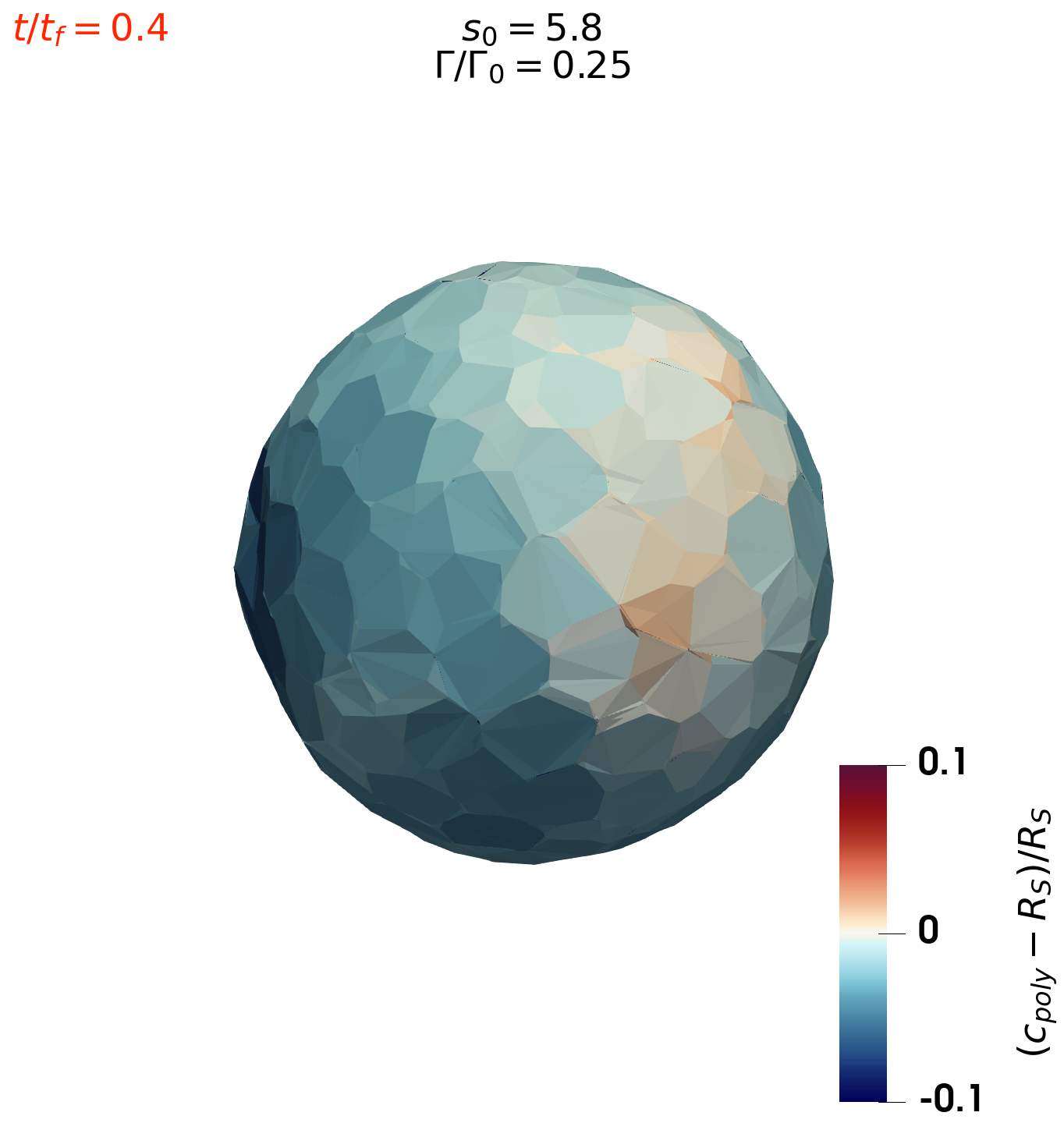}
    \hspace{0.5cm}
    \includegraphics[width=0.45\textwidth]{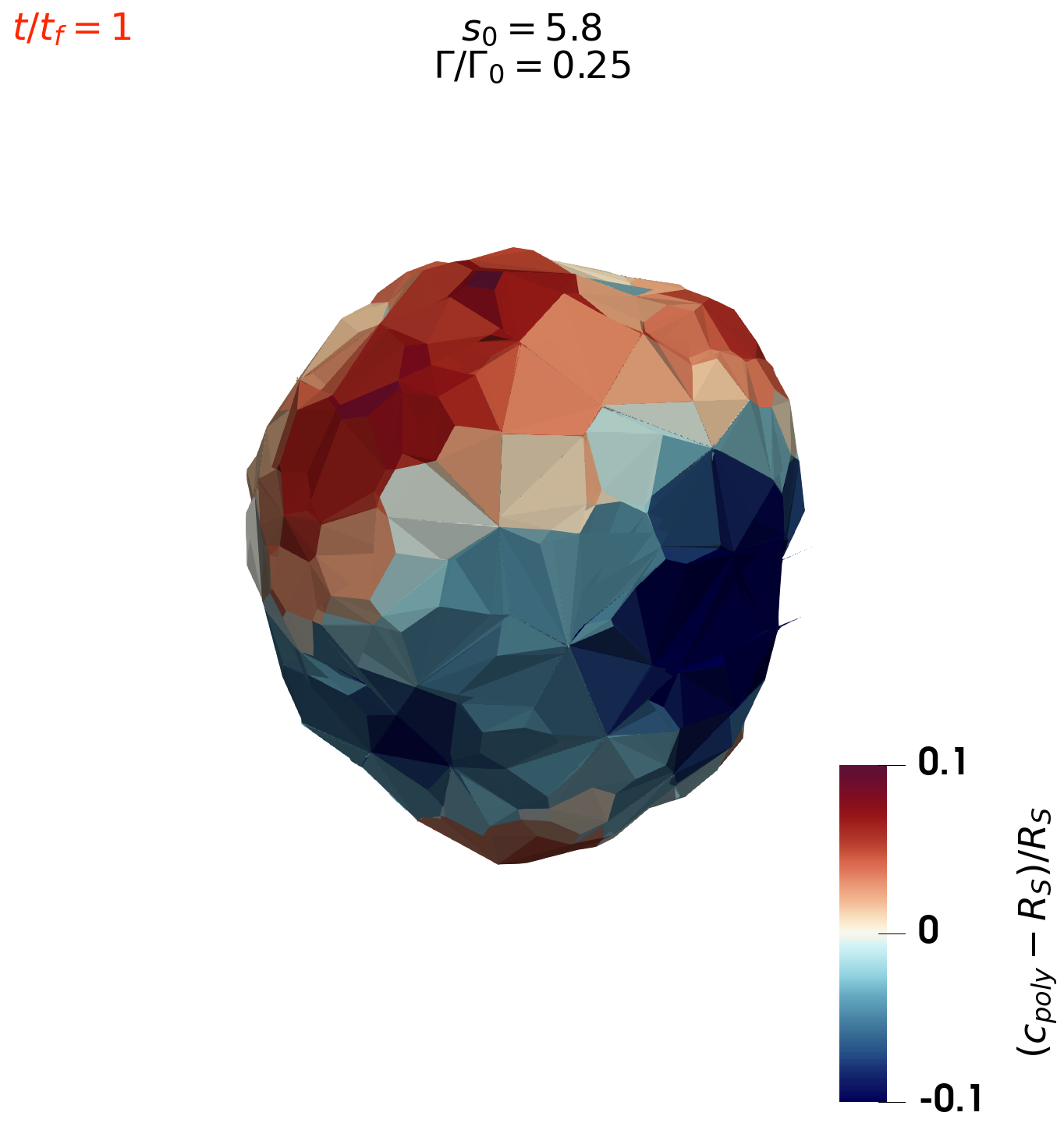}
    \caption{{\it Fluid-like spheroid shape changes after the initial contraction of the fiber network.} The eventual increase in spheroid shape index for fluid-like spheroids corresponds to local deviations from the initial spherical surface. Here, such surface deviations are quantified by the difference between polygon center to spheroid center distance ($c_{poly}$), and the initial radius of the spheroid ($R_S$), i.e., red regions denote outward bulges and blue regions denote inward bulges (or wrinkles).} 
    \label{fig:spheroid_shape_examples}
\end{figure*}

\begin{figure*}[!htbp]
    \centering
    \includegraphics[width=0.45\textwidth]{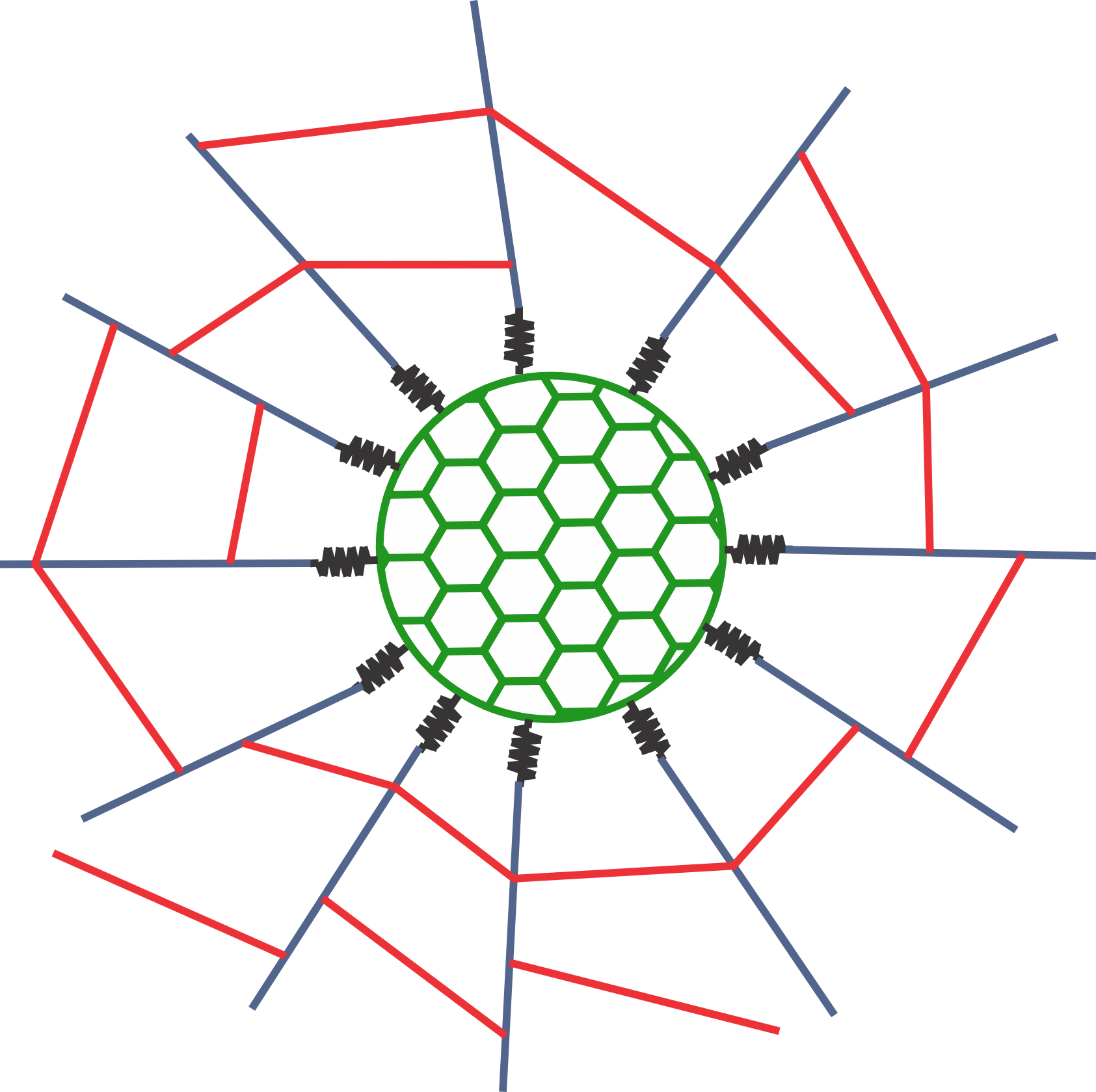}
    \hspace{0.5cm}
    \includegraphics[width=0.45\textwidth]{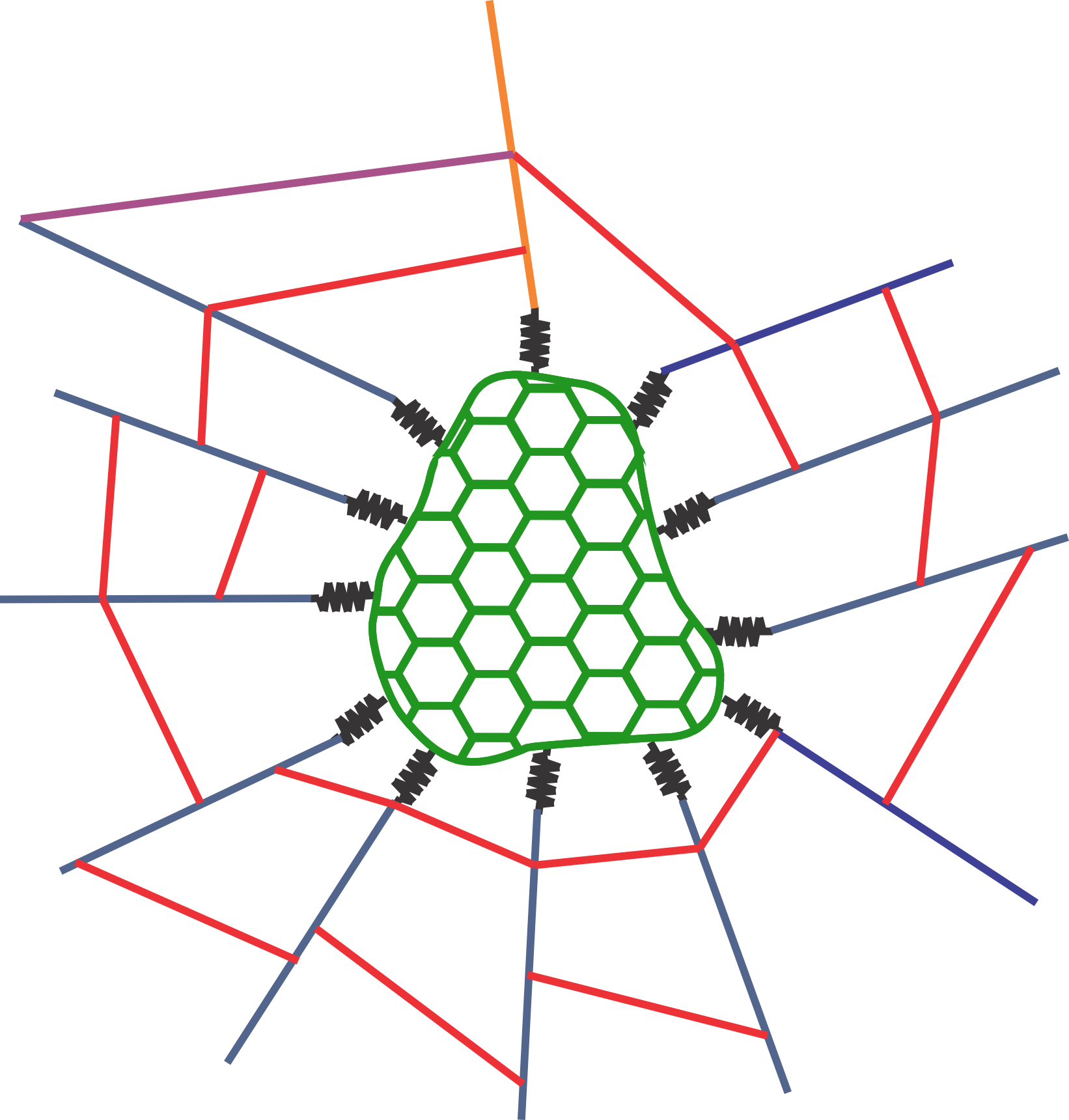}
    \caption{\textit{Schematic illustrating how spheroid shape changes can drive fiber network remodeling.} If one considers a spheroid as a spherical contractile force monopole with some radius (left figure), then the fiber network remodeling will typically consist of radial, high-tension ropes with circumferential compressed fibers. However, with spheroid shape changes (in the fluid phase) leading to changes in surface curvature (right figure), where the spheroid shape bulges outward, the radial, high-tension ropes may be compressed to lead to more intricate remodeling of the fiber network (as compared to a spherical contractile force monopole with some radius). Moreover, spheroidal shape changes break radial symmetry, which then manifests in a less radially oriented fiber network that is particularly prominent in the high-strain fibers (see Figure 6). The remodeling of the fiber network by the spheroid can subsequently affect the spheroid to induce additional cellular rearrangements. Thus, the spheroid shape fluctuations create a positive morphology-mechanics feedback loop between the spheroid and the fiber network to ultimately enhance remodeling in both structures.}
    \label{fig:schematic}
\end{figure*}

{\it Forces exerted by the active linker springs}: To further understand how the fiber network is remodeled, in Figure 8, we show the average total active linker spring force as a function of time $t/t_f$ for different target cell shape indices. For smaller target cell shape indices, after the initial contraction phase of the active linker springs, the average total active linker spring force remains constant as a function of time. However, for the larger target cell shape indices, after approximately $t/t_f=0.4$ for $s_0=5.8$, for example, the average total active linker spring force begins to increase with time. Given that the total active linker spring force is approximately the same at t$/t_f=0.2$, we have indeed checked that the fiber network remodeling for the different target cell shape index spheroids is the same. It is at later times that the fiber network remodeling continues, at least for the more fluid-like spheroids.  In other words, it is the increase in the average total active linker spring force that generates the enhanced amount of fiber network remodeling over time.  Figure S5 shows the average linker spring force as a function of time for the smaller interfacial tension case resulting in a smaller active linker spring force on average.

{\it Cell motion}: What then is inducing the increase in the total average linker spring force at later times in the simulations? To more readily answer this question, we track cell movement. More precisely, we study cell motion by tracking the average displacement of a cell center as a function of time for different target cell shape indices and for two different interfacial tensions (see Figs. 9 and S6). We find that the displacement of the cell centers, beyond the initial displacement due to the initial contraction of the active linker springs, does not increase with time for the more solid-like spheroids, while for more fluid-like spheroids the cell center displacement does increase over time. While we have used the cell shape index distribution to delineate between solid-like and fluid-like spheroids, here we present evidence for such differences in rheology in terms of cell motion.  Indeed, there does appear to be a delineation occurring between $s_0=5.6$ and $s_0=5.7$, as indicated earlier.  Prior bulk analysis led to a rigidity transition location at $s_0^*=5.4$. However, recent two-dimensional analysis of a vertex model coupled to a spring network found that the presence of the interfacial tension did alter the location of the rigidity transition~\cite{Parker2020}.  More specifically, a spheroid with interfacial tension can be mapped to a bulk model with a shift in the dimensionless area and area spring stiffness so that the location of the transition shifts~\cite{Parker2020}. Our three-dimensional results are consistent with the prior two-dimensional results, particularly as Figure S6 shows that for smaller interfacial tension, the delineation between fluid-like and solid-like occurs at a larger $s_0$.

{\it Spheroid shape index fluctuations}: We now ask what are the consequences of cells moving in the fluid-like spheroids, particularly after $t/t_f=0.2$ (see Figure 4)? One consequence is that the cell fluidity leads to changes in spheroid shape with larger fluctuations, as evidenced in Figure 10. Indeed, for the more solid-like spheroids, the spheroid shape, defined as $A_{sp}/V^{2/3}_{sp}$, where $A_{sp}$ and $V_{sp}$ denote the spheroid surface area and volume respectively, does not change with time, while for the more fluid-like spheroids, the spheroid shape does change with time. Note that the spheroid shape is more spherical for the larger interfacial tension, which is expected.  It is the change in spheroid shape that then moves the active linker springs, which continue to remodel the fiber network to displace the fibers. Interestingly, even though there is presumably a lack of bias in the spheroid shape fluctuations, meaning the shape can distort either "inwards" or "outwards", the fluctuations lead to enhanced fiber densification.  See Figure 11. For spheroid deformations that move inward, the fiber network is strain stiffened.  For spheroid deformations that move outward, the fiber network compression softens. Given the asymmetry of the response of strain stiffening and compression softening, the strain stiffening dominates to lead to more overall tension in the network as indicated in the tension/compression distribution of the fiber network.

Given the internal degrees of freedom of the spheroid, it is the spheroid shape fluctuations that give rise to a more complex response than a simple contractile monopole, or even a less simple elastically deformable object.  The fluctuations are due to the dynamics/movement of the cells within the spheroid that then modify the interactions between the spheroid and the fiber network by making the system more active in a fluctuation sense to allow for more remodeling of the fiber network, as opposed to the solid-like spheroids where such spheroid shape fluctuations do not occur as readily.  With the spheroidal shape distortions comes a breaking of the spheroidal radial symmetry. This breaking of radial symmetry leads to high-tensioned fibers that are less radial symmetric and so explains why the radial alignment of the high-tensioned (or high-strained) fibers decreases with increasing $s_0$ as the spheroid becomes more fluid-like. Moreover, the spheroid shape fluctuations also lead to additional remodeling of the fiber network, which, in turn, affects the spheroid to lead to further remodeling of the spheroid and so further remodeling of the fiber network. Thus, we find an emergent feedback loop even with this simple model in which there is no explicit feedback between the fluid-like spheroid and the fiber network. Moreover, our computational studies reveal that the strength of a contractile object does not alone determine the remodeling of the fiber network. There is a key additional contribution by the active shape fluctuations due to active cellular rearrangements. As cellular rearrangements are not significant in solid-like spheroids, this emergent feedback between the solid-like spheroid and the fiber network is not a prominent feature.

\section{Discussion}

We have developed a one-of-a-kind three-dimensional computational model for a spheroid embedded in a fiber network mimicking the extracellular matrix.  In addition to cellular-level resolution, we incorporate explicit focal adhesion attachment between the cells and the fiber network in the form of an active linker spring whose equilibrium spring length decreases with time. After the initial contraction of the active linker springs, we find that the fluidity of the spheroid can drive spheroid shape changes, which allows for enhanced fiber network modeling as regions of the spheroid that move inward can strain stiffen the fiber network, while regions of the spheroid that move outwards can compression soften the fiber network.  Given the force asymmetry between compression softening and strain stiffening, the strain stiffening dominates leading to an asymmetry in the distribution of tension and compression in the fiber network.  Moreover, as a fluid-like spheroid distorts itself due to cellular rearrangements, there is additional remodeling of the fiber network, which, in turns, affects the spheroid, to create a positive morphology-mechanics feedback loop that breaks spheroidal radial symmetry.  Therefore, counterintuitively, a fluid-like spheroid more readily remodels the fiber network than a solid-like spheroid. So while the strength of a contractile spheroid also impacts fiber network remodeling, we have demonstrated that strength alone, is not the only determinant, as both the fluid-like and solid-like spheroids have the same number and stiffness of active linker springs. See Figure 12 for a schematic illustration summarizing this effect. In addition, an increase in spheroid interfacial surface tension compactifies the spheroid and so makes it a stronger contractile puller. However, an increase in interfacial surface tension decreases the spheroid shape fluctuations, thereby decreasing the postive feedback between the spheroid and the fiber network. In turns out that fluid-like spheroids remodeling the fiber network more than solid-like spheroids is qualitatively consistent with recent observations~\cite{Pandey2023}. Yet, for a direct, quantitative comparison, more work is needed on the modeling side in terms of tuning the fiber network's characteristics to match the experiments as well as the invasion of the cells. 

How do our results compare with a single-cell interacting with a fiber network? Prior work demonstrates that as cells contract the fiber (collagen) network, there emerge two opposing sides of the cells that dominate~\cite{Hall2016}. The cell then forms stress fibers between these two opposing sides to lead to elongation of the cell along the polarization direction~\cite{Hall2016}. As the elongation occurs, presumably there are fewer focal adhesions between the cell and the fiber network in directions orthogonal to the polarization direction, and the cell now mimics more of the fiber network by becoming "long and skinny", i.e., there exists mechanoreciprocity between the cell and the fiber network~\cite{Friedl2011}. When the cells exist in a collective, such as a spheroid, the mechanoreciprocity between an individual cell and the fiber network is not as pronounced, at least in the pre-invasion stages, such that the cellular program, to build stress fibers along a polarization direction does not yet exist. The spheroid can, therefore, manipulate the fiber network in ways that can differ from an individual cell. For instance, the spheroid can still randomly pick out more than one polarization direction, and the shape of the spheroid need not conform to the morphology of the fibers (see Figure 11). The cell-cell reciprocity and the cell-ECM reciprocity thus compete to give a spectrum of possible emergent behavior. We suspect that once cells along the boundary of the spheroid become sufficiently elongated and, therefore, exhibit higher stress, they will be the candidate break-out cells.

Given the richness of our computational model, many questions remain and will be explored in future work. For instance, what does the model predict for denser fiber networks? And what happens when we incorporate cellular-based forms of explicit, mechanical feedback between the cells and the fiber network, such as cells becoming more contractile the higher the strain in a fiber? For the former question, we anticipate that a denser fiber network will be remodeled to a lesser extent. For the latter question, we anticipate a more heterogeneous distribution of cell shape indices, which could presumably enhance the breakout potential of tumor cells. One can also explore more detailed models of the active linker springs that bind and unbind depending on the change in strain. This detail may contribute to understanding the twitching phenomenon observed in embedded spheroids, where the contraction of the fiber network oscillates with time~\cite{Mark2020}, though we already observe hints of strain in the fiber network increasing and decreasing for the fluid-like spheroids that depends on the timescale for the morphology-mechanics feedback to occur. See Figure S7. 

Finally, what are the implications of our model for the cancer?  Indeed, we are presenting a minimal model from which to build upon in terms of allowing for cell escape to ultimately predict whether a tumor will invade or not given the cells in the tumor and the structure of the microenvironment. However, as we see here, much goes into the mechanical crosstalk to help set the stage for potential cell breakout that must be understood. We need such starting points that contain information beyond automaton models~\cite{Ilina2020,Beunk2022,Kato2023} that include mechanics and chemical signaling~\cite{Tserunyan2022} to begin to make quantitative predictions for cell breakout. While cell breakout is the obvious next step, we must also consider the multiscale aspect of cells~\cite{Zhang2023} as well as the adaptability of cells and their ability to "train" the fiber network to be able to escape within a physical learning framework~\cite{Anisetti2022,Anisetti2023} just as neural networks are trained to perform a specific function. Moreover, cancer cells interact with other types of cells, such as immune cells, providing an entire cellular ecology as a backdrop with cancer cells trying to train immune cells and vice versa~\cite{Galeano2020,Mukherjee2022}. All such elements will help provide a more accurate, quantitative picture of the complexities of cancer and other diseases and biological processes more generally. 

\section{Acknowledgements}
Tao Zhang acknowledges financial support from the NSFC/China via award 22303051. This work was also supported by NIH R35 GM142963 awarded to Alison E. Patteson. Minging Wu thanks the support from NIH NCI  R01 CA 221346. JMS and Mingming Wu acknowledge financial support from NSF-PoLS-2014192. 

\bibliography{ref}

\begin{thebibliography}{49}%
\makeatletter
\providecommand \@ifxundefined [1]{%
 \@ifx{#1\undefined}
}%
\providecommand \@ifnum [1]{%
 \ifnum #1\expandafter \@firstoftwo
 \else \expandafter \@secondoftwo
 \fi
}%
\providecommand \@ifx [1]{%
 \ifx #1\expandafter \@firstoftwo
 \else \expandafter \@secondoftwo
 \fi
}%
\providecommand \natexlab [1]{#1}%
\providecommand \enquote  [1]{``#1''}%
\providecommand \bibnamefont  [1]{#1}%
\providecommand \bibfnamefont [1]{#1}%
\providecommand \citenamefont [1]{#1}%
\providecommand \href@noop [0]{\@secondoftwo}%
\providecommand \href [0]{\begingroup \@sanitize@url \@href}%
\providecommand \@href[1]{\@@startlink{#1}\@@href}%
\providecommand \@@href[1]{\endgroup#1\@@endlink}%
\providecommand \@sanitize@url [0]{\catcode `\\12\catcode `\$12\catcode
  `\&12\catcode `\#12\catcode `\^12\catcode `\_12\catcode `\%12\relax}%
\providecommand \@@startlink[1]{}%
\providecommand \@@endlink[0]{}%
\providecommand \url  [0]{\begingroup\@sanitize@url \@url }%
\providecommand \@url [1]{\endgroup\@href {#1}{\urlprefix }}%
\providecommand \urlprefix  [0]{URL }%
\providecommand \Eprint [0]{\href }%
\providecommand \doibase [0]{https://doi.org/}%
\providecommand \selectlanguage [0]{\@gobble}%
\providecommand \bibinfo  [0]{\@secondoftwo}%
\providecommand \bibfield  [0]{\@secondoftwo}%
\providecommand \translation [1]{[#1]}%
\providecommand \BibitemOpen [0]{}%
\providecommand \bibitemStop [0]{}%
\providecommand \bibitemNoStop [0]{.\EOS\space}%
\providecommand \EOS [0]{\spacefactor3000\relax}%
\providecommand \BibitemShut  [1]{\csname bibitem#1\endcsname}%
\let\auto@bib@innerbib\@empty
%</preamble>
\bibitem [{\citenamefont {Haeger}\ \emph {et~al.}(2014)\citenamefont {Haeger},
  \citenamefont {Krause}, \citenamefont {Wolf},\ and\ \citenamefont
  {Friedl}}]{Haeger2014}%
  \BibitemOpen
  \bibfield  {author} {\bibinfo {author} {\bibfnamefont {A.}~\bibnamefont
  {Haeger}}, \bibinfo {author} {\bibfnamefont {M.}~\bibnamefont {Krause}},
  \bibinfo {author} {\bibfnamefont {K.}~\bibnamefont {Wolf}},\ and\ \bibinfo
  {author} {\bibfnamefont {P.}~\bibnamefont {Friedl}},\ }\bibfield  {title}
  {\bibinfo {title} {Cell jamming: collective invasion of mesenchymal tumor
  cells imposed by tissue confinement},\ }\href@noop {} {\bibfield  {journal}
  {\bibinfo  {journal} {Biochimica et Biophysica Acta (BBA)-General Subjects}\
  }\textbf {\bibinfo {volume} {1840}},\ \bibinfo {pages} {2386} (\bibinfo
  {year} {2014})}\BibitemShut {NoStop}%
\bibitem [{\citenamefont {Tevis}\ \emph {et~al.}(2017)\citenamefont {Tevis},
  \citenamefont {Colson},\ and\ \citenamefont {Grinstaff}}]{Tevis2017}%
  \BibitemOpen
  \bibfield  {author} {\bibinfo {author} {\bibfnamefont {K.~M.}\ \bibnamefont
  {Tevis}}, \bibinfo {author} {\bibfnamefont {Y.~L.}\ \bibnamefont {Colson}},\
  and\ \bibinfo {author} {\bibfnamefont {M.~W.}\ \bibnamefont {Grinstaff}},\
  }\bibfield  {title} {\bibinfo {title} {Embedded spheroids as models of the
  cancer microenvironment},\ }\href@noop {} {\bibfield  {journal} {\bibinfo
  {journal} {Advanced Biosystems}\ }\textbf {\bibinfo {volume} {1}},\ \bibinfo
  {pages} {1700083} (\bibinfo {year} {2017})}\BibitemShut {NoStop}%
\bibitem [{\citenamefont {Suh}\ \emph {et~al.}(2019{\natexlab{a}})\citenamefont
  {Suh}, \citenamefont {Hall}, \citenamefont {Huang}, \citenamefont {Moon},
  \citenamefont {Song}, \citenamefont {Ma}, \citenamefont {Bonassar},
  \citenamefont {Segall},\ and\ \citenamefont {Wu}}]{Suh_2019}%
  \BibitemOpen
  \bibfield  {author} {\bibinfo {author} {\bibfnamefont {Y.~J.}\ \bibnamefont
  {Suh}}, \bibinfo {author} {\bibfnamefont {M.~S.}\ \bibnamefont {Hall}},
  \bibinfo {author} {\bibfnamefont {Y.~L.}\ \bibnamefont {Huang}}, \bibinfo
  {author} {\bibfnamefont {S.~Y.}\ \bibnamefont {Moon}}, \bibinfo {author}
  {\bibfnamefont {W.}~\bibnamefont {Song}}, \bibinfo {author} {\bibfnamefont
  {M.}~\bibnamefont {Ma}}, \bibinfo {author} {\bibfnamefont {L.~J.}\
  \bibnamefont {Bonassar}}, \bibinfo {author} {\bibfnamefont {J.~E.}\
  \bibnamefont {Segall}},\ and\ \bibinfo {author} {\bibfnamefont
  {M.}~\bibnamefont {Wu}},\ }\bibfield  {title} {\bibinfo {title} {{Glycation
  of collagen matrices promotes breast tumor cell invasion}},\ }\href@noop {}
  {\bibfield  {journal} {\bibinfo  {journal} {Integrative Biology}\ }\textbf
  {\bibinfo {volume} {11}},\ \bibinfo {pages} {109} (\bibinfo {year}
  {2019}{\natexlab{a}})}\BibitemShut {NoStop}%
\bibitem [{\citenamefont {Pandey}\ \emph {et~al.}(2023)\citenamefont {Pandey},
  \citenamefont {Suh}, \citenamefont {Kim}, \citenamefont {Davis},
  \citenamefont {Segall},\ and\ \citenamefont {Wu}}]{Pandey2023}%
  \BibitemOpen
  \bibfield  {author} {\bibinfo {author} {\bibfnamefont {M.}~\bibnamefont
  {Pandey}}, \bibinfo {author} {\bibfnamefont {Y.~J.}\ \bibnamefont {Suh}},
  \bibinfo {author} {\bibfnamefont {M.}~\bibnamefont {Kim}}, \bibinfo {author}
  {\bibfnamefont {H.~J.}\ \bibnamefont {Davis}}, \bibinfo {author}
  {\bibfnamefont {J.~E.}\ \bibnamefont {Segall}},\ and\ \bibinfo {author}
  {\bibfnamefont {M.}~\bibnamefont {Wu}},\ }\bibfield  {title} {\bibinfo
  {title} {Mechanical compression regulates tumor spheroid invasion into a 3d
  collagen matrix},\ }\href@noop {} {\bibfield  {journal} {\bibinfo  {journal}
  {arXiv preprint arXiv:2307.01289}\ } (\bibinfo {year} {2023})}\BibitemShut
  {NoStop}%
\bibitem [{\citenamefont {Huang}\ \emph {et~al.}(2020)\citenamefont {Huang},
  \citenamefont {Ma}, \citenamefont {Wu}, \citenamefont {Shiau}, \citenamefont
  {Segall},\ and\ \citenamefont {Wu}}]{Huang2020}%
  \BibitemOpen
  \bibfield  {author} {\bibinfo {author} {\bibfnamefont {Y.~L.}\ \bibnamefont
  {Huang}}, \bibinfo {author} {\bibfnamefont {Y.}~\bibnamefont {Ma}}, \bibinfo
  {author} {\bibfnamefont {C.}~\bibnamefont {Wu}}, \bibinfo {author}
  {\bibfnamefont {C.}~\bibnamefont {Shiau}}, \bibinfo {author} {\bibfnamefont
  {J.~E.}\ \bibnamefont {Segall}},\ and\ \bibinfo {author} {\bibfnamefont
  {M.}~\bibnamefont {Wu}},\ }\bibfield  {title} {\bibinfo {title} {Tumor
  spheroids under perfusion within a 3d microfluidic platform reveal critical
  roles of cell-cell adhesion in tumor invasion},\ }\href@noop {} {\bibfield
  {journal} {\bibinfo  {journal} {Scientific reports}\ }\textbf {\bibinfo
  {volume} {10}},\ \bibinfo {pages} {9648} (\bibinfo {year}
  {2020})}\BibitemShut {NoStop}%
\bibitem [{\citenamefont {Baskaran}\ \emph {et~al.}(2020)\citenamefont
  {Baskaran}, \citenamefont {Weldy}, \citenamefont {Guarin}, \citenamefont
  {Munoz}, \citenamefont {Shpilker}, \citenamefont {Kotlik}, \citenamefont
  {Subbiah}, \citenamefont {Wishart}, \citenamefont {Peng}, \citenamefont
  {Miller} \emph {et~al.}}]{Baskaran2020}%
  \BibitemOpen
  \bibfield  {author} {\bibinfo {author} {\bibfnamefont {J.~P.}\ \bibnamefont
  {Baskaran}}, \bibinfo {author} {\bibfnamefont {A.}~\bibnamefont {Weldy}},
  \bibinfo {author} {\bibfnamefont {J.}~\bibnamefont {Guarin}}, \bibinfo
  {author} {\bibfnamefont {G.}~\bibnamefont {Munoz}}, \bibinfo {author}
  {\bibfnamefont {P.~H.}\ \bibnamefont {Shpilker}}, \bibinfo {author}
  {\bibfnamefont {M.}~\bibnamefont {Kotlik}}, \bibinfo {author} {\bibfnamefont
  {N.}~\bibnamefont {Subbiah}}, \bibinfo {author} {\bibfnamefont
  {A.}~\bibnamefont {Wishart}}, \bibinfo {author} {\bibfnamefont
  {Y.}~\bibnamefont {Peng}}, \bibinfo {author} {\bibfnamefont {M.~A.}\
  \bibnamefont {Miller}}, \emph {et~al.},\ }\bibfield  {title} {\bibinfo
  {title} {Cell shape, and not 2d migration, predicts extracellular
  matrix-driven 3d cell invasion in breast cancer},\ }\href@noop {} {\bibfield
  {journal} {\bibinfo  {journal} {APL bioengineering}\ }\textbf {\bibinfo
  {volume} {4}} (\bibinfo {year} {2020})}\BibitemShut {NoStop}%
\bibitem [{\citenamefont {Hall}\ \emph {et~al.}(2016)\citenamefont {Hall},
  \citenamefont {Alisafaei}, \citenamefont {Ban}, \citenamefont {Feng},
  \citenamefont {Hui}, \citenamefont {Shenoy},\ and\ \citenamefont
  {Wu}}]{Hall2016}%
  \BibitemOpen
  \bibfield  {author} {\bibinfo {author} {\bibfnamefont {M.~S.}\ \bibnamefont
  {Hall}}, \bibinfo {author} {\bibfnamefont {F.}~\bibnamefont {Alisafaei}},
  \bibinfo {author} {\bibfnamefont {E.}~\bibnamefont {Ban}}, \bibinfo {author}
  {\bibfnamefont {X.}~\bibnamefont {Feng}}, \bibinfo {author} {\bibfnamefont
  {C.-Y.}\ \bibnamefont {Hui}}, \bibinfo {author} {\bibfnamefont {V.~B.}\
  \bibnamefont {Shenoy}},\ and\ \bibinfo {author} {\bibfnamefont
  {M.}~\bibnamefont {Wu}},\ }\bibfield  {title} {\bibinfo {title} {Fibrous
  nonlinear elasticity enables positive mechanical feedback between cells and
  ecms},\ }\href@noop {} {\bibfield  {journal} {\bibinfo  {journal}
  {Proceedings of the National Academy of Sciences}\ }\textbf {\bibinfo
  {volume} {113}},\ \bibinfo {pages} {14043} (\bibinfo {year}
  {2016})}\BibitemShut {NoStop}%
\bibitem [{\citenamefont {Steinwachs}\ \emph {et~al.}(2016)\citenamefont
  {Steinwachs}, \citenamefont {Metzner}, \citenamefont {Skodzek}, \citenamefont
  {Lang}, \citenamefont {Thievessen}, \citenamefont {Mark}, \citenamefont
  {M{\"u}nster}, \citenamefont {Aifantis},\ and\ \citenamefont
  {Fabry}}]{Steinwachs2016}%
  \BibitemOpen
  \bibfield  {author} {\bibinfo {author} {\bibfnamefont {J.}~\bibnamefont
  {Steinwachs}}, \bibinfo {author} {\bibfnamefont {C.}~\bibnamefont {Metzner}},
  \bibinfo {author} {\bibfnamefont {K.}~\bibnamefont {Skodzek}}, \bibinfo
  {author} {\bibfnamefont {N.}~\bibnamefont {Lang}}, \bibinfo {author}
  {\bibfnamefont {I.}~\bibnamefont {Thievessen}}, \bibinfo {author}
  {\bibfnamefont {C.}~\bibnamefont {Mark}}, \bibinfo {author} {\bibfnamefont
  {S.}~\bibnamefont {M{\"u}nster}}, \bibinfo {author} {\bibfnamefont {K.~E.}\
  \bibnamefont {Aifantis}},\ and\ \bibinfo {author} {\bibfnamefont
  {B.}~\bibnamefont {Fabry}},\ }\bibfield  {title} {\bibinfo {title}
  {Three-dimensional force microscopy of cells in biopolymer networks},\
  }\href@noop {} {\bibfield  {journal} {\bibinfo  {journal} {Nature Methods}\
  }\textbf {\bibinfo {volume} {13}},\ \bibinfo {pages} {171} (\bibinfo {year}
  {2016})}\BibitemShut {NoStop}%
\bibitem [{\citenamefont {Kim}\ \emph {et~al.}(2018)\citenamefont {Kim},
  \citenamefont {Silberberg}, \citenamefont {Abeyaratne}, \citenamefont
  {Kamm},\ and\ \citenamefont {Asada}}]{Kim_2018}%
  \BibitemOpen
  \bibfield  {author} {\bibinfo {author} {\bibfnamefont {M.-C.}\ \bibnamefont
  {Kim}}, \bibinfo {author} {\bibfnamefont {Y.~R.}\ \bibnamefont {Silberberg}},
  \bibinfo {author} {\bibfnamefont {R.}~\bibnamefont {Abeyaratne}}, \bibinfo
  {author} {\bibfnamefont {R.~D.}\ \bibnamefont {Kamm}},\ and\ \bibinfo
  {author} {\bibfnamefont {H.~H.}\ \bibnamefont {Asada}},\ }\bibfield  {title}
  {\bibinfo {title} {Computational modeling of three-dimensional ecm-rigidity
  sensing to guide directed cell migration},\ }\href@noop {} {\bibfield
  {journal} {\bibinfo  {journal} {Proceedings of the National Academy of
  Sciences}\ }\textbf {\bibinfo {volume} {115}},\ \bibinfo {pages} {E390}
  (\bibinfo {year} {2018})}\BibitemShut {NoStop}%
\bibitem [{\citenamefont {Han}\ \emph {et~al.}(2018)\citenamefont {Han},
  \citenamefont {Ronceray}, \citenamefont {Xu}, \citenamefont {Malandrino},
  \citenamefont {Kamm}, \citenamefont {Lenz}, \citenamefont {Broedersz},\ and\
  \citenamefont {Guo}}]{Han_2018}%
  \BibitemOpen
  \bibfield  {author} {\bibinfo {author} {\bibfnamefont {Y.~L.}\ \bibnamefont
  {Han}}, \bibinfo {author} {\bibfnamefont {P.}~\bibnamefont {Ronceray}},
  \bibinfo {author} {\bibfnamefont {G.}~\bibnamefont {Xu}}, \bibinfo {author}
  {\bibfnamefont {A.}~\bibnamefont {Malandrino}}, \bibinfo {author}
  {\bibfnamefont {R.~D.}\ \bibnamefont {Kamm}}, \bibinfo {author}
  {\bibfnamefont {M.}~\bibnamefont {Lenz}}, \bibinfo {author} {\bibfnamefont
  {C.~P.}\ \bibnamefont {Broedersz}},\ and\ \bibinfo {author} {\bibfnamefont
  {M.}~\bibnamefont {Guo}},\ }\bibfield  {title} {\bibinfo {title} {Cell
  contraction induces long-ranged stress stiffening in the extracellular
  matrix},\ }\href {https://doi.org/10.1073/pnas.1722619115} {\bibfield
  {journal} {\bibinfo  {journal} {Proc. Nat. Acad. Sci.}\ }\textbf {\bibinfo
  {volume} {115}},\ \bibinfo {pages} {4075} (\bibinfo {year}
  {2018})}\BibitemShut {NoStop}%
\bibitem [{\citenamefont {Doyle}\ \emph {et~al.}(2021)\citenamefont {Doyle},
  \citenamefont {Sykora}, \citenamefont {Pacheco}, \citenamefont {Kutys},\ and\
  \citenamefont {Yamada}}]{Doyle2021}%
  \BibitemOpen
  \bibfield  {author} {\bibinfo {author} {\bibfnamefont {A.~D.}\ \bibnamefont
  {Doyle}}, \bibinfo {author} {\bibfnamefont {D.~J.}\ \bibnamefont {Sykora}},
  \bibinfo {author} {\bibfnamefont {G.~G.}\ \bibnamefont {Pacheco}}, \bibinfo
  {author} {\bibfnamefont {M.~L.}\ \bibnamefont {Kutys}},\ and\ \bibinfo
  {author} {\bibfnamefont {K.~M.}\ \bibnamefont {Yamada}},\ }\bibfield  {title}
  {\bibinfo {title} {3d mesenchymal cell migration is driven by anterior
  cellular contraction that generates an extracellular matrix prestrain},\
  }\href@noop {} {\bibfield  {journal} {\bibinfo  {journal} {Developmental
  Cell}\ }\textbf {\bibinfo {volume} {56}},\ \bibinfo {pages} {826} (\bibinfo
  {year} {2021})}\BibitemShut {NoStop}%
\bibitem [{\citenamefont {Nagai}\ and\ \citenamefont
  {Honda}(2001)}]{Honda_2001}%
  \BibitemOpen
  \bibfield  {author} {\bibinfo {author} {\bibfnamefont {T.}~\bibnamefont
  {Nagai}}\ and\ \bibinfo {author} {\bibfnamefont {H.}~\bibnamefont {Honda}},\
  }\bibfield  {title} {\bibinfo {title} {A dynamic cell model for the formation
  of epithelial tissues},\ }\href@noop {} {\bibfield  {journal} {\bibinfo
  {journal} {Philosophical Magazine B - Physics of Condensed Matter,
  Statistical Mechanics, Electronic, Optical, and Magnetic Properties}\
  }\textbf {\bibinfo {volume} {81}},\ \bibinfo {pages} {699} (\bibinfo {year}
  {2001})}\BibitemShut {NoStop}%
\bibitem [{\citenamefont {Farhadifar}\ \emph {et~al.}(2007)\citenamefont
  {Farhadifar}, \citenamefont {Roper},\ and\ \citenamefont
  {et~al}}]{Farhadifar_2007}%
  \BibitemOpen
  \bibfield  {author} {\bibinfo {author} {\bibfnamefont {R.}~\bibnamefont
  {Farhadifar}}, \bibinfo {author} {\bibfnamefont {J.-C.}\ \bibnamefont
  {Roper}},\ and\ \bibinfo {author} {\bibnamefont {et~al}},\ }\bibfield
  {title} {\bibinfo {title} {The influence of cell mechanics, cell-cell
  interactions, and proliferation on epithelial packing},\ }\href@noop {}
  {\bibfield  {journal} {\bibinfo  {journal} {Current Biology}\ }\textbf
  {\bibinfo {volume} {17}},\ \bibinfo {pages} {2095} (\bibinfo {year}
  {2007})}\BibitemShut {NoStop}%
\bibitem [{\citenamefont {Bi}\ \emph {et~al.}(2015{\natexlab{a}})\citenamefont
  {Bi}, \citenamefont {Lopez}, \citenamefont {Schwarz},\ and\ \citenamefont
  {Manning}}]{Bi_2015}%
  \BibitemOpen
  \bibfield  {author} {\bibinfo {author} {\bibfnamefont {D.}~\bibnamefont
  {Bi}}, \bibinfo {author} {\bibfnamefont {J.}~\bibnamefont {Lopez}}, \bibinfo
  {author} {\bibfnamefont {J.~M.}\ \bibnamefont {Schwarz}},\ and\ \bibinfo
  {author} {\bibfnamefont {M.~L.}\ \bibnamefont {Manning}},\ }\bibfield
  {title} {\bibinfo {title} {A density-independent rigidity transition in
  biological tissues},\ }\href@noop {} {\bibfield  {journal} {\bibinfo
  {journal} {Nature Physics}\ }\textbf {\bibinfo {volume} {11}},\ \bibinfo
  {pages} {1074–1079} (\bibinfo {year} {2015}{\natexlab{a}})}\BibitemShut
  {NoStop}%
\bibitem [{\citenamefont {Okuda}\ \emph {et~al.}(2013)\citenamefont {Okuda},
  \citenamefont {Inoue}, \citenamefont {Eiraku}, \citenamefont {Sasai},\ and\
  \citenamefont {Adachi}}]{Okuda2013}%
  \BibitemOpen
  \bibfield  {author} {\bibinfo {author} {\bibfnamefont {S.}~\bibnamefont
  {Okuda}}, \bibinfo {author} {\bibfnamefont {Y.}~\bibnamefont {Inoue}},
  \bibinfo {author} {\bibfnamefont {M.}~\bibnamefont {Eiraku}}, \bibinfo
  {author} {\bibfnamefont {Y.}~\bibnamefont {Sasai}},\ and\ \bibinfo {author}
  {\bibfnamefont {T.}~\bibnamefont {Adachi}},\ }\bibfield  {title} {\bibinfo
  {title} {{Reversible network reconnection model for simulating large
  deformation in dynamic tissue morphogenesis}},\ }\href
  {https://doi.org/10.1007/s10237-012-0430-7} {\bibfield  {journal} {\bibinfo
  {journal} {Biomechanics and Modeling in Mechanobiology}\ }\textbf {\bibinfo
  {volume} {12}},\ \bibinfo {pages} {627} (\bibinfo {year} {2013})}\BibitemShut
  {NoStop}%
\bibitem [{\citenamefont {Zhang}\ and\ \citenamefont
  {Schwarz}(2022)}]{Zhang2022}%
  \BibitemOpen
  \bibfield  {author} {\bibinfo {author} {\bibfnamefont {T.}~\bibnamefont
  {Zhang}}\ and\ \bibinfo {author} {\bibfnamefont {J.}~\bibnamefont
  {Schwarz}},\ }\bibfield  {title} {\bibinfo {title} {Topologically-protected
  interior for three-dimensional confluent cellular collectives},\ }\href@noop
  {} {\bibfield  {journal} {\bibinfo  {journal} {Physical Review Research}\
  }\textbf {\bibinfo {volume} {4}},\ \bibinfo {pages} {043148} (\bibinfo {year}
  {2022})}\BibitemShut {NoStop}%
\bibitem [{\citenamefont {Sahu}\ and\ \citenamefont {{\it et
  al.}}(2020)}]{Sahu_2019}%
  \BibitemOpen
  \bibfield  {author} {\bibinfo {author} {\bibfnamefont {P.}~\bibnamefont
  {Sahu}}\ and\ \bibinfo {author} {\bibnamefont {{\it et al.}}},\ }\bibfield
  {title} {\bibinfo {title} {Small-scale demixing in confluent biological
  tissues},\ }\href@noop {} {\bibfield  {journal} {\bibinfo  {journal} {Soft
  Matt.}\ }\textbf {\bibinfo {volume} {16}},\ \bibinfo {pages} {3325} (\bibinfo
  {year} {2020})}\BibitemShut {NoStop}%
\bibitem [{\citenamefont {Park}\ \emph {et~al.}(2015)\citenamefont {Park},
  \citenamefont {Kim}, \citenamefont {Bi},\ and\ \citenamefont
  {et~al}}]{Park_2015}%
  \BibitemOpen
  \bibfield  {author} {\bibinfo {author} {\bibfnamefont {J.}~\bibnamefont
  {Park}}, \bibinfo {author} {\bibfnamefont {J.}~\bibnamefont {Kim}}, \bibinfo
  {author} {\bibfnamefont {D.}~\bibnamefont {Bi}},\ and\ \bibinfo {author}
  {\bibnamefont {et~al}},\ }\bibfield  {title} {\bibinfo {title} {Unjamming and
  cell shape in the asthmatic airway epithelium},\ }\href@noop {} {\bibfield
  {journal} {\bibinfo  {journal} {Nature Materials}\ }\textbf {\bibinfo
  {volume} {14}},\ \bibinfo {pages} {1040–1048} (\bibinfo {year}
  {2015})}\BibitemShut {NoStop}%
\bibitem [{\citenamefont {Malinverno}\ and\ \citenamefont {{\it et
  al.}}(2017)}]{Malinverno_2017}%
  \BibitemOpen
  \bibfield  {author} {\bibinfo {author} {\bibfnamefont {C.}~\bibnamefont
  {Malinverno}}\ and\ \bibinfo {author} {\bibnamefont {{\it et al.}}},\
  }\bibfield  {title} {\bibinfo {title} {{Endocytic reawakening of motility in
  jammed epithelia}},\ }\href {https://doi.org/10.1038/nmat4848} {\bibfield
  {journal} {\bibinfo  {journal} {Nature Mat.}\ }\textbf {\bibinfo {volume}
  {16}},\ \bibinfo {pages} {587} (\bibinfo {year} {2017})}\BibitemShut
  {NoStop}%
\bibitem [{\citenamefont {Licup}\ \emph {et~al.}(2015)\citenamefont {Licup},
  \citenamefont {Münster}, \citenamefont {Sharma}, \citenamefont {Sheinman},
  \citenamefont {Jawerth}, \citenamefont {Fabry}, \citenamefont {Weitz},\ and\
  \citenamefont {MacKintosh}}]{Licup_2015}%
  \BibitemOpen
  \bibfield  {author} {\bibinfo {author} {\bibfnamefont {A.~J.}\ \bibnamefont
  {Licup}}, \bibinfo {author} {\bibfnamefont {S.}~\bibnamefont {Münster}},
  \bibinfo {author} {\bibfnamefont {A.}~\bibnamefont {Sharma}}, \bibinfo
  {author} {\bibfnamefont {M.}~\bibnamefont {Sheinman}}, \bibinfo {author}
  {\bibfnamefont {L.~M.}\ \bibnamefont {Jawerth}}, \bibinfo {author}
  {\bibfnamefont {B.}~\bibnamefont {Fabry}}, \bibinfo {author} {\bibfnamefont
  {D.~A.}\ \bibnamefont {Weitz}},\ and\ \bibinfo {author} {\bibfnamefont
  {F.~C.}\ \bibnamefont {MacKintosh}},\ }\bibfield  {title} {\bibinfo {title}
  {Stress controls the mechanics of collagen networks},\ }\href
  {https://doi.org/10.1073/pnas.1504258112} {\bibfield  {journal} {\bibinfo
  {journal} {Proceedings of the National Academy of Sciences}\ }\textbf
  {\bibinfo {volume} {112}},\ \bibinfo {pages} {9573} (\bibinfo {year}
  {2015})}\BibitemShut {NoStop}%
\bibitem [{\citenamefont {Sharma}\ \emph {et~al.}(2016)\citenamefont {Sharma},
  \citenamefont {Licup}, \citenamefont {Jansen},\ and\ \citenamefont
  {et~al.}}]{Sharma_2016}%
  \BibitemOpen
  \bibfield  {author} {\bibinfo {author} {\bibfnamefont {A.}~\bibnamefont
  {Sharma}}, \bibinfo {author} {\bibfnamefont {A.}~\bibnamefont {Licup}},
  \bibinfo {author} {\bibfnamefont {K.}~\bibnamefont {Jansen}},\ and\ \bibinfo
  {author} {\bibnamefont {et~al.}},\ }\bibfield  {title} {\bibinfo {title}
  {train-controlled criticality governs the nonlinear mechanics of fibre
  networks},\ }\href {https://doi.org/10.1038/nphys3628} {\bibfield  {journal}
  {\bibinfo  {journal} {Nature Phys}\ }\textbf {\bibinfo {volume} {12}},\
  \bibinfo {pages} {584} (\bibinfo {year} {2016})}\BibitemShut {NoStop}%
\bibitem [{\citenamefont {Jansen}\ \emph {et~al.}(2018)\citenamefont {Jansen},
  \citenamefont {Licup}, \citenamefont {Sharma}, \citenamefont {Rens},
  \citenamefont {MacKintosh},\ and\ \citenamefont {Koenderink}}]{Jansen_2018}%
  \BibitemOpen
  \bibfield  {author} {\bibinfo {author} {\bibfnamefont {K.~A.}\ \bibnamefont
  {Jansen}}, \bibinfo {author} {\bibfnamefont {A.~J.}\ \bibnamefont {Licup}},
  \bibinfo {author} {\bibfnamefont {A.}~\bibnamefont {Sharma}}, \bibinfo
  {author} {\bibfnamefont {R.}~\bibnamefont {Rens}}, \bibinfo {author}
  {\bibfnamefont {F.~C.}\ \bibnamefont {MacKintosh}},\ and\ \bibinfo {author}
  {\bibfnamefont {G.~H.}\ \bibnamefont {Koenderink}},\ }\bibfield  {title}
  {\bibinfo {title} {The role of network architecture in collagen mechanics},\
  }\href {https://doi.org/https://doi.org/10.1016/j.bpj.2018.04.043} {\bibfield
   {journal} {\bibinfo  {journal} {Biophysical Journal}\ }\textbf {\bibinfo
  {volume} {114}},\ \bibinfo {pages} {2665 } (\bibinfo {year}
  {2018})}\BibitemShut {NoStop}%
\bibitem [{\citenamefont {Parker}\ \emph {et~al.}(2020)\citenamefont {Parker},
  \citenamefont {Marchetti}, \citenamefont {Manning},\ and\ \citenamefont
  {Schwarz}}]{Parker2020}%
  \BibitemOpen
  \bibfield  {author} {\bibinfo {author} {\bibfnamefont {A.}~\bibnamefont
  {Parker}}, \bibinfo {author} {\bibfnamefont {M.~C.}\ \bibnamefont
  {Marchetti}}, \bibinfo {author} {\bibfnamefont {M.~L.}\ \bibnamefont
  {Manning}},\ and\ \bibinfo {author} {\bibfnamefont {J.~M.}\ \bibnamefont
  {Schwarz}},\ }\bibfield  {title} {\bibinfo {title} {How does the
  extracellular matrix affect the rigidity of an embedded spheroid?},\
  }\href@noop {} {\bibfield  {journal} {\bibinfo  {journal} {arXiv preprint
  arXiv:2006.16203}\ } (\bibinfo {year} {2020})}\BibitemShut {NoStop}%
\bibitem [{\citenamefont {Tong}\ \emph {et~al.}(2022)\citenamefont {Tong},
  \citenamefont {Singh}, \citenamefont {Sknepnek},\ and\ \citenamefont
  {Ko{\v{s}}mrlj}}]{Tong2022}%
  \BibitemOpen
  \bibfield  {author} {\bibinfo {author} {\bibfnamefont {S.}~\bibnamefont
  {Tong}}, \bibinfo {author} {\bibfnamefont {N.~K.}\ \bibnamefont {Singh}},
  \bibinfo {author} {\bibfnamefont {R.}~\bibnamefont {Sknepnek}},\ and\
  \bibinfo {author} {\bibfnamefont {A.}~\bibnamefont {Ko{\v{s}}mrlj}},\
  }\bibfield  {title} {\bibinfo {title} {Linear viscoelastic properties of the
  vertex model for epithelial tissues},\ }\href@noop {} {\bibfield  {journal}
  {\bibinfo  {journal} {PLoS Computational Biology}\ }\textbf {\bibinfo
  {volume} {18}},\ \bibinfo {pages} {e1010135} (\bibinfo {year}
  {2022})}\BibitemShut {NoStop}%
\bibitem [{\citenamefont {Barbazan}\ \emph {et~al.}(2023)\citenamefont
  {Barbazan}, \citenamefont {P{\'e}rez-Gonz{\'a}lez}, \citenamefont
  {G{\'o}mez-Gonz{\'a}lez}, \citenamefont {Dedenon}, \citenamefont {Richon},
  \citenamefont {Latorre}, \citenamefont {Serra}, \citenamefont {Mariani},
  \citenamefont {Descroix}, \citenamefont {Sens} \emph
  {et~al.}}]{Barbazan2023}%
  \BibitemOpen
  \bibfield  {author} {\bibinfo {author} {\bibfnamefont {J.}~\bibnamefont
  {Barbazan}}, \bibinfo {author} {\bibfnamefont {C.}~\bibnamefont
  {P{\'e}rez-Gonz{\'a}lez}}, \bibinfo {author} {\bibfnamefont {M.}~\bibnamefont
  {G{\'o}mez-Gonz{\'a}lez}}, \bibinfo {author} {\bibfnamefont {M.}~\bibnamefont
  {Dedenon}}, \bibinfo {author} {\bibfnamefont {S.}~\bibnamefont {Richon}},
  \bibinfo {author} {\bibfnamefont {E.}~\bibnamefont {Latorre}}, \bibinfo
  {author} {\bibfnamefont {M.}~\bibnamefont {Serra}}, \bibinfo {author}
  {\bibfnamefont {P.}~\bibnamefont {Mariani}}, \bibinfo {author} {\bibfnamefont
  {S.}~\bibnamefont {Descroix}}, \bibinfo {author} {\bibfnamefont
  {P.}~\bibnamefont {Sens}}, \emph {et~al.},\ }\bibfield  {title} {\bibinfo
  {title} {Cancer-associated fibroblasts actively compress cancer cells and
  modulate mechanotransduction},\ }\href@noop {} {\bibfield  {journal}
  {\bibinfo  {journal} {Nature Communications}\ }\textbf {\bibinfo {volume}
  {14}},\ \bibinfo {pages} {6966} (\bibinfo {year} {2023})}\BibitemShut
  {NoStop}%
\bibitem [{\citenamefont {Parker}\ and\ \citenamefont
  {Schwarz}(2024)}]{Parker2024}%
  \BibitemOpen
  \bibfield  {author} {\bibinfo {author} {\bibfnamefont {A.}~\bibnamefont
  {Parker}}\ and\ \bibinfo {author} {\bibfnamefont {J.~M.}\ \bibnamefont
  {Schwarz}},\ }\href@noop {} {\bibfield  {journal} {\bibinfo  {journal} {in
  preparation}\ } (\bibinfo {year} {2024})}\BibitemShut {NoStop}%
\bibitem [{\citenamefont {Chan}\ and\ \citenamefont {Odde}(2008)}]{Chan2008}%
  \BibitemOpen
  \bibfield  {author} {\bibinfo {author} {\bibfnamefont {C.~E.}\ \bibnamefont
  {Chan}}\ and\ \bibinfo {author} {\bibfnamefont {D.~J.}\ \bibnamefont
  {Odde}},\ }\bibfield  {title} {\bibinfo {title} {Traction dynamics of
  filopodia on compliant substrates},\ }\href@noop {} {\bibfield  {journal}
  {\bibinfo  {journal} {Science}\ }\textbf {\bibinfo {volume} {322}},\ \bibinfo
  {pages} {1687} (\bibinfo {year} {2008})}\BibitemShut {NoStop}%
\bibitem [{\citenamefont {Mark}\ \emph {et~al.}(2020)\citenamefont {Mark},
  \citenamefont {Grundy}, \citenamefont {Strissel}, \citenamefont
  {B{\"o}hringer}, \citenamefont {Grummel}, \citenamefont {Gerum},
  \citenamefont {Steinwachs}, \citenamefont {Hack}, \citenamefont {Beckmann},
  \citenamefont {Eckstein} \emph {et~al.}}]{Mark2020}%
  \BibitemOpen
  \bibfield  {author} {\bibinfo {author} {\bibfnamefont {C.}~\bibnamefont
  {Mark}}, \bibinfo {author} {\bibfnamefont {T.~J.}\ \bibnamefont {Grundy}},
  \bibinfo {author} {\bibfnamefont {P.~L.}\ \bibnamefont {Strissel}}, \bibinfo
  {author} {\bibfnamefont {D.}~\bibnamefont {B{\"o}hringer}}, \bibinfo {author}
  {\bibfnamefont {N.}~\bibnamefont {Grummel}}, \bibinfo {author} {\bibfnamefont
  {R.}~\bibnamefont {Gerum}}, \bibinfo {author} {\bibfnamefont
  {J.}~\bibnamefont {Steinwachs}}, \bibinfo {author} {\bibfnamefont {C.~C.}\
  \bibnamefont {Hack}}, \bibinfo {author} {\bibfnamefont {M.~W.}\ \bibnamefont
  {Beckmann}}, \bibinfo {author} {\bibfnamefont {M.}~\bibnamefont {Eckstein}},
  \emph {et~al.},\ }\bibfield  {title} {\bibinfo {title} {Collective forces of
  tumor spheroids in three-dimensional biopolymer networks},\ }\href@noop {}
  {\bibfield  {journal} {\bibinfo  {journal} {Elife}\ }\textbf {\bibinfo
  {volume} {9}},\ \bibinfo {pages} {e51912} (\bibinfo {year}
  {2020})}\BibitemShut {NoStop}%
\bibitem [{\citenamefont {Hoffman}\ and\ \citenamefont
  {Yap}(2015)}]{Hoffman2015}%
  \BibitemOpen
  \bibfield  {author} {\bibinfo {author} {\bibfnamefont {B.~D.}\ \bibnamefont
  {Hoffman}}\ and\ \bibinfo {author} {\bibfnamefont {A.~S.}\ \bibnamefont
  {Yap}},\ }\bibfield  {title} {\bibinfo {title} {Towards a dynamic
  understanding of cadherin-based mechanobiology},\ }\href@noop {} {\bibfield
  {journal} {\bibinfo  {journal} {Trends in Cell Biology}\ }\textbf {\bibinfo
  {volume} {25}},\ \bibinfo {pages} {803} (\bibinfo {year} {2015})}\BibitemShut
  {NoStop}%
\bibitem [{\citenamefont {Sahu}\ \emph {et~al.}(2020)\citenamefont {Sahu},
  \citenamefont {Sussman}, \citenamefont {R{\"{u}}bsam}, \citenamefont {Mertz},
  \citenamefont {Horsley}, \citenamefont {Dufresne}, \citenamefont {Niessen},
  \citenamefont {Marchetti}, \citenamefont {Manning},\ and\ \citenamefont
  {Schwarz}}]{Sahu2020a}%
  \BibitemOpen
  \bibfield  {author} {\bibinfo {author} {\bibfnamefont {P.}~\bibnamefont
  {Sahu}}, \bibinfo {author} {\bibfnamefont {D.~M.}\ \bibnamefont {Sussman}},
  \bibinfo {author} {\bibfnamefont {M.}~\bibnamefont {R{\"{u}}bsam}}, \bibinfo
  {author} {\bibfnamefont {A.~F.}\ \bibnamefont {Mertz}}, \bibinfo {author}
  {\bibfnamefont {V.}~\bibnamefont {Horsley}}, \bibinfo {author} {\bibfnamefont
  {E.~R.}\ \bibnamefont {Dufresne}}, \bibinfo {author} {\bibfnamefont {C.~M.}\
  \bibnamefont {Niessen}}, \bibinfo {author} {\bibfnamefont {M.~C.}\
  \bibnamefont {Marchetti}}, \bibinfo {author} {\bibfnamefont {M.~L.}\
  \bibnamefont {Manning}},\ and\ \bibinfo {author} {\bibfnamefont {J.~M.}\
  \bibnamefont {Schwarz}},\ }\bibfield  {title} {\bibinfo {title} {{Small-scale
  demixing in confluent biological tissues}},\ }\href@noop {} {\bibfield
  {journal} {\bibinfo  {journal} {Soft Matter}\ }\textbf {\bibinfo {volume}
  {16}},\ \bibinfo {pages} {3325} (\bibinfo {year} {2020})}\BibitemShut
  {NoStop}%
\bibitem [{\citenamefont {Warmt}\ \emph {et~al.}(2021)\citenamefont {Warmt},
  \citenamefont {Grosser}, \citenamefont {Blauth}, \citenamefont {Xie},
  \citenamefont {Kubitschke}, \citenamefont {Stange}, \citenamefont {Sauer},
  \citenamefont {Schnau{\ss}}, \citenamefont {Tomm}, \citenamefont {von Bergen}
  \emph {et~al.}}]{Warmt2021}%
  \BibitemOpen
  \bibfield  {author} {\bibinfo {author} {\bibfnamefont {E.}~\bibnamefont
  {Warmt}}, \bibinfo {author} {\bibfnamefont {S.}~\bibnamefont {Grosser}},
  \bibinfo {author} {\bibfnamefont {E.}~\bibnamefont {Blauth}}, \bibinfo
  {author} {\bibfnamefont {X.}~\bibnamefont {Xie}}, \bibinfo {author}
  {\bibfnamefont {H.}~\bibnamefont {Kubitschke}}, \bibinfo {author}
  {\bibfnamefont {R.}~\bibnamefont {Stange}}, \bibinfo {author} {\bibfnamefont
  {F.}~\bibnamefont {Sauer}}, \bibinfo {author} {\bibfnamefont
  {J.}~\bibnamefont {Schnau{\ss}}}, \bibinfo {author} {\bibfnamefont {J.~M.}\
  \bibnamefont {Tomm}}, \bibinfo {author} {\bibfnamefont {M.}~\bibnamefont {von
  Bergen}}, \emph {et~al.},\ }\bibfield  {title} {\bibinfo {title} {Differences
  in cortical contractile properties between healthy epithelial and cancerous
  mesenchymal breast cells},\ }\href@noop {} {\bibfield  {journal} {\bibinfo
  {journal} {New Journal of Physics}\ }\textbf {\bibinfo {volume} {23}},\
  \bibinfo {pages} {103020} (\bibinfo {year} {2021})}\BibitemShut {NoStop}%
\bibitem [{\citenamefont {Lopez}\ \emph {et~al.}(2014)\citenamefont {Lopez},
  \citenamefont {Das},\ and\ \citenamefont {Schwarz}}]{Lopez2014}%
  \BibitemOpen
  \bibfield  {author} {\bibinfo {author} {\bibfnamefont {J.~H.}\ \bibnamefont
  {Lopez}}, \bibinfo {author} {\bibfnamefont {M.}~\bibnamefont {Das}},\ and\
  \bibinfo {author} {\bibfnamefont {J.~M.}\ \bibnamefont {Schwarz}},\
  }\bibfield  {title} {\bibinfo {title} {Active elastic dimers: Cells moving on
  rigid tracks},\ }\href@noop {} {\bibfield  {journal} {\bibinfo  {journal}
  {Physical Review E}\ }\textbf {\bibinfo {volume} {90}},\ \bibinfo {pages}
  {032707} (\bibinfo {year} {2014})}\BibitemShut {NoStop}%
\bibitem [{\citenamefont {Mayett}\ \emph {et~al.}(2017)\citenamefont {Mayett},
  \citenamefont {Bitten}, \citenamefont {Das},\ and\ \citenamefont
  {Schwarz}}]{Mayett2017}%
  \BibitemOpen
  \bibfield  {author} {\bibinfo {author} {\bibfnamefont {D.}~\bibnamefont
  {Mayett}}, \bibinfo {author} {\bibfnamefont {N.}~\bibnamefont {Bitten}},
  \bibinfo {author} {\bibfnamefont {M.}~\bibnamefont {Das}},\ and\ \bibinfo
  {author} {\bibfnamefont {J.~M.}\ \bibnamefont {Schwarz}},\ }\bibfield
  {title} {\bibinfo {title} {Chase-and-run dynamics in cell motility and the
  molecular rupture of interacting active elastic dimers},\ }\href@noop {}
  {\bibfield  {journal} {\bibinfo  {journal} {Physical Review E}\ }\textbf
  {\bibinfo {volume} {96}},\ \bibinfo {pages} {032407} (\bibinfo {year}
  {2017})}\BibitemShut {NoStop}%
\bibitem [{\citenamefont {Bi}\ \emph {et~al.}(2015{\natexlab{b}})\citenamefont
  {Bi}, \citenamefont {Lopez}, \citenamefont {Schwarz},\ and\ \citenamefont
  {Manning}}]{Bi2015}%
  \BibitemOpen
  \bibfield  {author} {\bibinfo {author} {\bibfnamefont {D.}~\bibnamefont
  {Bi}}, \bibinfo {author} {\bibfnamefont {J.~H.}\ \bibnamefont {Lopez}},
  \bibinfo {author} {\bibfnamefont {J.~M.}\ \bibnamefont {Schwarz}},\ and\
  \bibinfo {author} {\bibfnamefont {M.~L.}\ \bibnamefont {Manning}},\
  }\bibfield  {title} {\bibinfo {title} {{A density-independent rigidity
  transition in biological tissues}},\ }\href
  {https://doi.org/10.1038/nphys3471} {\bibfield  {journal} {\bibinfo
  {journal} {Nature Physics}\ }\textbf {\bibinfo {volume} {11}},\ \bibinfo
  {pages} {1074} (\bibinfo {year} {2015}{\natexlab{b}})}\BibitemShut {NoStop}%
\bibitem [{\citenamefont {Sarkar}\ and\ \citenamefont
  {Krajnc}(2023)}]{Sarkar2023}%
  \BibitemOpen
  \bibfield  {author} {\bibinfo {author} {\bibfnamefont {T.}~\bibnamefont
  {Sarkar}}\ and\ \bibinfo {author} {\bibfnamefont {M.}~\bibnamefont
  {Krajnc}},\ }\bibfield  {title} {\bibinfo {title} {Graph vertex model},\
  }\href@noop {} {\bibfield  {journal} {\bibinfo  {journal} {arXiv preprint
  arXiv:2309.04818}\ } (\bibinfo {year} {2023})}\BibitemShut {NoStop}%
\bibitem [{\citenamefont {Suh}\ \emph {et~al.}(2019{\natexlab{b}})\citenamefont
  {Suh}, \citenamefont {Hall}, \citenamefont {Huang}, \citenamefont {Moon},
  \citenamefont {Song}, \citenamefont {Ma}, \citenamefont {Bonassar},
  \citenamefont {Segall},\ and\ \citenamefont {Wu}}]{Suh2019}%
  \BibitemOpen
  \bibfield  {author} {\bibinfo {author} {\bibfnamefont {Y.~J.}\ \bibnamefont
  {Suh}}, \bibinfo {author} {\bibfnamefont {M.~S.}\ \bibnamefont {Hall}},
  \bibinfo {author} {\bibfnamefont {Y.~L.}\ \bibnamefont {Huang}}, \bibinfo
  {author} {\bibfnamefont {S.~Y.}\ \bibnamefont {Moon}}, \bibinfo {author}
  {\bibfnamefont {W.}~\bibnamefont {Song}}, \bibinfo {author} {\bibfnamefont
  {M.}~\bibnamefont {Ma}}, \bibinfo {author} {\bibfnamefont {L.~J.}\
  \bibnamefont {Bonassar}}, \bibinfo {author} {\bibfnamefont {J.~E.}\
  \bibnamefont {Segall}},\ and\ \bibinfo {author} {\bibfnamefont
  {M.}~\bibnamefont {Wu}},\ }\bibfield  {title} {\bibinfo {title} {Glycation of
  collagen matrices promotes breast tumor cell invasion},\ }\href@noop {}
  {\bibfield  {journal} {\bibinfo  {journal} {Integrative Biology}\ }\textbf
  {\bibinfo {volume} {11}},\ \bibinfo {pages} {109} (\bibinfo {year}
  {2019}{\natexlab{b}})}\BibitemShut {NoStop}%
\bibitem [{\citenamefont {Delarue}\ \emph {et~al.}(2014)\citenamefont
  {Delarue}, \citenamefont {Montel}, \citenamefont {Vignjevic}, \citenamefont
  {Prost}, \citenamefont {Joanny},\ and\ \citenamefont
  {Cappello}}]{Delarue2014}%
  \BibitemOpen
  \bibfield  {author} {\bibinfo {author} {\bibfnamefont {M.}~\bibnamefont
  {Delarue}}, \bibinfo {author} {\bibfnamefont {F.}~\bibnamefont {Montel}},
  \bibinfo {author} {\bibfnamefont {D.}~\bibnamefont {Vignjevic}}, \bibinfo
  {author} {\bibfnamefont {J.}~\bibnamefont {Prost}}, \bibinfo {author}
  {\bibfnamefont {J.-F.}\ \bibnamefont {Joanny}},\ and\ \bibinfo {author}
  {\bibfnamefont {G.}~\bibnamefont {Cappello}},\ }\bibfield  {title} {\bibinfo
  {title} {Compressive stress inhibits proliferation in tumor spheroids through
  a volume limitation},\ }\href@noop {} {\bibfield  {journal} {\bibinfo
  {journal} {Biophysical journal}\ }\textbf {\bibinfo {volume} {107}},\
  \bibinfo {pages} {1821} (\bibinfo {year} {2014})}\BibitemShut {NoStop}%
\bibitem [{\citenamefont {Tsingos}\ \emph {et~al.}(2023)\citenamefont
  {Tsingos}, \citenamefont {Bakker}, \citenamefont {Keijzer}, \citenamefont
  {Hupkes},\ and\ \citenamefont {Merks}}]{Tsingos2023}%
  \BibitemOpen
  \bibfield  {author} {\bibinfo {author} {\bibfnamefont {E.}~\bibnamefont
  {Tsingos}}, \bibinfo {author} {\bibfnamefont {B.~H.}\ \bibnamefont {Bakker}},
  \bibinfo {author} {\bibfnamefont {K.~A.}\ \bibnamefont {Keijzer}}, \bibinfo
  {author} {\bibfnamefont {H.~J.}\ \bibnamefont {Hupkes}},\ and\ \bibinfo
  {author} {\bibfnamefont {R.~M.}\ \bibnamefont {Merks}},\ }\bibfield  {title}
  {\bibinfo {title} {Hybrid cellular potts and bead-spring modeling of cells in
  fibrous extracellular matrix},\ }\href@noop {} {\bibfield  {journal}
  {\bibinfo  {journal} {Biophysical Journal}\ } (\bibinfo {year}
  {2023})}\BibitemShut {NoStop}%
\bibitem [{\citenamefont {Ronceray}\ \emph {et~al.}(2016)\citenamefont
  {Ronceray}, \citenamefont {Broedersz},\ and\ \citenamefont
  {Lenz}}]{Ronceray2016}%
  \BibitemOpen
  \bibfield  {author} {\bibinfo {author} {\bibfnamefont {P.}~\bibnamefont
  {Ronceray}}, \bibinfo {author} {\bibfnamefont {C.~P.}\ \bibnamefont
  {Broedersz}},\ and\ \bibinfo {author} {\bibfnamefont {M.}~\bibnamefont
  {Lenz}},\ }\bibfield  {title} {\bibinfo {title} {Fiber networks amplify
  active stress},\ }\href@noop {} {\bibfield  {journal} {\bibinfo  {journal}
  {Proceedings of the National Academy of Sciences}\ }\textbf {\bibinfo
  {volume} {113}},\ \bibinfo {pages} {2827} (\bibinfo {year}
  {2016})}\BibitemShut {NoStop}%
\bibitem [{\citenamefont {Friedl}\ and\ \citenamefont
  {Alexander}(2011)}]{Friedl2011}%
  \BibitemOpen
  \bibfield  {author} {\bibinfo {author} {\bibfnamefont {P.}~\bibnamefont
  {Friedl}}\ and\ \bibinfo {author} {\bibfnamefont {S.}~\bibnamefont
  {Alexander}},\ }\bibfield  {title} {\bibinfo {title} {Cancer invasion and the
  microenvironment: plasticity and reciprocity},\ }\href@noop {} {\bibfield
  {journal} {\bibinfo  {journal} {Cell}\ }\textbf {\bibinfo {volume} {147}},\
  \bibinfo {pages} {992} (\bibinfo {year} {2011})}\BibitemShut {NoStop}%
\bibitem [{\citenamefont {Ilina}\ \emph {et~al.}(2020)\citenamefont {Ilina},
  \citenamefont {Gritsenko}, \citenamefont {Syga}, \citenamefont {Lippoldt},
  \citenamefont {La~Porta}, \citenamefont {Chepizhko}, \citenamefont {Grosser},
  \citenamefont {Vullings}, \citenamefont {Bakker}, \citenamefont {Starru{\ss}}
  \emph {et~al.}}]{Ilina2020}%
  \BibitemOpen
  \bibfield  {author} {\bibinfo {author} {\bibfnamefont {O.}~\bibnamefont
  {Ilina}}, \bibinfo {author} {\bibfnamefont {P.~G.}\ \bibnamefont
  {Gritsenko}}, \bibinfo {author} {\bibfnamefont {S.}~\bibnamefont {Syga}},
  \bibinfo {author} {\bibfnamefont {J.}~\bibnamefont {Lippoldt}}, \bibinfo
  {author} {\bibfnamefont {C.~A.}\ \bibnamefont {La~Porta}}, \bibinfo {author}
  {\bibfnamefont {O.}~\bibnamefont {Chepizhko}}, \bibinfo {author}
  {\bibfnamefont {S.}~\bibnamefont {Grosser}}, \bibinfo {author} {\bibfnamefont
  {M.}~\bibnamefont {Vullings}}, \bibinfo {author} {\bibfnamefont {G.-J.}\
  \bibnamefont {Bakker}}, \bibinfo {author} {\bibfnamefont {J.}~\bibnamefont
  {Starru{\ss}}}, \emph {et~al.},\ }\bibfield  {title} {\bibinfo {title}
  {Cell--cell adhesion and 3d matrix confinement determine jamming transitions
  in breast cancer invasion},\ }\href@noop {} {\bibfield  {journal} {\bibinfo
  {journal} {Nature Cell Biology}\ }\textbf {\bibinfo {volume} {22}},\ \bibinfo
  {pages} {1103} (\bibinfo {year} {2020})}\BibitemShut {NoStop}%
\bibitem [{\citenamefont {Beunk}\ \emph {et~al.}(2022)\citenamefont {Beunk},
  \citenamefont {van Helvert}, \citenamefont {Bekker}, \citenamefont {Ran},
  \citenamefont {Kang}, \citenamefont {Paulat}, \citenamefont {Syga},
  \citenamefont {Deutsch}, \citenamefont {Friedl},\ and\ \citenamefont
  {Wolf}}]{Beunk2022}%
  \BibitemOpen
  \bibfield  {author} {\bibinfo {author} {\bibfnamefont {L.}~\bibnamefont
  {Beunk}}, \bibinfo {author} {\bibfnamefont {S.}~\bibnamefont {van Helvert}},
  \bibinfo {author} {\bibfnamefont {B.}~\bibnamefont {Bekker}}, \bibinfo
  {author} {\bibfnamefont {L.}~\bibnamefont {Ran}}, \bibinfo {author}
  {\bibfnamefont {R.}~\bibnamefont {Kang}}, \bibinfo {author} {\bibfnamefont
  {T.}~\bibnamefont {Paulat}}, \bibinfo {author} {\bibfnamefont
  {S.}~\bibnamefont {Syga}}, \bibinfo {author} {\bibfnamefont {A.}~\bibnamefont
  {Deutsch}}, \bibinfo {author} {\bibfnamefont {P.}~\bibnamefont {Friedl}},\
  and\ \bibinfo {author} {\bibfnamefont {K.}~\bibnamefont {Wolf}},\ }\bibfield
  {title} {\bibinfo {title} {Extracellular matrix guidance determines
  proteolytic and non-proteolytic cancer cell patterning},\ }\href@noop {}
  {\bibfield  {journal} {\bibinfo  {journal} {bioRxiv}\ ,\ \bibinfo {pages}
  {2022}} (\bibinfo {year} {2022})}\BibitemShut {NoStop}%
\bibitem [{\citenamefont {Kato}\ \emph {et~al.}(2023)\citenamefont {Kato},
  \citenamefont {Jenkins}, \citenamefont {Derzsi}, \citenamefont {Tozluoglu},
  \citenamefont {Rullan}, \citenamefont {Hooper}, \citenamefont {Chaleil},
  \citenamefont {Joyce}, \citenamefont {Fu}, \citenamefont {Thavaraj} \emph
  {et~al.}}]{Kato2023}%
  \BibitemOpen
  \bibfield  {author} {\bibinfo {author} {\bibfnamefont {T.}~\bibnamefont
  {Kato}}, \bibinfo {author} {\bibfnamefont {R.~P.}\ \bibnamefont {Jenkins}},
  \bibinfo {author} {\bibfnamefont {S.}~\bibnamefont {Derzsi}}, \bibinfo
  {author} {\bibfnamefont {M.}~\bibnamefont {Tozluoglu}}, \bibinfo {author}
  {\bibfnamefont {A.}~\bibnamefont {Rullan}}, \bibinfo {author} {\bibfnamefont
  {S.}~\bibnamefont {Hooper}}, \bibinfo {author} {\bibfnamefont {R.~A.}\
  \bibnamefont {Chaleil}}, \bibinfo {author} {\bibfnamefont {H.}~\bibnamefont
  {Joyce}}, \bibinfo {author} {\bibfnamefont {X.}~\bibnamefont {Fu}}, \bibinfo
  {author} {\bibfnamefont {S.}~\bibnamefont {Thavaraj}}, \emph {et~al.},\
  }\bibfield  {title} {\bibinfo {title} {Interplay of adherens junctions and
  matrix proteolysis determines the invasive pattern and growth of squamous
  cell carcinoma},\ }\href@noop {} {\bibfield  {journal} {\bibinfo  {journal}
  {Elife}\ }\textbf {\bibinfo {volume} {12}},\ \bibinfo {pages} {e76520}
  (\bibinfo {year} {2023})}\BibitemShut {NoStop}%
\bibitem [{\citenamefont {Tserunyan}\ and\ \citenamefont
  {Finley}(2022)}]{Tserunyan2022}%
  \BibitemOpen
  \bibfield  {author} {\bibinfo {author} {\bibfnamefont {V.}~\bibnamefont
  {Tserunyan}}\ and\ \bibinfo {author} {\bibfnamefont {S.~D.}\ \bibnamefont
  {Finley}},\ }\bibfield  {title} {\bibinfo {title} {Modelling predicts
  differences in chimeric antigen receptor t-cell signalling due to biological
  variability},\ }\href@noop {} {\bibfield  {journal} {\bibinfo  {journal}
  {Royal Society Open Science}\ }\textbf {\bibinfo {volume} {9}},\ \bibinfo
  {pages} {220137} (\bibinfo {year} {2022})}\BibitemShut {NoStop}%
\bibitem [{\citenamefont {Zhang}\ \emph {et~al.}(2023)\citenamefont {Zhang},
  \citenamefont {Gupta}, \citenamefont {Lancaster},\ and\ \citenamefont
  {Schwarz}}]{Zhang2023}%
  \BibitemOpen
  \bibfield  {author} {\bibinfo {author} {\bibfnamefont {T.}~\bibnamefont
  {Zhang}}, \bibinfo {author} {\bibfnamefont {S.}~\bibnamefont {Gupta}},
  \bibinfo {author} {\bibfnamefont {M.~A.}\ \bibnamefont {Lancaster}},\ and\
  \bibinfo {author} {\bibfnamefont {J.~M.}\ \bibnamefont {Schwarz}},\
  }\bibfield  {title} {\bibinfo {title} {How human-derived brain organoids are
  built differently from brain organoids derived of genetically-close
  relatives: A multi-scale hypothesis},\ }\href@noop {} {\bibfield  {journal}
  {\bibinfo  {journal} {bioRxiv}\ ,\ \bibinfo {pages} {2023}} (\bibinfo {year}
  {2023})}\BibitemShut {NoStop}%
\bibitem [{\citenamefont {Anisetti}\ \emph {et~al.}(2022)\citenamefont
  {Anisetti}, \citenamefont {Kandala}, \citenamefont {Scellier},\ and\
  \citenamefont {Schwarz}}]{Anisetti2022}%
  \BibitemOpen
  \bibfield  {author} {\bibinfo {author} {\bibfnamefont {V.~R.}\ \bibnamefont
  {Anisetti}}, \bibinfo {author} {\bibfnamefont {A.}~\bibnamefont {Kandala}},
  \bibinfo {author} {\bibfnamefont {B.}~\bibnamefont {Scellier}},\ and\
  \bibinfo {author} {\bibfnamefont {J.~M.}\ \bibnamefont {Schwarz}},\
  }\bibfield  {title} {\bibinfo {title} {Frequency propagation: Multi-mechanism
  learning in nonlinear physical networks},\ }\href@noop {} {\bibfield
  {journal} {\bibinfo  {journal} {arXiv preprint arXiv:2208.08862 (accepted to
  Neural Computation)}\ } (\bibinfo {year} {2022})}\BibitemShut {NoStop}%
\bibitem [{\citenamefont {Anisetti}\ \emph {et~al.}(2023)\citenamefont
  {Anisetti}, \citenamefont {Scellier},\ and\ \citenamefont
  {Schwarz}}]{Anisetti2023}%
  \BibitemOpen
  \bibfield  {author} {\bibinfo {author} {\bibfnamefont {V.~R.}\ \bibnamefont
  {Anisetti}}, \bibinfo {author} {\bibfnamefont {B.}~\bibnamefont {Scellier}},\
  and\ \bibinfo {author} {\bibfnamefont {J.~M.}\ \bibnamefont {Schwarz}},\
  }\bibfield  {title} {\bibinfo {title} {Learning by non-interfering feedback
  chemical signaling in physical networks},\ }\href@noop {} {\bibfield
  {journal} {\bibinfo  {journal} {Physical Review Research}\ }\textbf {\bibinfo
  {volume} {5}},\ \bibinfo {pages} {023024} (\bibinfo {year}
  {2023})}\BibitemShut {NoStop}%
\bibitem [{\citenamefont {Galeano~Ni{\~n}o}\ \emph {et~al.}(2020)\citenamefont
  {Galeano~Ni{\~n}o}, \citenamefont {Pageon}, \citenamefont {Tay},
  \citenamefont {Colakoglu}, \citenamefont {Kempe}, \citenamefont {Hywood},
  \citenamefont {Mazalo}, \citenamefont {Cremasco}, \citenamefont {Govendir},
  \citenamefont {Dagley} \emph {et~al.}}]{Galeano2020}%
  \BibitemOpen
  \bibfield  {author} {\bibinfo {author} {\bibfnamefont {J.~L.}\ \bibnamefont
  {Galeano~Ni{\~n}o}}, \bibinfo {author} {\bibfnamefont {S.~V.}\ \bibnamefont
  {Pageon}}, \bibinfo {author} {\bibfnamefont {S.~S.}\ \bibnamefont {Tay}},
  \bibinfo {author} {\bibfnamefont {F.}~\bibnamefont {Colakoglu}}, \bibinfo
  {author} {\bibfnamefont {D.}~\bibnamefont {Kempe}}, \bibinfo {author}
  {\bibfnamefont {J.}~\bibnamefont {Hywood}}, \bibinfo {author} {\bibfnamefont
  {J.~K.}\ \bibnamefont {Mazalo}}, \bibinfo {author} {\bibfnamefont
  {J.}~\bibnamefont {Cremasco}}, \bibinfo {author} {\bibfnamefont {M.~A.}\
  \bibnamefont {Govendir}}, \bibinfo {author} {\bibfnamefont {L.~F.}\
  \bibnamefont {Dagley}}, \emph {et~al.},\ }\bibfield  {title} {\bibinfo
  {title} {Cytotoxic t cells swarm by homotypic chemokine signalling},\
  }\href@noop {} {\bibfield  {journal} {\bibinfo  {journal} {Elife}\ }\textbf
  {\bibinfo {volume} {9}},\ \bibinfo {pages} {e56554} (\bibinfo {year}
  {2020})}\BibitemShut {NoStop}%
\bibitem [{\citenamefont {Mukherjee}\ \emph {et~al.}(2022)\citenamefont
  {Mukherjee}, \citenamefont {Chepizhko}, \citenamefont {Lionetti},
  \citenamefont {Zapperi}, \citenamefont {La~Porta},\ and\ \citenamefont
  {Levine}}]{Mukherjee2022}%
  \BibitemOpen
  \bibfield  {author} {\bibinfo {author} {\bibfnamefont {M.}~\bibnamefont
  {Mukherjee}}, \bibinfo {author} {\bibfnamefont {O.}~\bibnamefont
  {Chepizhko}}, \bibinfo {author} {\bibfnamefont {M.~C.}\ \bibnamefont
  {Lionetti}}, \bibinfo {author} {\bibfnamefont {S.}~\bibnamefont {Zapperi}},
  \bibinfo {author} {\bibfnamefont {C.~A.}\ \bibnamefont {La~Porta}},\ and\
  \bibinfo {author} {\bibfnamefont {H.}~\bibnamefont {Levine}},\ }\bibfield
  {title} {\bibinfo {title} {Infiltration of tumor spheroids by activated
  immune cells},\ }\href@noop {} {\bibfield  {journal} {\bibinfo  {journal}
  {Physical Biology}\ } (\bibinfo {year} {2022})}\BibitemShut {NoStop}%
\end{thebibliography}%
\pagebreak

\widetext

\section{Supplemental Material}

\setcounter{secnumdepth}{3}
\setcounter{equation}{0}
\setcounter{figure}{0}  
\setcounter{table}{0}

\makeatletter 
\renewcommand{\thefigure}{S\@arabic\c@figure}
\makeatother

\makeatletter 
\renewcommand{\thetable}{S\@Roman\c@table}
\makeatother

\makeatletter 
\renewcommand{\theequation}{S\@arabic\c@equation}
\makeatother

\begin{figure}[!htbp]
    \centering
    \includegraphics[width=0.45\textwidth]{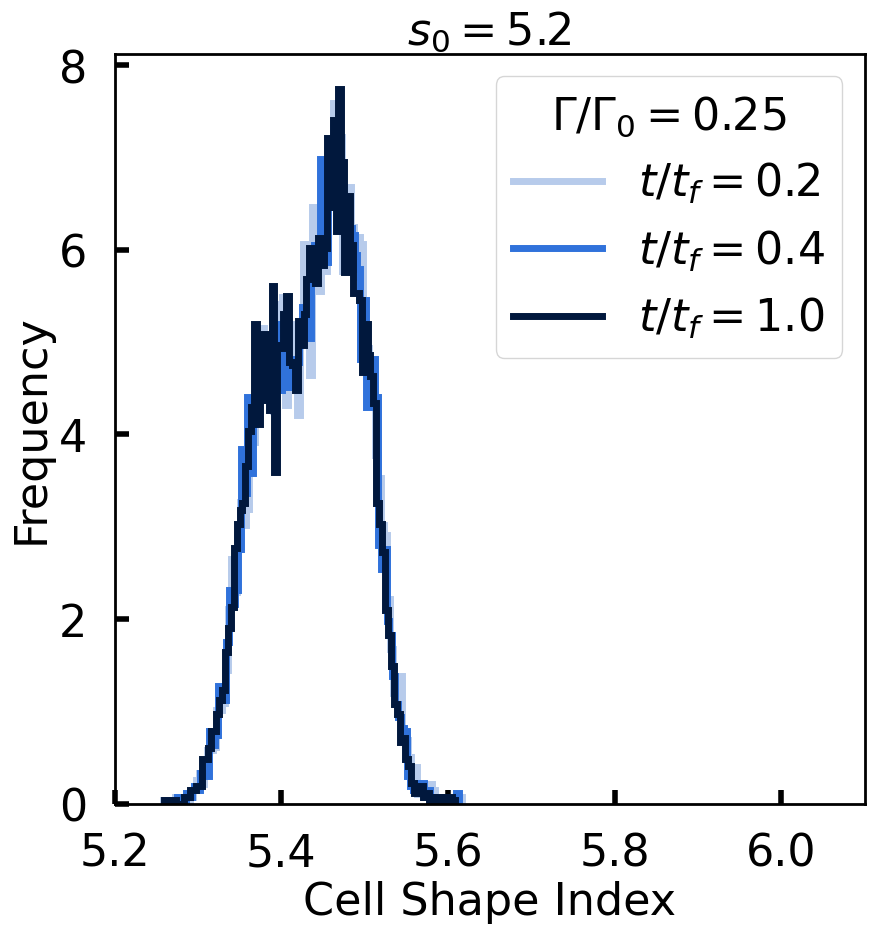}
    \includegraphics[width=0.45\textwidth]{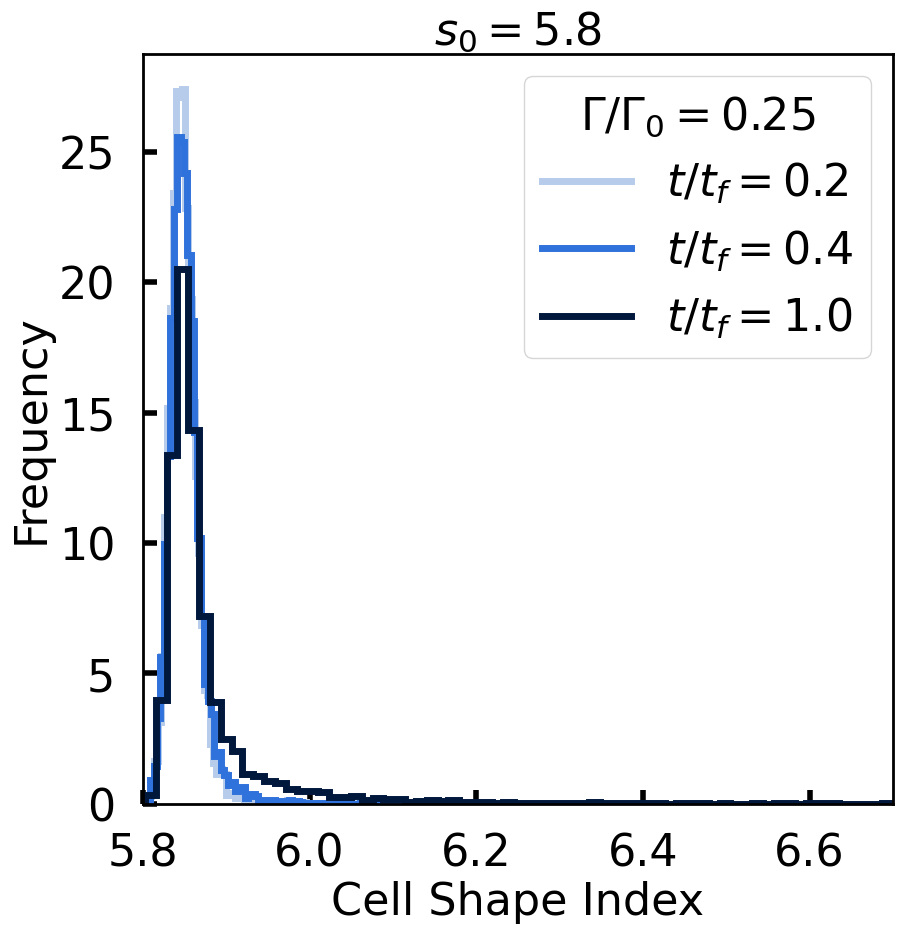}
    \caption{\textit{Individual cell shape index histogram is two-peaked for solid-like spheroids and single-peaked for fluid-like spheroids} Top: Cell shape index histogram for $s_0=5.2$ at different time points in the simulations. Bottom: Cell shape index histogram for $s_0=5.8$ at the same time points in the simulations as above.  } 
    \label{fig:cell_shape_index_distribution_025}
\end{figure}

\begin{figure}[!htbp]
    \centering
    \includegraphics[width=0.45\textwidth]{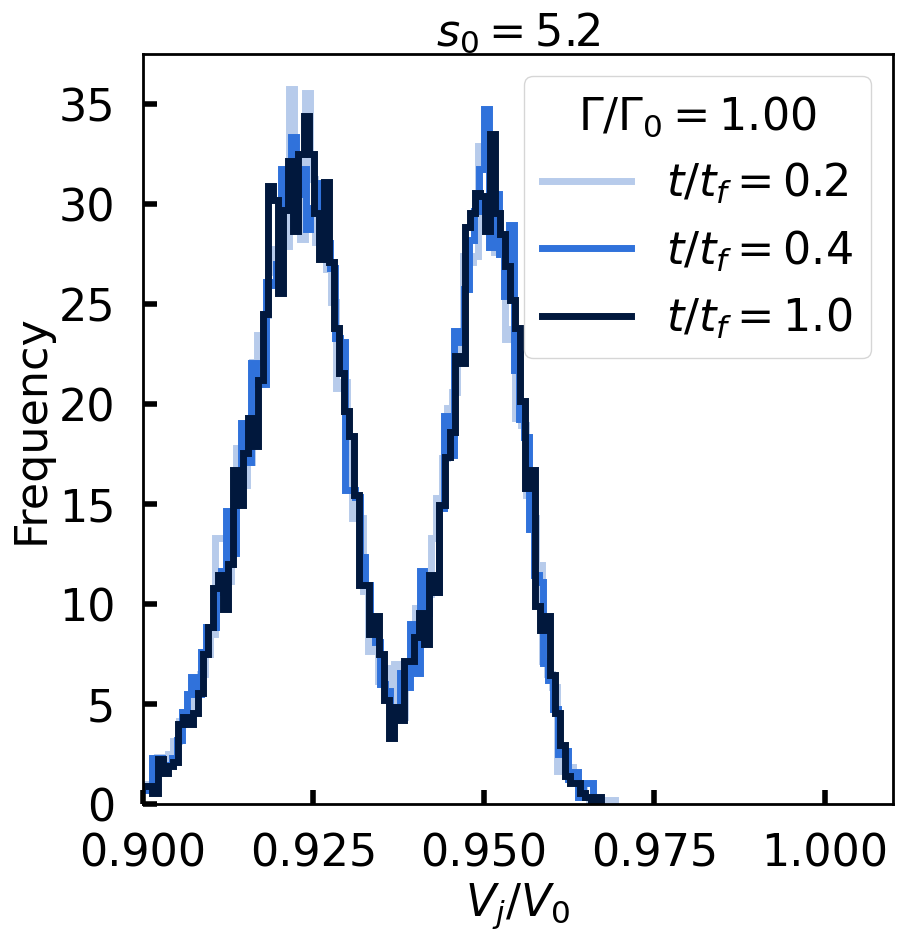}
    \includegraphics[width=0.45\textwidth]{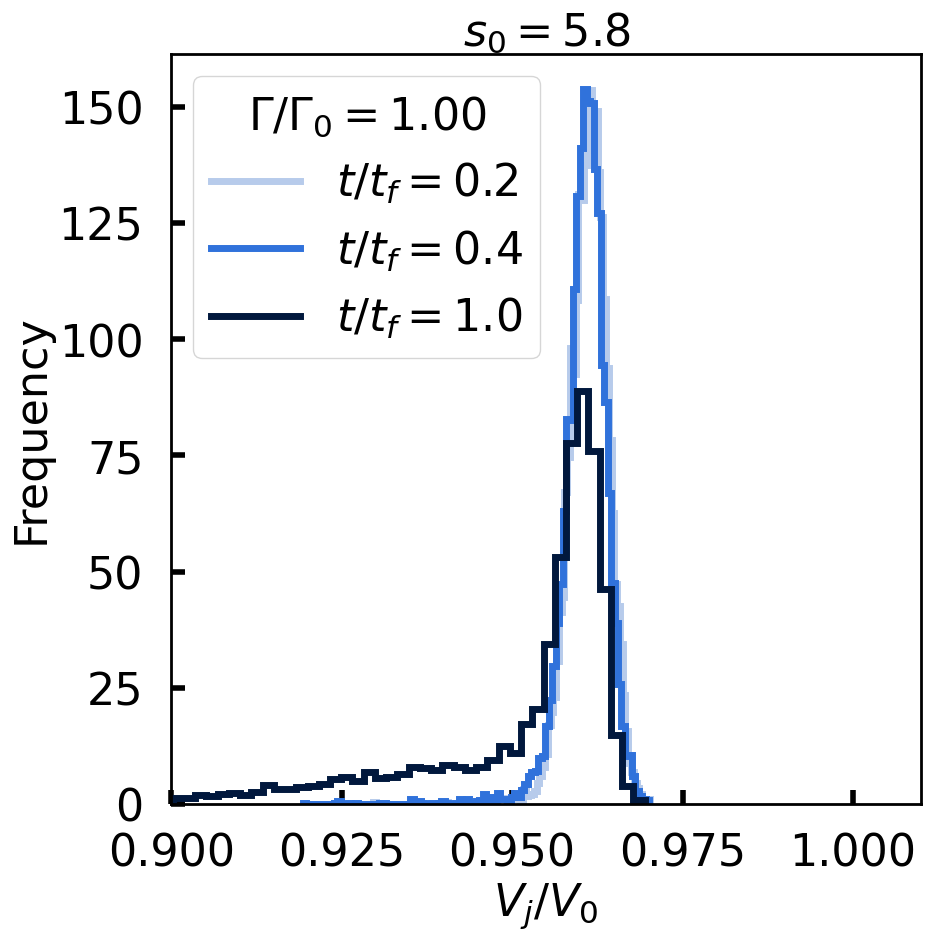}
    \caption{\textit{Individual cell volume histogram is two-peaked for solid-like spheroids and single-peaked for fluid-like spheroids}. Top: Individual cell volume histogram for $s_0=5.2$ at different times throughout the simulation. Bottom: Individual cell volume histogram for $s_0=5.8$ at different times throughout the simulation. Note that the volume of the cells are approximately percent less than their target volume for the fluid-like spheroids. For the solid-like spheroids, the deviation is larger.  } 
    \label{fig:cell_volumes_appendix}
\end{figure}

\begin{figure*}[!htbp]
    \centering
		\begin{tabular}[t]{c c p{0.4\textwidth}}
			
			\raisebox{14mm}{\includegraphics[height=0.27\textwidth]{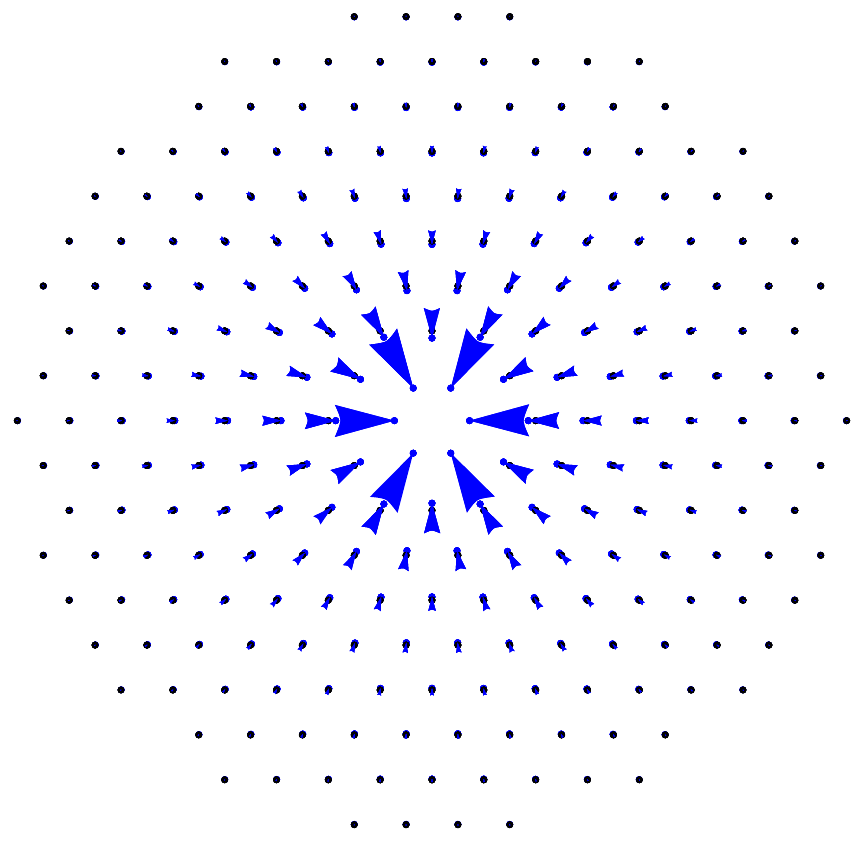}}
			&
			\raisebox{14mm}{\includegraphics[height=0.27\textwidth]{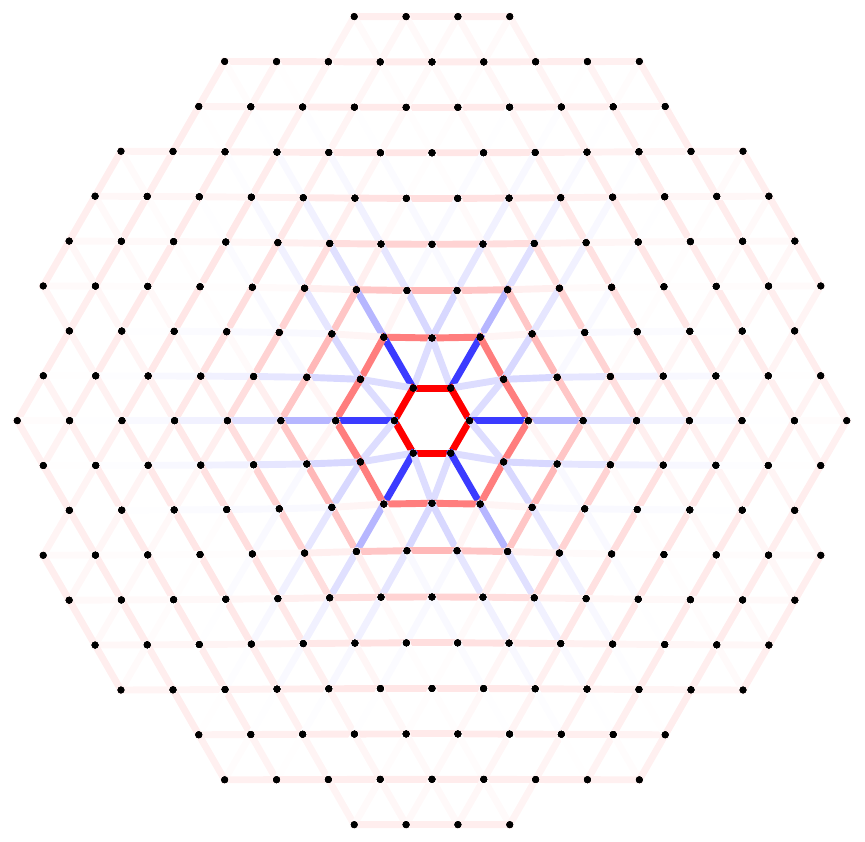}}
            &
            \includegraphics[height=0.4\textwidth] {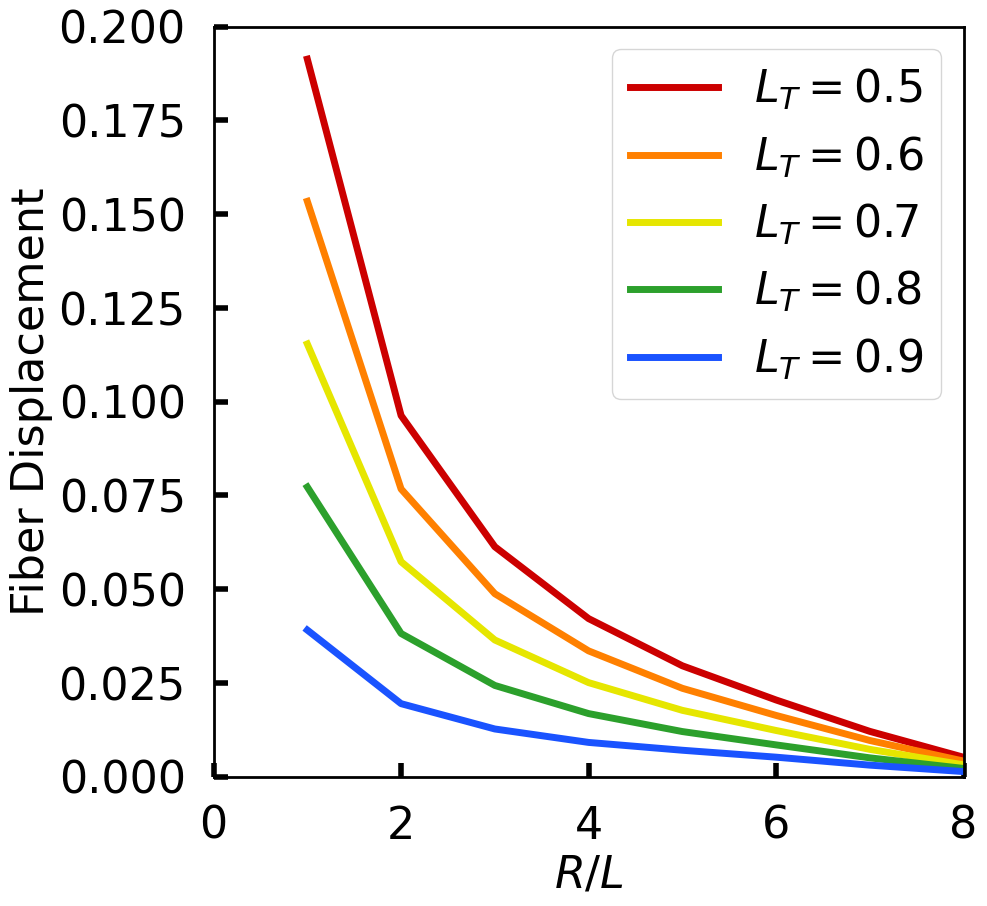}

			  \\
			(a) Vector map&
			(b) Tension map&
            (c) Average displacement of fiber network nodes from the center.\\
		\end{tabular}
    \caption{\textit{Simpler two-dimensional model for fiber displacement with a spheroid represented by six active, contractile linker springs in the center of the lattice}. The energy functional is a two-dimensional version of the fiber network described in the manuscript but with no phantom vertices. The spheroid is idealized as six central active, contractile linker springs with some target length that is smaller than the lattice spacing. The results here are obtained from energy minimization using Mathematica. As for boundary conditions, the outer nodes of the network are fixed. (a) Vector map showing position changes from the initial points for a final target length $L_T$ of 0.5 initial lattice spacing. (b) Tension map where fibers with blue indicating extension and red indicating compression. (c) Average displacement plot from the center based on the Euclidean distance from 1 to 8 initial lattice spacings for different final target spring lengths $L_T$ of the six active linker springs. }
    \label{fig:2d_model}
\end{figure*}

\begin{figure*}[!htbp]
    \centering
    \includegraphics[width=0.457\textwidth]{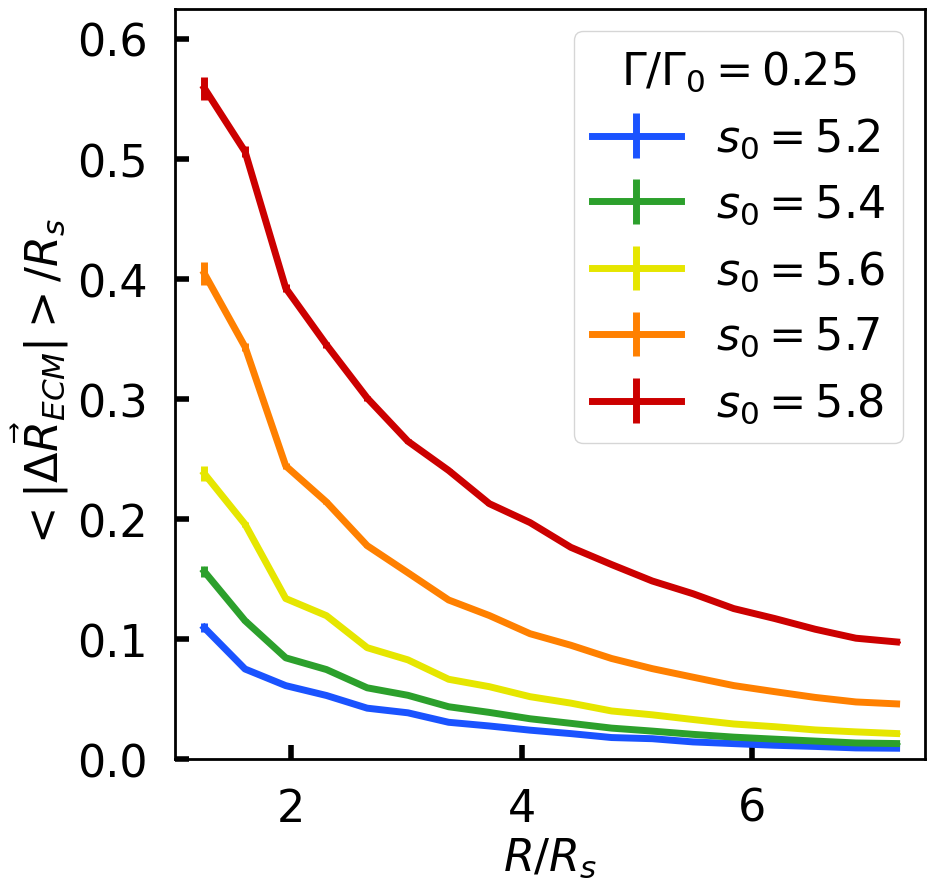}
    \includegraphics[width=0.45\textwidth]{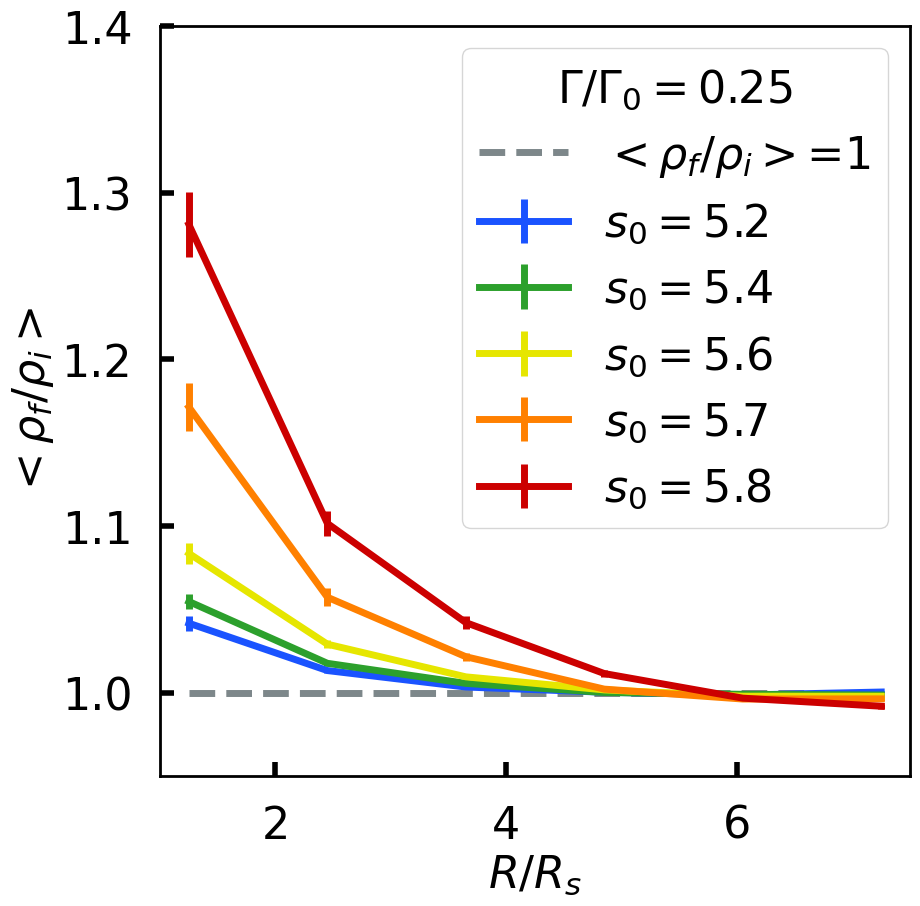}
    \caption{\textit{Fluid-like spheroids displace and densify the fiber network near the spheroid more than solid-like spheroids} Top: Magnitude of fiber displacement as a function of radial distance $R$ from the center of mass of the spheroid. As more fluid-like spheroids are quantified by a larger target cell shape index, $s_0$s, the fiber positions are displaced more the larger the target cell shape index. Bottom: The larger the target cell shape index $s_0$, the more the fibers are densified towards the center of the system. Note that $\rho_i$ denotes the initial fiber density, $\rho_f$ denotes the final fiber density, and $R_s$ denotes the radius of the embedded spheroid system. These results are a dimensionless interfacial surface tension, $\Gamma/\Gamma_0=0.25$.  } 
    \label{fig:fiber_displacement_density_025}
\end{figure*}

\begin{figure}[!htbp]
    \centering
    \includegraphics[width=0.45\textwidth]{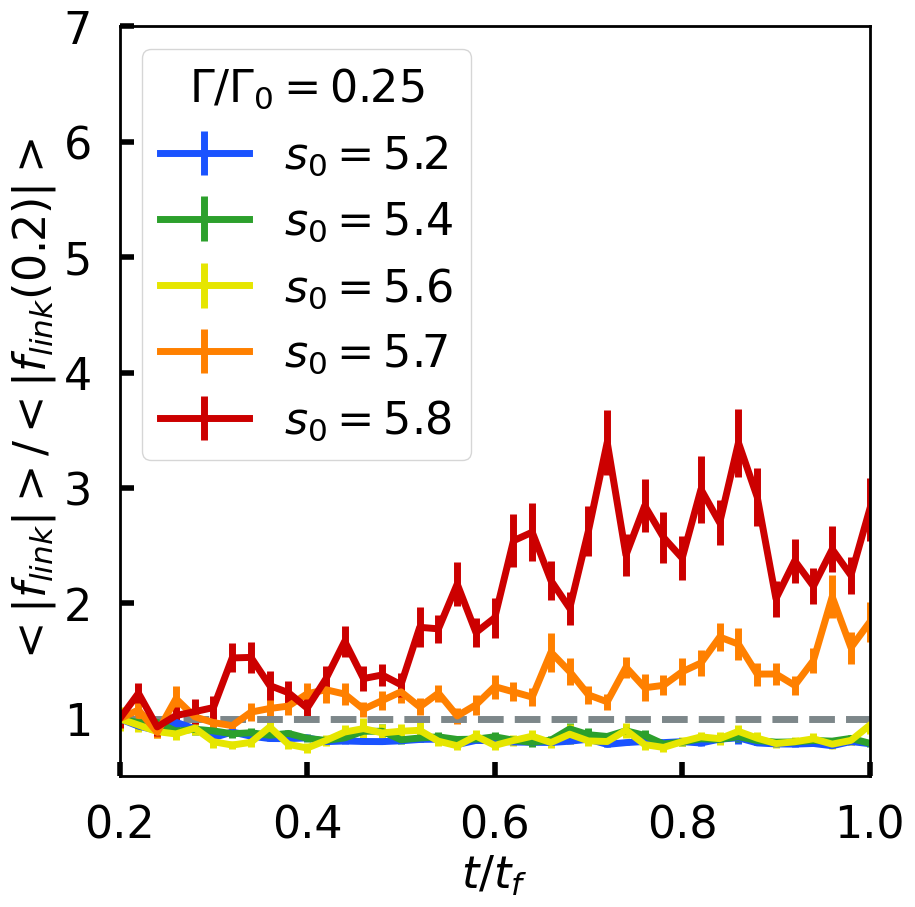}
    \caption{\textit{Average linker spring force increases at later times more significantly for fluid-like spheroids as compared to solid-like spheroids}: Plot of the average linker spring force as a function of time for different $s_0$s.  } 
    \label{fig:average_linker_spring_force_025}
\end{figure}

\begin{figure}[!htbp]
    \centering
    \includegraphics[width=0.45\textwidth]{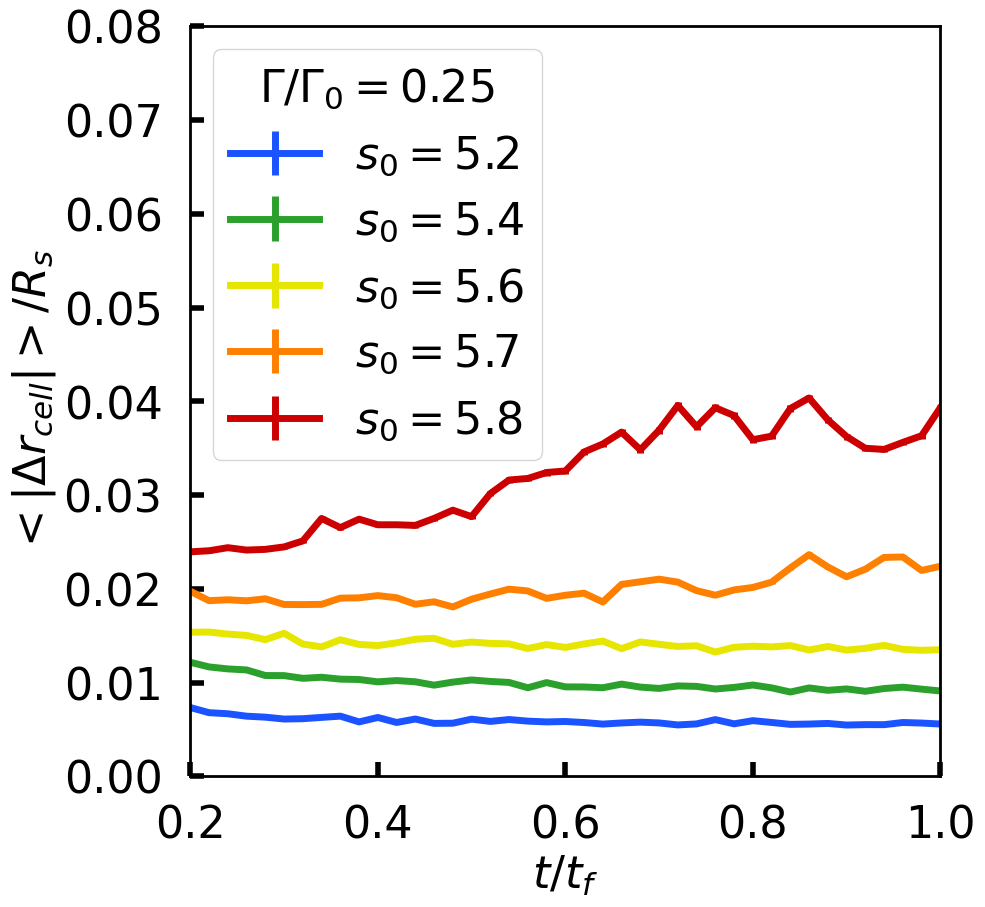}
    \caption{\textit{Cells displace more in the fluid-like spheroids than in the solid-like spheroids with the difference accentuated for $t/t_f>0.6$}. The magnitude of the cell displacement $|\Delta r_{cell}|$ over a fixed time window as a function of time for different values of the target shape index. Here, $\Gamma/\Gamma_0=0.25$ and should be compared with Figure 9, where $\Gamma/\Gamma_0=1$. Note that the increase in cell displacement for the $s_0=5.8$ spheroids is slightly larger for the $\Gamma/\Gamma_0=1$, as compared to the case plotted here.} 
    \label{fig:cell_displacements_appendix}
\end{figure}

\begin{figure}[!htbp]
    \centering
    \includegraphics[width=0.45\textwidth]{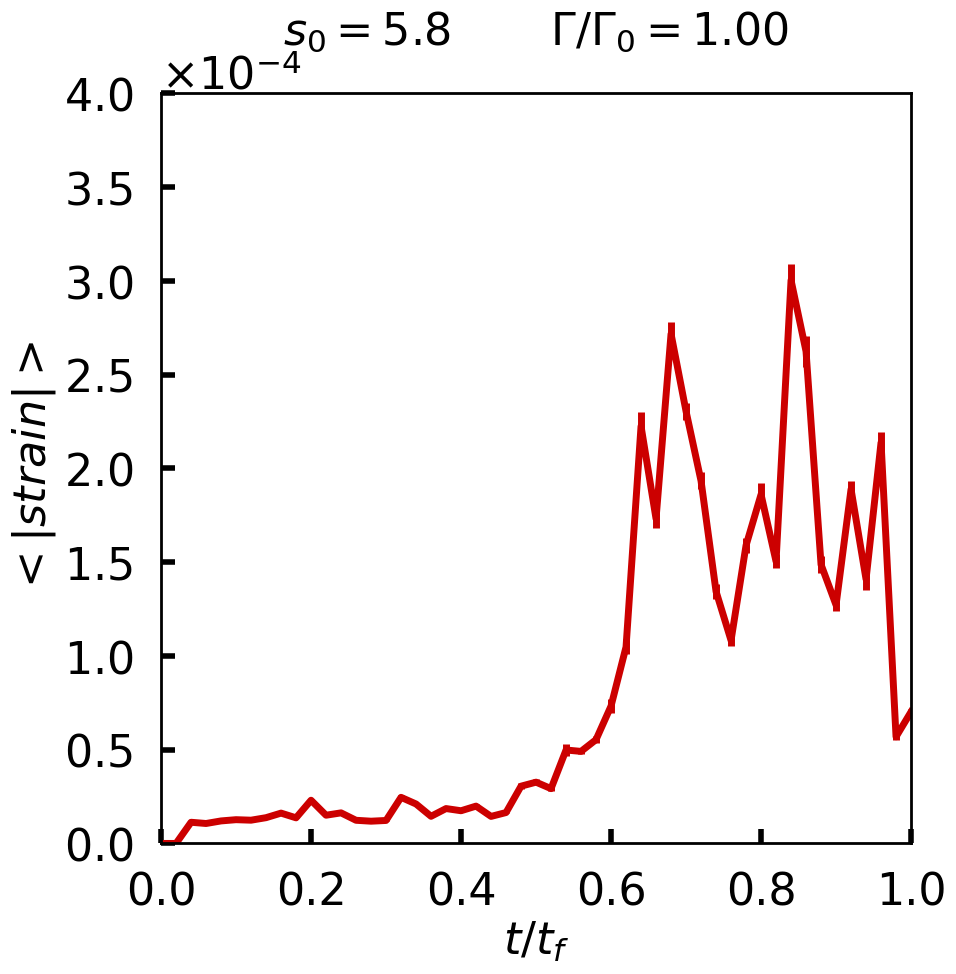}
    \includegraphics[width=0.45\textwidth]{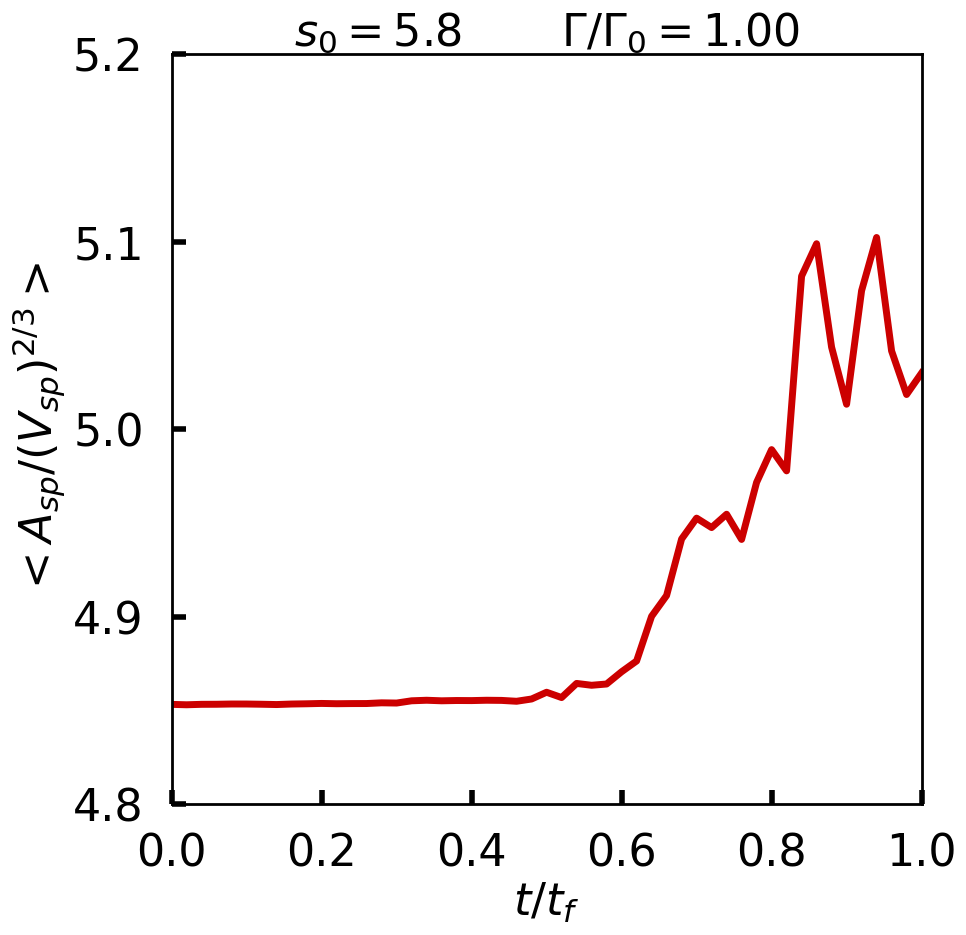}
    \caption{\textit{Fluctuations in the absolute average strain of the fiber network that correlate with spheroid shape changes}. Left: The absolute value of the average strain in the fiber network, denoted as $<|strain|>$ over time, for one realization. Right: The spheroid shape index over time for the same realization. Note the increase and decrease and subsequent increase in the total strain in the fiber network. A more sophisticated model with explicit mechanical feedback between the spheroid and the fiber network and/or binding and unbinding of the active linker springs may readily lead to the oscillatory twitching phenomenon reported in Ref.~\cite{Mark2020}.} 
    \label{fig:spheroid_shape_index_appendix}
\end{figure}

\end{document}